\documentclass[preprintnumbers,amsmath,amssymb,floatfix,10pt,prd,onecolumn,superscriptaddress,nofootinbib]{revtex4}
\usepackage{latexsym}
\usepackage{epsfig}
\usepackage{epstopdf}
\usepackage{graphicx}
\usepackage{amssymb}
\usepackage{amsmath}
\usepackage{amsfonts}
\usepackage{subfigure}
\usepackage{dcolumn}
\usepackage{bm}
\usepackage{color}
\usepackage{comment}
\usepackage[utf8]{inputenc}
\usepackage[dvipsnames]{xcolor}
\usepackage{hyperref}
\hypersetup{colorlinks=true,linkcolor=blue,citecolor=red,urlcolor=magenta}

\begin{document}
\title{\bf Rotating Black Hole in Kalb-Ramond Gravity: Constraining Parameters by Comparison with EHT Observations of Sgr A* and M87*}
\author{M. Zubair}
\email{mzubairkk@gmail.com;drmzubair@cuilahore.edu.pk}\affiliation{Department of Mathematics, COMSATS University Islamabad, Lahore Campus, Lahore, Pakistan}
\author{Muhammad Ali Raza}
\email{maliraza01234@gmail.com}\affiliation{Department of Mathematics, COMSATS University Islamabad, Lahore Campus, Lahore, Pakistan}
\author{Eiman Maqsood}
\email{eiman1482000@gmail.com}\affiliation{Department of Mathematics, COMSATS University Islamabad, Lahore Campus, Lahore, Pakistan}

\begin{abstract}
This article deals with the study of some properties of the static and rotating black holes in Kalb-Ramond gravity in four dimensional spacetime. First, we discuss the action of the corresponding theory and the static black hole metric. Then we investigate the light sphere for the static black hole by using the Hamiltonian formalism and the corresponding linear radius of the shadow, angular velocity and Lyapunov exponent. For the rotating black hole, we discuss the horizon structure. Moreover, we study the effective potential to discuss the structure of null sphere and unstable circular null orbits around the rotating black hole. The properties such as energy emission rate and distortion are calculated and analyzed by using the numerical data for the shadows calculated by appropriately chosen parametric values for two different angular locations of the observer off the equatorial plane. We also obtain the constraints on the black hole parameters by comparing the shadow sizes of the black hole in Kalb-Ramond gravity and the supermassive black holes M87* and Sgr A*. Finally, we investigate the effect of mass, energy, angular momentum and the black hole parameters on the center of mass energy of two colliding particles that are accelerated in the vicinity of the black hole.
\end{abstract}
\maketitle
\date{\today}

\section{Introduction}
It is believed that the year $1915$ is one the most important years in terms of scientific revolution. In this year, Albert Einstein extended and generalized the Special Theory of Relativity and developed a theory of gravity known as General Relativity (GR). According to GR, the gravitational interaction causing the gravitational effect is a consequence of the warping of spacetime around a massive object. Several objections were presented against GR, however, mathematically, it remained a well-defined theory. To counter those objections, the weak \cite{1} and strong field tests have been presented \cite{2,3,4}. A black hole (BH) is a strong field region in the space with an immense gravity that can swallow everything; even the light will never come out if it falls into a BH. In the gravitational field of a BH, a particle can move in curved path. The field strength of a BH is so strong that it can deflect the trajectory of a photon as well. It was first discovered by a group of astronomers led by Arthur Eddington \cite{5} and consequently, the observations verified GR. These observations also laid the foundation of the subject known as gravitational lensing which has been comprehensively studied over the years \cite{6,7,8,9,10,11,12,13}.

Usually, the capturing of the photons in our eye gives the image of a certain object. Contrarily, the disappearance of the photons from our sight is responsible for the possible visual appearance of the BH because there is no direct method to see or observe a BH itself. Therefore, we measure the inner boundary of the photon sphere to estimate the image of a BH. A photon sphere is the region around a BH in which photons are trapped that keep orbiting the BH. These photons will remain in the photon sphere unless the orbits are unstable. The photons in the unstable orbits will either fall into the event horizon or will travel to infinity. The falling in photons being absent from our sight will give a dark image called shadow. The boundary of the shadow is determined from the interface of the falling in and scattering trajectories \cite{14}. The photons are not bound to move in the same direction around a BH. Instead, they can move in any arbitrary direction. If we assume two photons moving in opposite directions around a rotating BH, then the trajectory along the direction of the spin of the BH can be considered at smaller radius from the origin. Whereas, the trajectory along the opposite direction of the spin of the BH can be considered at larger radius from the origin. This gives rise to the flattened shadow on one side which is quantitatively studied in terms of distortion. Kerr BH with the maximum value of spin parameter gives a flat silhouette on one side \cite{15}.

The visual appearance of a BH being a fundamental question, is widely investigated over the past few decades. Apart from the basic BHs such as Schwarzschild, Kerr, Reissner-Nordstr\"{o}m and Kerr-Newman BHs, various complicated BHs have been taken into account to calculate their visual image, see Refs. \cite{16,17,18,19,20,21,22,23,24,25,26,27,27a,27b,27c,27d}. Sharif and Iftikhar \cite{16} considered the noncommutative extension of the charged rotating BH and found that by decreasing the noncommutative charge, the shadow is observed different from the circular image. Lee et al. \cite{19} worked for the shadow of rotating BH in anisotropic matter and found that the anisotropy has a significant impact on shadow observables. In another study, Amarilla and Eiroa \cite{23} focused on a braneworld BH in Randall-Sundrum model. They discussed the shadow of the BH and proposed that the term with tidal charge and the angular momentum distort the shadow. It also shows that the shadow size grows with negative tidal charge, however, the deformation is diminished as compared to Kerr BH and a converse result is observed for the positive tidal charge. Khodadi \cite{27b} investigated the shadow of rotating asymptotically flat BH surrounded by magnetized axion-plasmon cloud. It is found that the fixed axion-plasmon medium highly influences the shape and size of the shadow as compared to that for non-magnetized plasma and vacuum cases. Parbin et al. \cite{27c} investigated the influence of axionic parameter on the shadow of slowly rotating BH in Chern-Simons modified theory of gravity. It is found that the shadow size and the distortion in shadow increase as the spin parameter increases. Moreover, the Chern-Simons coupling parameter behaves contrarily to the spin parameter. The distortion in the shadow is diminished as the Chern-Simons parameter increases. Karmakar et al. \cite{27d} considered a Schwarzschild-like BH which is corrected by the generalized uncertainty principle with topological defects in the framework of Bumblebee gravity. They investigated the shadow of the BH and found that the shadow is unaffected by the Lorentz symmetry violation. It is also worth mentioning some of the important studies on BH shadows along with other aspects such as deflection angle and quasinormal modes etc., given in Refs. \cite{27e,27f,27g,27h,27i,27j,27k,27l,27m}. \"{O}vg\"{u}n et al. \cite{27e} studied the shadow and deflection angle of Kerr-Newman-Kasuya BH. In another study, Junior et al. \cite{27h} obtained the conditions for two BHs having same shadow for both static and rotating BH cases. Okyay and \"{O}vg\"{u}n \cite{27i} investigated the effects of nonlinear electrodynamics field on the shadow, deflection angle, quasinormal modes and greybody factors of a static BH.

Another fundamental question is related to the realistic and physical existence of the BH in our universe for which efforts have been made for the physical detection of such a compact object. In 1999, Falke et al. \cite{27n} showed that by using very long-baseline interferometry at sub-millimeter wavelengths, it is possible to observe the shadow image of Sgr A* BH. Therefore, they proposed the visualization of the event horizon of a BH in the future. Recently in 2015, the gravitational waves were detected by Laser Interferometer Gravitational Wave Observatory (LIGO) \cite{28}. These waves were emitted by the collision of two BHs and remains a pioneering discovery in the field of Astronomy. Later in 2019, the first ever direct visualization of a BH in terms of shadow was captured by the Event Horizon Telescope (EHT) \cite{29,30,31,32,33,34}. This BH is known as M87* and is residing at the center of Messier 87 galaxy. Two years later, in 2021, a polarization image of M87* was shown by the EHT. It suggests that the BH comprises the magnetic field and it became possible to explain the emerging jets from the BH \cite{35,36}. Finally in 2022, the EHT captured the image of Sagittarius A* (Sgr A*) BH located at the center of our Milky Way galaxy \cite{37}. As we know, the images taken under the said observations are actually the shadow images and that in the neighborhood of the BH horizon, give some worthwhile information about the BH mass, spin and the spacetime structure around the BH \cite{38}. Using this data related to the observations of EHT, various studies have been accomplished to compare the shadow size of the theoretical and the physically observed BHs, especially M87* and Sgr A*. In these studies, parameters of the theoretical BHs are restricted in order to obtain the size of the shadow equal to that of M87* and Sgr A*. Under this analysis, it can be proposed that the theoretical BH might be one of either M87* or Sgr A*, see for example \cite{39,40,41,42,43,44}.

Soon after the pioneering discovery of the image of M87* BH by EHT collaboration, Bambi et al. \cite{44a} tested the rotational behavior of M87* BH. They obtained the constraints on the dimensionless spin parameter by assuming M87* BH as a Kerr BH and a superspinar, where a superspinar is an object that is described by Kerr metric but rotates so fast such that it violates Kerr BH's spin bound. As spin is an important parameter along with mass in the no-hair theorem. Therefore, testing the no-hair theorem using the observations have also been studied in Refs. \cite{44l,44e}. Using the EHT observations and their results, various studies have been accomplished to test the gravity theories and BH solutions in addition to the constraining of the BH parameters and the comparison of the shadow results from the EHT observations for M87* and Sgr A* BHs \cite{44b,44c,44d,44f,44g,44h,44i,44j,44k}. Allahyari et al. \cite{44b} investigated the shadows of Einstein-Euler-Heisenberg BH and Einstein-Bronnikov BH and compared the results with the EHT data obtained for M87* BH. They obtained some novel constraints on magnetic charge parameter. Khodadi et al. \cite{44c} considered a minimally coupled charged BH with scalar hair and a conformally coupled charged BH with scalar hair to discuss their shadows and obtained the constraints on scalar hair by comparing the BH shadows with the shadow of M87* BH. The bounds on hairy parameter for conformally coupled BH are in agreement with the EHT data. Banerjee et al. \cite{44h} investigated the shadows of an ultra-compact object on brane and a braneworld wormhole and presented an analysis on the comparison of their results with EHT data for Sgr A* to answer the existence of extra dimensions. Afrin et al. \cite{44j} exploited the shadow of Sgr A* in constraining the rotating loop quantum BHs and found that with the increase in quantum effects, the size of the shadow and distortion increases that allows to obtain the bound on the fundamental loop quantum parameter.

The technological advancement enabled us to perform experiments at a high resolution and precision to obtain the observations for testing GR \cite{45,46}. It further enabled us to modify GR and to construct alternative gravitational theories to meet the scientific necessities. To modify GR, we need to modify the Einstein action and one way to do this is to introduce the Kalb-Ramond (KR) field \cite{47}. KR field is a quantum field ascribed as the closed string excitation that is related to the heterotic string theory \cite{48} and transforms as a two-form. It is observed that the spontaneous Lorentz invariance is violated due to the existence of nonminimal coupling between the gravity sector and non-zero expectation value of the tensor field in vacuum \cite{49}. The effect of spontaneous Lorentz symmetry violation has also been studied by Khodadi \cite{49a} on superradiance scattering and instability for a rotating BH in Einstein-bumblebee gravity. He found that at low frequencies of scalar wave, the superradiance scattering will get enhanced for the negative Lorentz-violating parameter and will get weakened for the positive values. An improved limit for the instability regime has also been obtained. In another study, Khodadi \cite{49b} considered a fastly rotating BH being different from Kerr BH due to the Lorentz symmetry violation parameter and studied the energy extraction caused by magnetic reconnection in the ergosphere. It is found that the negative Lorentz symmetry violation parameter favors the extraction of energy via magnetic reconnection for the given BH immersed in weakly magnetized plasma. Khodadi et al. \cite{49c} considered a slowly rotating BH modified by Lorentz invariance violation. In the equatorial plane of the BH, they studied the energy deposition rate for gamma-rays produced by the annihilation of the pairs of neutrinos. They found that for positive Lorentz symmetry violation parameter, the energy deposition rate is enhanced. Lorentz symmetry violation has also been studied in the cosmological perspective in Refs. \cite{49d,49e}. Several interesting features have been inferred for the presence of the KR field such as the antisymmetric tensor of rank $3$ that can be derived and is expounded as a source of torsion of spacetime \cite{50}, intrinsic angular momentum of the objects and various structures in far galaxies arising from the topological defects \cite{51}. The gravitational aspects of the KR field and its influence on particles have been studied in \cite{52,53,54,55}. Kumar et al. \cite{55} have also studied the optical image and strong lensing for the rotating BH in KR gravity.

As mentioned above, the field strength of KR gravity may influence the particles. Therefore, it is interesting to study the motion and collision of particles near the BHs in KR gravity. As we are discussing the gravitational aspects of the KR gravity, so it is interesting to discuss the collision of particles around the BH in KR gravity. As a result of collision, energy is released that can be calculated in terms of center of mass energy (CME) because the particles under motion possess momentum and hence the energy. Such BHs are termed as particle accelerators. Various studies are available in this context, see \cite{44,56,57,58,59,60,61,62}.

This motivates us to study various properties of the BHs in KR gravity such as the shadow and related physical observables, parametric bounds arising from the comparison of the shadow size with the physical BHs and CME of colliding particles. The paper is presented as follows: In the following section we will present a brief discussion on the action of the theory, static BH in KR gravity, null sphere, shadow radius, angular velocity and Lyapunov exponent. The third section will comprise a discussion on the rotating BH and the horizon structure. We will also present an analysis on effective potential and will calculate shadows for certain parametric values that will be used to investigate the distortion and energy emission rate. In fourth section, the comparison of the shadows with EHT data will be given to obtain the parametric bounds. In fifth section, the CME of two colliding particles will be presented. The paper will be concluded in the sixth section. Moreover, in our calculations, we will use $G=c=M=1$, where $G$, $c$ and $M$ are Newton's constant, cosmic speed limit and BH mass.

\section{The Static Black Hole in Kalb-Ramond Gravity}
We begin with the discussion on the action of KR gravity. The nonminimal coupling of the gravity with a self-interacting KR field is given as \cite{49,63}
\begin{eqnarray}
S=\int\sqrt{-g}d^4x\bigg[\frac{R}{16\pi G}-\frac{1}{12}H_{\alpha\mu\nu}H^{\alpha\mu\nu}-V\big(B_{\mu\nu}B^{\mu\nu}\pm b_{\mu\nu}b^{\mu\nu}\big)+\frac{1}{16\pi G}\big(\xi_2B^{\mu\lambda}B^\nu_\lambda R_{\mu\nu}+\xi_3B_{\mu\nu}B^{\mu\nu}R\big)\bigg], \label{1}
\end{eqnarray}
where $g$ is the determinant of the metric tensor and $R$ is Ricci curvature scalar. Moreover, the tensor field $B_{\mu\nu}$ defines the KR field that drives the Lorentz invariance violation such that $B_{\mu\nu}=-B_{\nu\mu}$ with the expectation value in vacuum to be $<B_{\mu\nu}>=b_{\mu\nu}\neq0$ \cite{47}. With such an expectation value, it is possible to decompose the tensor field $B_{\mu\nu}$ into a timelike vector and two spacelike vectors, having the resemblance with the decomposition of electromagnetic field tensor $F_{\mu\nu}$ \cite{49}. The tensor $H_{\mu\nu\gamma}=\partial_{\mu[}B_{\nu\gamma]}$ is an antisymmetric tensor that can be written in such a way that it provides an analogy with the electromagnetic field tensor $F_{\mu\nu}$ and the tensor $B_{\mu\nu}$ provides an analogy with the vector potential. Thus the KR action can be constructed in analogical way to the electrodynamics. The potential $V$ is related to the above mentioned expectation value of the tensor field $B_{\mu\nu}$. The symbols $\xi_2$ and $\xi_3$ denote the nonminimal coupling constants. In Ref. \cite{63}, in order to study the influence of KR vacuum expectation value on the gravitational field, the authors considered the KR field in vacuum, i.e., $B_{\mu\nu}B^{\mu\nu}=b_{\mu\nu}b^{\mu\nu}$. The Lorentz violating vacuum expectation value $b_{\mu\nu}$ is constant in a flat spacetime, i.e., $\partial_\sigma b_{\mu\nu}=0$, admitting a constant norm $j^2=\eta^{\alpha\beta}\eta^{\mu\nu}b_{\alpha\mu}b_{\beta\nu}$. Moreover, the KR field strength vanishes under the influence of a constant vacuum expectation value \cite{49}. Therefore, the KR vacuum expectation value $b_{\mu\nu}$ is considered as constant tensor with vanishing Hamiltonian. Analogically, in the curved spacetime, the vacuum expectation value can be considered constant, i.e., $\nabla_\sigma b_{\mu\nu}=0$, ensuring the vanishing of KR field strength and the Hamiltonian. In deriving the BH solution in KR gravity, it is assumed that the KR vacuum expectation value $b_{\mu\nu}$ has a constant norm and a vanishing Hamiltonian. The modified gravitational field equations are
\begin{eqnarray}
R_{\mu\nu}-\frac{1}{2}Rg_{\mu\nu}=\kappa T_{\mu\nu}^{\xi_2}, \label{2}
\end{eqnarray}
where $R_{\mu\nu}$ is the Ricci tensor, $T_{\mu\nu}^{\xi_2}$ is the energy-momentum tensor and the four dimensional static spacetime in symmetric and spherical geometry is written as
\begin{eqnarray}
ds^2=-f(r)dt^2 +\frac{dr^2}{g(r)}+r^2\big(d\theta^2+\sin^2\theta d\phi^2\big). \label{3}
\end{eqnarray}
The ansatz for KR vacuum expectation value can be written as
\begin{eqnarray}
b_2=-E(r)dt\wedge dr, \label{4}
\end{eqnarray}
where $b_{tr}=-E$. As mentioned before, $b^2=g^{\alpha\beta}g^{\mu\nu}b_{\alpha\mu}b_{\beta\nu}$, the norm of ansatz for KR vacuum expectation value is a constant $b^2$ with metric (\ref{3}) such that
\begin{eqnarray}
E(r)=|b|\sqrt{\frac{f(r)}{2g(r)}}, \label{5}
\end{eqnarray}
where $b$ is a constant. It must be noted that the function $E(r)$ in Eq. (\ref{5}) describes a static pseudo-electric field in radial direction, i.e., $E^\mu=(0,E,0,0)$. Then expanding the gravitational field equations (\ref{2}) for the spacetime metric (\ref{3}) and solving for the metric functions we get $f(r)=g(r)$ such that
\begin{equation}
f(r)=1-\frac{2M}{r}+\frac{\gamma}{r^{\frac{2}{\lambda}}}. \label{6}
\end{equation}
Therefore, the BH metric in KR gravity can be written as
\begin{eqnarray}
ds^2=-\bigg(1-\frac{2M}{r}+\frac{\gamma}{r^{\frac{2}{\lambda}}}\bigg)dt^2 +\frac{dr^2}{1-\frac{2M}{r}+\frac{\gamma}{r^{\frac{2}{\lambda}}}}+r^2\big(d\theta^2+\sin^2\theta d\phi^2\big), \label{7}
\end{eqnarray}
where $\gamma$ and $\lambda$ are the associated parameters of spontaneous Lorentz symmetry breaking that are related to the expectation value in vacuum in KR field and nonminimal coupling parameters, specifically, $\lambda=|b|^2\xi_2$ such that $b^2=b_{\mu\nu}b^{\mu\nu}$, whereas $\gamma$ is an integration constant. It may be noted that in the currently appearing units, $M$ has dimension of $[length]$, whereas $\gamma$ admits the dimensions of $[length]^{\frac{2}{\lambda}}$. The above static BH solution can be coined as power-law hairy BH solution that reduces to Schwarzschild metric when $\lambda\rightarrow0$, i.e., for $|b|^2=0$ or $\xi_2=0$ for a given value of $\gamma$. In this way, for a non-zero finite value of $\lambda$, one can study BH in KR gravity and the relation between vacuum expectation value and the coupling constant $\xi_2$. Since the Lorentz symmetry violation has very minor effects on the gravitational field, therefore the coupling constant must be small. However, it may be expected that the Lorentz symmetry violation may occur close to the Planck scale. Therefore, the vacuum expectation value $b^2$ that causes the couplings of Lorentz symmetry violation, can also be expected of the order of Planck scale. Consequently, the violation of Lorentz invariance spontaneously gives rise to the configurations described by a large vacuum expectation value and a small coupling constant. Moreover, the BH metric (\ref{7}) reduces to Schwarzschild dS and Reissner-Nordstr\"{o}m solution accordingly as we consider $\lambda=-1$ and $\lambda=1$, respectively. In our work, we will consider $\lambda>0$ because $\lambda\leq0$ gives asymptotically non-flat metric.

\subsection{Radius of Photon Sphere}
\begin{figure}[t]
\centering
\subfigure{
\includegraphics[width=0.45\textwidth]{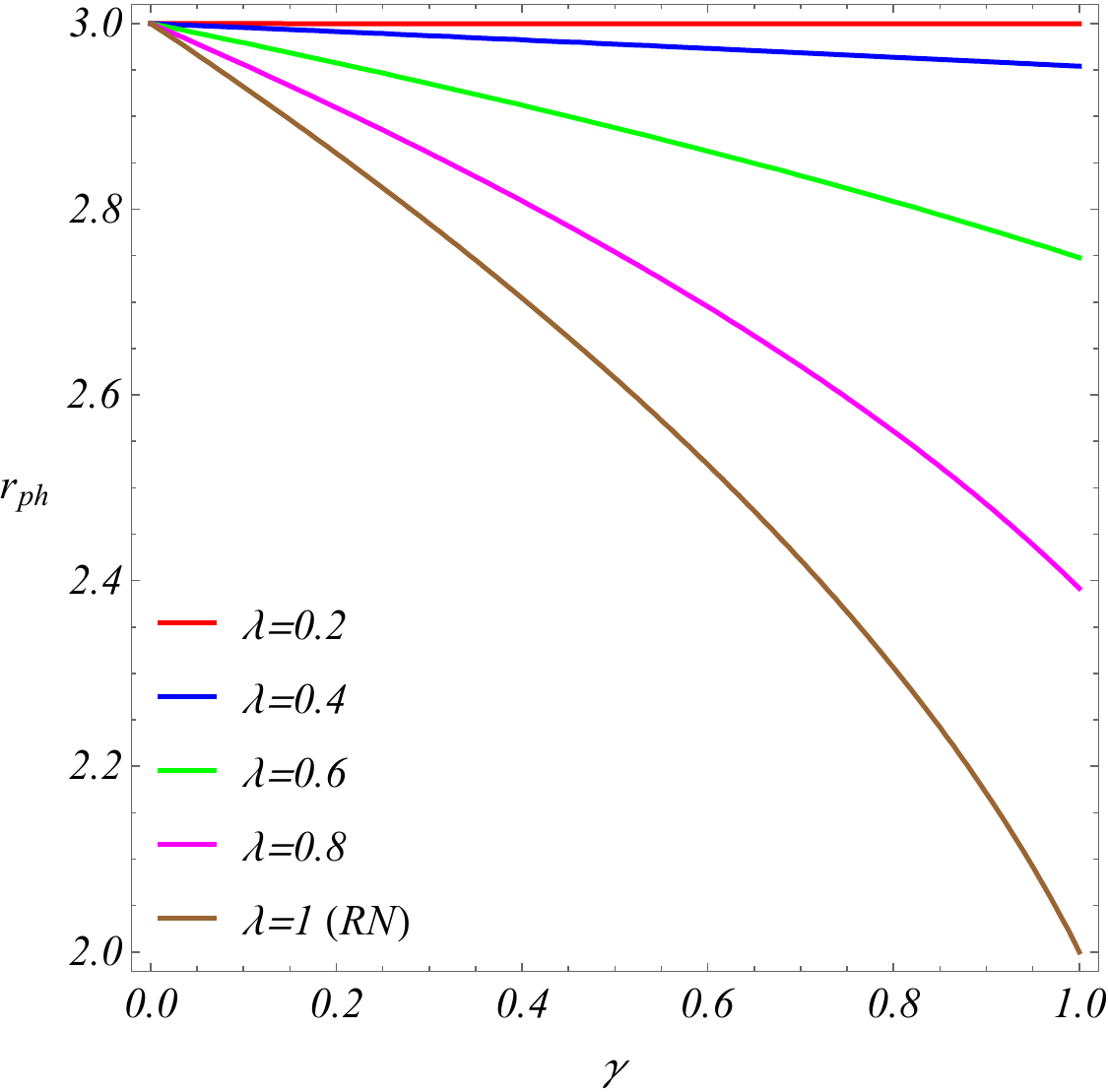}}~~~~~
\subfigure{
\includegraphics[width=0.45\textwidth]{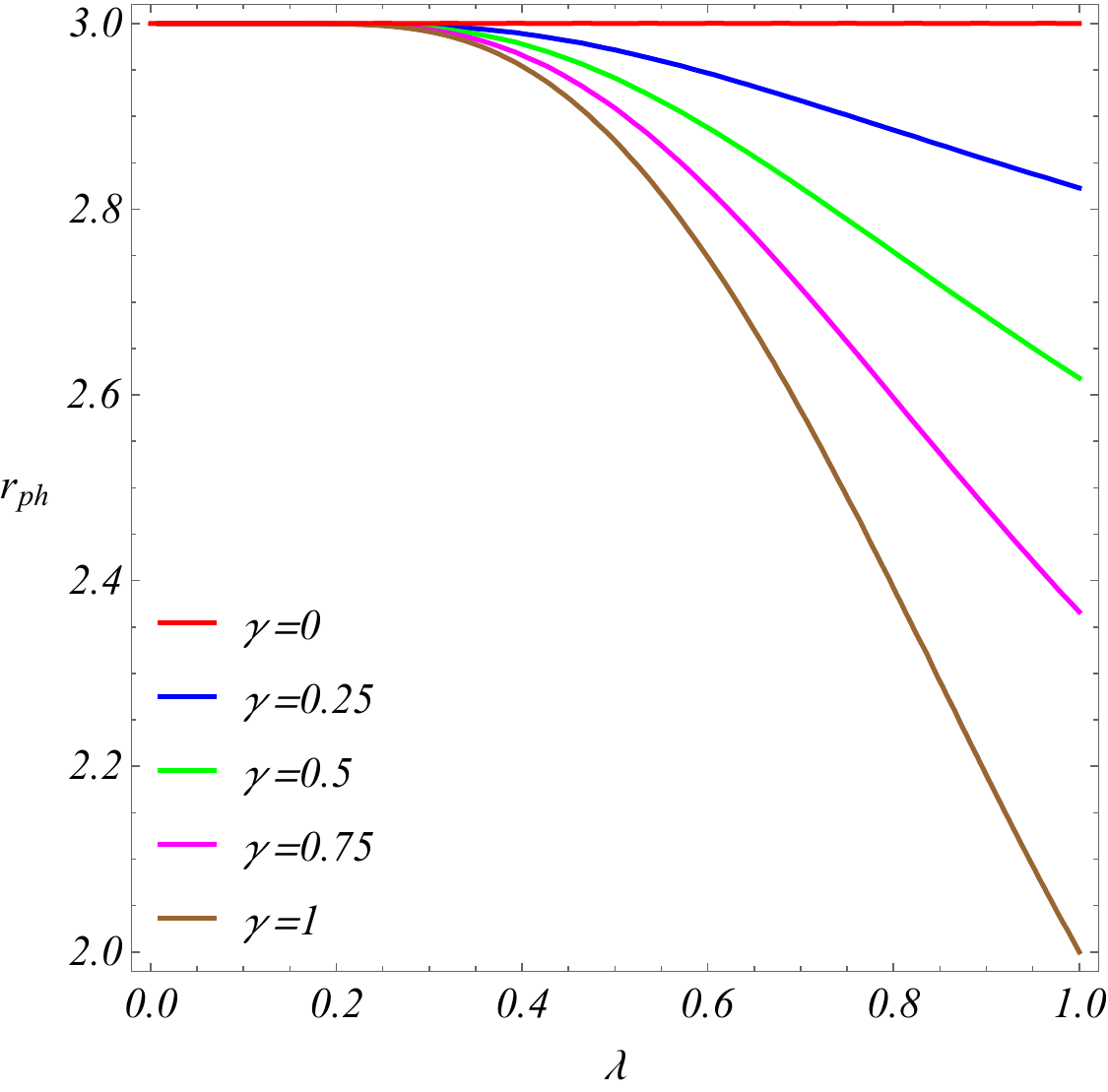}}
\caption{Plots showing the behaviour of photon sphere's radius for different values of $\gamma$ and $\lambda$ for the case of static BH.} \label{RPS}
\end{figure}

Let us consider the photons are emitted from a source and move towards the BH assuming an ideal case such that there is no other source between the observer and the BH. The photons moving close enough to the event horizon get trapped into the circular orbits forming a sphere of null trajectories. We study the size of this sphere formed by the trapped photons near the event horizon. The size of null sphere is characterized by the radius or diameter of the sphere, however, we calculate only the radius. For this, we can employ either Lagrangian or Hamiltonian formalism. Here, we incorporate the Lagrangian formalism \cite{14} for which the Lagrangian
\begin{eqnarray}
\mathcal{L}(q,\dot{q})=\frac{1}{2}g_{\mu\nu}\dot{q}^\mu\dot{q}^\nu \label{8}
\end{eqnarray}
takes the form in the equatorial plane with $\theta=\frac{\pi}{2}$ and $\dot{\theta}=0$ as
\begin{eqnarray}
\mathcal{L}(q,\dot{q})=\frac{1}{2}\Big(-f(r)\dot{t}^2+\frac{\dot{r}^2}{f(r)}+r^2\dot{\phi}^2\Big). \label{9}
\end{eqnarray}
Note that the derivative with respect to the affine parameter $\tau$ is denoted by the dot. Solving the Euler-Lagrange equation
\begin{eqnarray}
\frac{\partial\mathcal{L}}{\partial q^\mu}-\frac{d}{d\tau}\bigg(\frac{\partial\mathcal{L}}{\partial\dot{q}^\mu}\bigg)=0 \label{10}
\end{eqnarray}
for the $t$ and $\phi$ components, we obtain the two constants of the motion
\begin{eqnarray}
E=f(r)\dot{t}, \qquad L=r^2\dot{\phi} \label{11}
\end{eqnarray}
characterized as the energy and $z$-component of angular momentum, respectively. The radial equation becomes complicated by solving for $r$ component of the Euler-Lagrange equation. Therefore, we solve the first integral for null geodesics $\mathcal{L}=0$, i.e.,
\begin{eqnarray}
-f(r)\dot{t}^2+\frac{\dot{r}^2}{g(r)}+r^2\dot{\phi}^2=0. \label{12}
\end{eqnarray}
By using the values of $\dot{t}$ and $\dot{\phi}$ from Eq. (\ref{11}) in (\ref{12}), we get the orbital equation for null geodesics as
\begin{eqnarray}
\frac{dr}{d\phi}=\pm r\sqrt{f(r)}\sqrt{h(r)^2\frac{E^2}{L^2}-1}, \label{13}
\end{eqnarray}
where
\begin{eqnarray}
h(r)^2=\frac{r^2}{f(r)}. \label{14}
\end{eqnarray}
By solving the conditions $\frac{dr}{d\phi}=0=\frac{d^2r}{d\phi^2}$, we get the required condition for the radius of circular null sphere as
\begin{eqnarray}
\frac{d}{dr}\Big(h(r)^2\Big)\bigg|_{r=r_{ph}}=0. \label{15}
\end{eqnarray}
The real and positive roots of Eq. (\ref{15}) with $r\rightarrow r_{ph}$ give the radius of null sphere. By using above condition (\ref{15}), we have plotted the photon sphere radius $r_{ph}$ versus the BH parameters $\gamma$ and $\lambda$ in Fig. \ref{RPS}. In the left plot, it can be seen that the curves for different values of $\lambda$ are starting from the same radius 3 that is because when $\gamma=0$, the BH metric function resembles Schwarzschild metric and with increasing value of $\gamma$, the size of photon sphere gradually starts decreasing. Moreover, for each curve, as $\lambda$ increases, the rate of decrease in the size of photon sphere increases. That is, the curve for $\lambda=0.2$ measures a very small decrease with increasing $\gamma$ and for $\lambda=1$, the curve decreases the fastest with increase in $\gamma$. All other curves have intermediate behavior. Note that, the curve for $\lambda=1$ corresponds to the case of Reissner-Nordstr\"{o}m BH. In the right plot, we find that for $\lambda\lessapprox0.2$, the radius of null sphere is unaffected whatever the value of $\gamma$ is considered. As we increase the value of $\lambda$, the radius starts decreasing. The flat horizontal curve has been observed for the case of $\gamma=0$ and as $\gamma$ increases, the drop rate of the photon sphere's size increases. In this manner, both parameters have somewhat identical behaviour.

\subsection{Radius of Black Hole Shadow}
\begin{figure}[t]
\centering
\subfigure{
\includegraphics[width=0.45\textwidth]{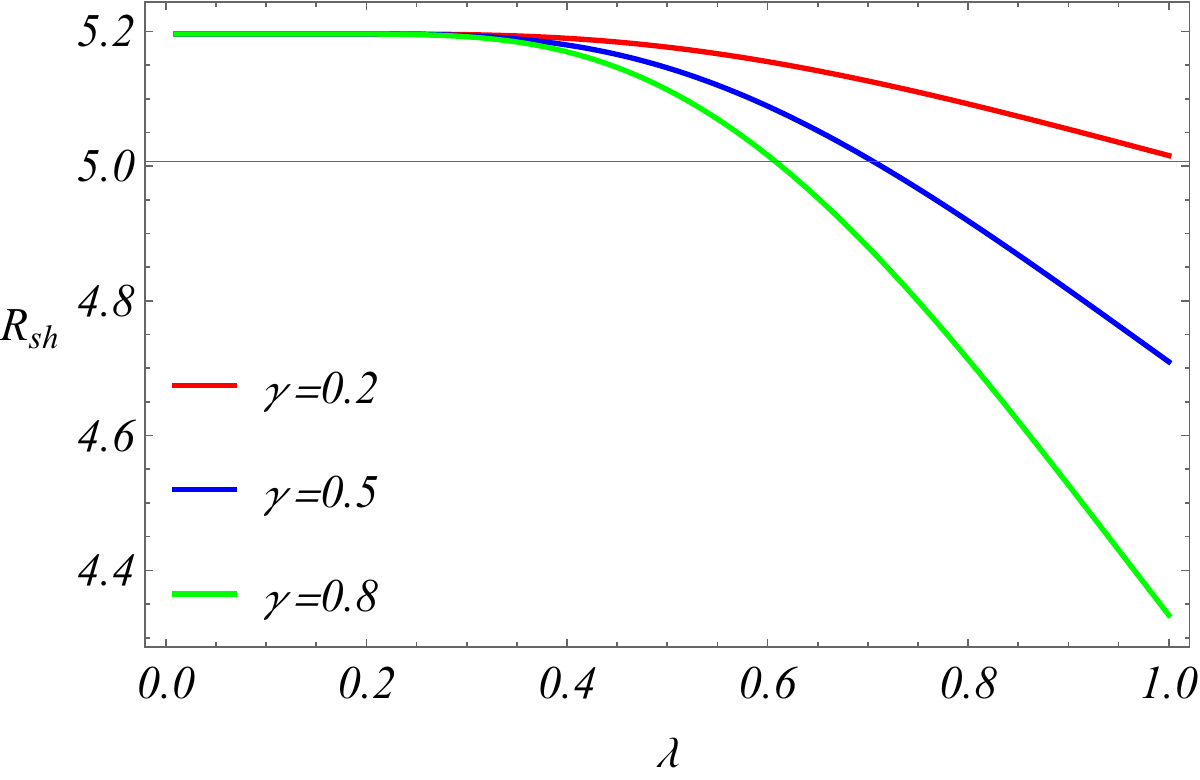}}~~~~~
\subfigure{
\includegraphics[width=0.45\textwidth]{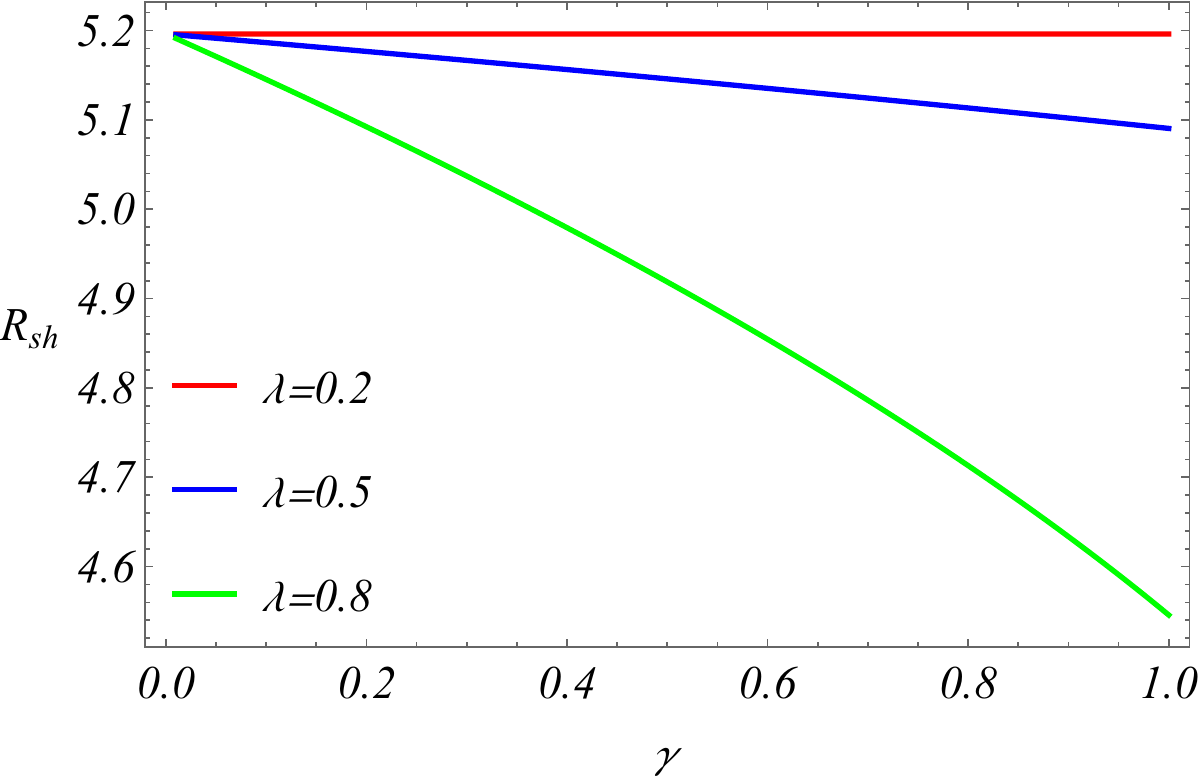}}
\caption{The behavior of linear radius of the BH shadow $R_{sh}$ for various values of $\gamma$ and $\lambda$ for the case of static BH.} \label{RSH}
\end{figure}

As we know that the boundary of the shadow is dependent on the radius of photon sphere. Therefore, we have calculated the photon region's size and will be used in obtaining the behavior of the shadow radius for the different values of $\lambda$ and $\gamma$. Suppose that an observer is located at a radial distance $r_0$ from the origin, throws light into the past. Then, the angle $\theta_s$ between the radial axis and the light beam is given by
\begin{eqnarray}
\tan\theta_s=r\sqrt{f(r)}\frac{d\phi}{dr}\bigg|_{r\rightarrow r_0}. \label{16}
\end{eqnarray}
By using Eq. (\ref{13}), we get
\begin{eqnarray}
\tan^2\theta_s=\frac{1}{h(r_0)^2\frac{E^2}{L^2}-1}, \label{17}
\end{eqnarray}
Or,
\begin{eqnarray}
\sin^2\theta_s=\frac{L^2}{E^2h(r_0)^2}, \label{18}
\end{eqnarray}
When the photon beam travels towards the central object and then goes out after attaining a least radius $R_p$ which is the turning point of the beam, the condition $\frac{dr}{d\phi}\big|_{R_p}=0$ needs to be satisfied. It gives
\begin{eqnarray}
h(R_p)=\frac{L}{E}. \label{19}
\end{eqnarray}
Then we get
\begin{eqnarray}
\sin^2\theta_s=\frac{h(R_p)^2}{h(r_0)^2}. \label{20}
\end{eqnarray}
Among all of the circular null orbits, the shadow curve corresponds to the unstable null orbit with radius $r_{ph}$. Therefore, $R_p\rightarrow r_{ph}$ and hence we get
\begin{eqnarray}
\sin^2\theta_s=\frac{h(r_{ph})^2}{h(r_0)^2}. \label{21}
\end{eqnarray}
In this case, the angle $\theta_s$ becomes the angular shadow radius. Following the Fig. $\textbf{7}$ in \cite{14}, when the observer is shifted to infinity, i.e., $r_0\rightarrow \infty$, the elementary trigonometry gives
\begin{eqnarray}
\sin\theta_s=\frac{R_{sh}}{r_0}. \label{22}
\end{eqnarray}
Note that we have used the triangle formed by the radial separation between the observer and the origin denoted by $r_0$ as the base, vertical shadow radius $R_{sh}$ as the perpendicular and the hypotenuse formed by the light ray. If the observer is at infinity, then the hypotenuse is approximately equal to $r_0$ and the angle $\theta_s$ becomes small such that $\sin\theta_s\approx\theta_s$. Using the fact that the metric (\ref{7}) is asymptotically flat, Eqs. (\ref{14}) and (\ref{22}) in (\ref{21}), we get
\begin{eqnarray}
R_{sh}=\frac{r_{ph}}{\sqrt{f(r_{ph})}}. \label{23}
\end{eqnarray}
Since, $r_{ph}$ cannot be derived analytically in terms of either $\lambda$ or $\gamma$, therefore we cannot use it directly in Eq. (\ref{23}) to obtain the plots for $R_{sh}$ vs $\lambda$ and $\gamma$. In order to plot $R_{sh}$ vs $\lambda$ and $\gamma$, we use point-wise numerical technique. According to this technique, by fixing $\lambda$, we calculate $r_{ph}$ for one value of $\gamma$ and use it further in obtaining $R_{sh}$. This gives one point in $\gamma$-$R_{sh}$ plane and plotting for many discrete points of $\gamma$ in the interval $[0,1]$, it gives a smooth and accurate curve. The behavior of $R_{sh}$ vs $\lambda$ and $\gamma$ is plotted in Fig. \ref{RSH}. It can be seen that the linear radius of the shadow decreases with increase in both $\lambda$ and $\gamma$. However, like the case of photon sphere, we see that for the values up to $\lambda\approx0.2$, the shadow radius has no significant change. Moreover, the shadow radius decreases more rapidly as $\lambda$ increases as compared to the increase in $\gamma$.

\subsection{Real and Imaginary Parts of Quasinormal Modes}
The unstable circular null geodesics in the vicinity of a spherically symmetric and static BH that is asymptotically flat, possess some parameters, e.g. the angular velocity $\Omega$ and the principal Lyapunov exponent $\Gamma$, related to the quasinormal modes \cite{63a,63b,63c,63d} that are emitted by the BH in the eikonal part of its spectrum. It is found that the eikonal quasinormal frequencies $\omega_{n}$ in $D\geq4$ dimensions can be written as \cite{63e}
\begin{eqnarray}
\omega_{n}=\Omega l_c-i\bigg(n+\frac{1}{2}\bigg)|\Gamma|, \label{24}
\end{eqnarray}
where $n$ and $l_c$ are overtone number and multipole number, respectively. Here, we would like to study the angular velocity $\Omega$ and Lyapunov exponent $\Gamma$ for the null sphere around the static BH in KR gravity. The significance of studying these parameters lie in the sense that in order to understand the thermodynamics of the BH, it can be useful to study the connection between the phase transition and quasinormal modes \cite{63f}. While leaving the thermodynamical properties to another project and focusing only on the real and imaginary parts of quasinormal frequencies, the relation for the angular velocity can be written as \cite{63a}
\begin{eqnarray}
\Omega=\frac{\dot{\phi}}{\dot{t}}=\frac{\sqrt{f(r_{ph})}}{r_{ph}}. \label{25}
\end{eqnarray}
To derive the Lyapunov exponent, we first need to derive an expression for the effective potential $V_{eff}$. Therefore, using $\theta=\frac{\pi}{2}$ and Eq. (\ref{11}) in (\ref{12}) gives
\begin{eqnarray}
\dot{r}^2+V_{eff}(r)=0 \label{26}
\end{eqnarray}
such that
\begin{eqnarray}
V_{eff}(r)=\frac{L^2f(r)}{r^2}-E^2. \label{27}
\end{eqnarray}
Therefore, the Lyapunov exponent can thus be written as \cite{63a}
\begin{eqnarray}
\Gamma=\sqrt{-\frac{1}{2\dot{t}^2}\frac{\partial^2 V_{eff}}{\partial r^2}}\Bigg|_{r=r_{ph}}. \label{28}
\end{eqnarray}
\begin{figure}[t]
\centering
\subfigure{
\includegraphics[width=0.45\textwidth]{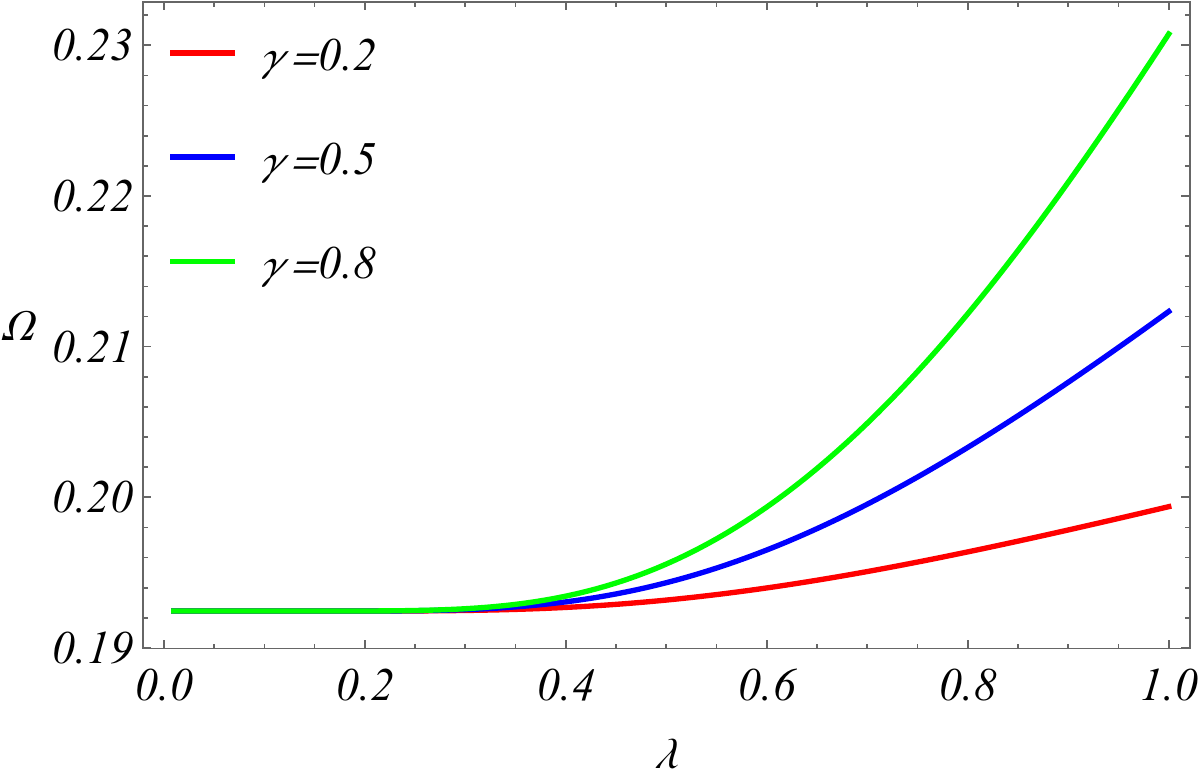}}~~~~~
\subfigure{
\includegraphics[width=0.45\textwidth]{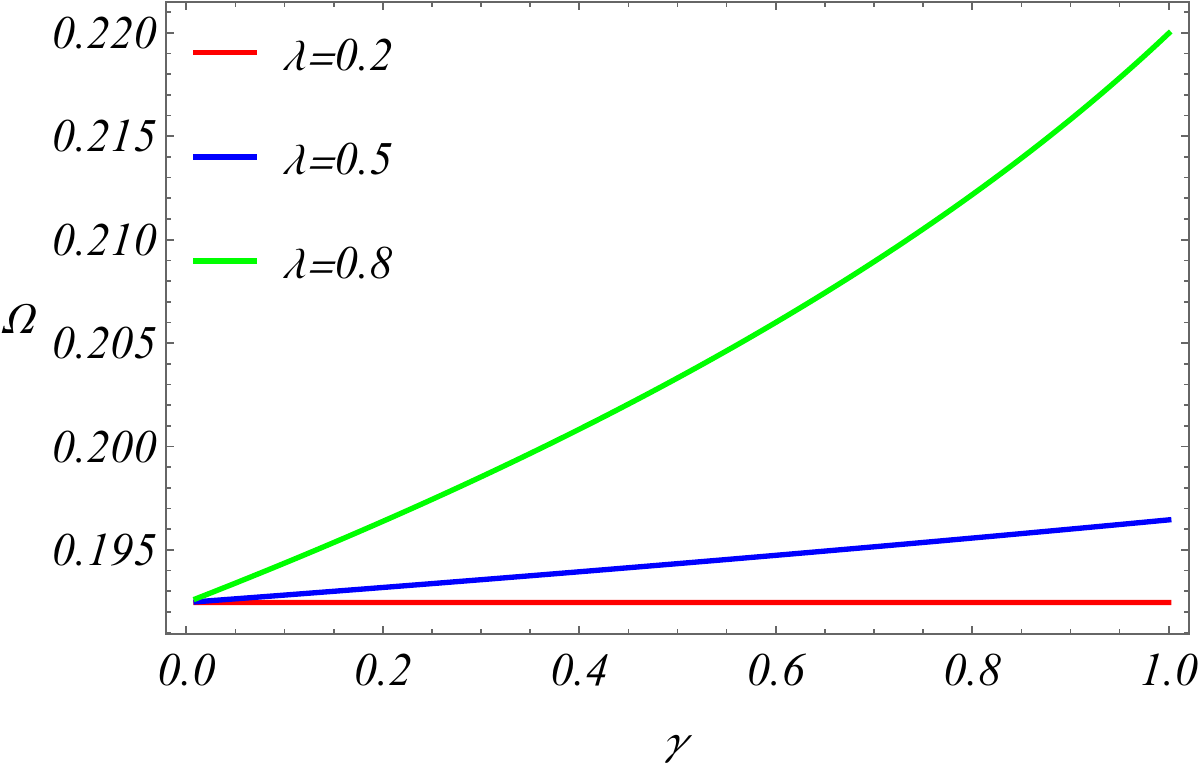}}\\
\subfigure{
\includegraphics[width=0.45\textwidth]{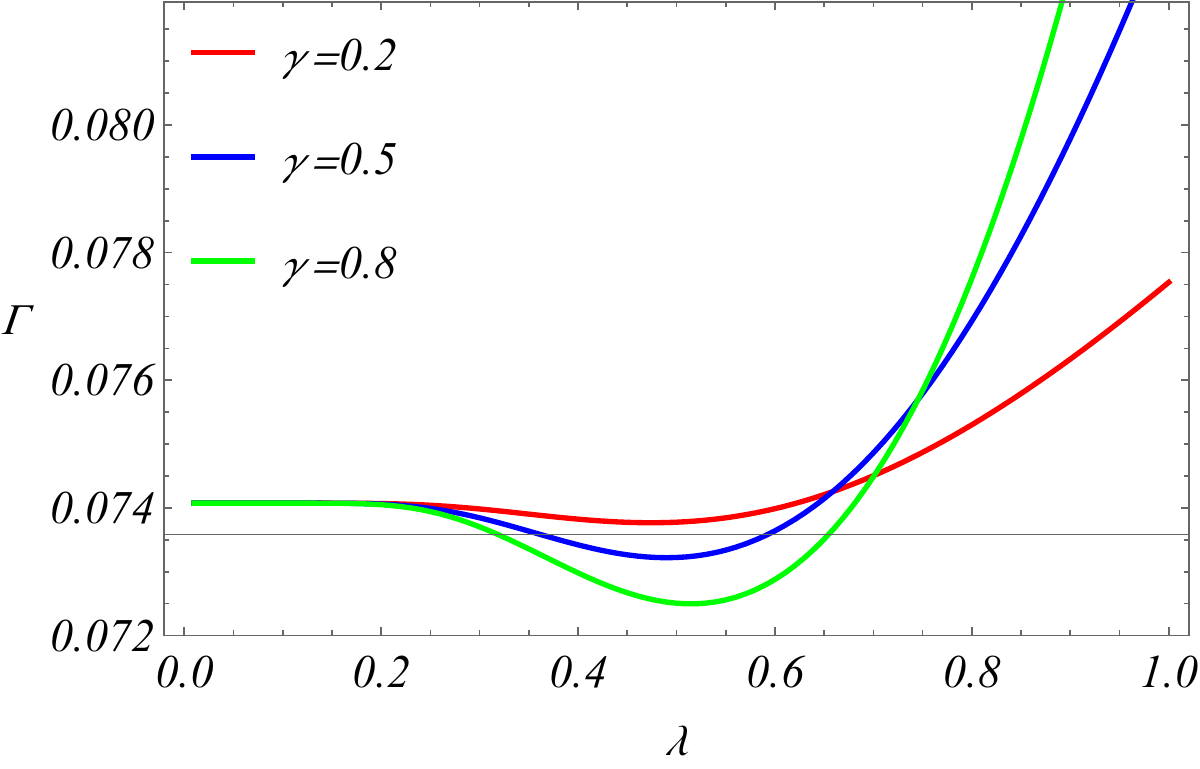}}~~~~~
\subfigure{
\includegraphics[width=0.45\textwidth]{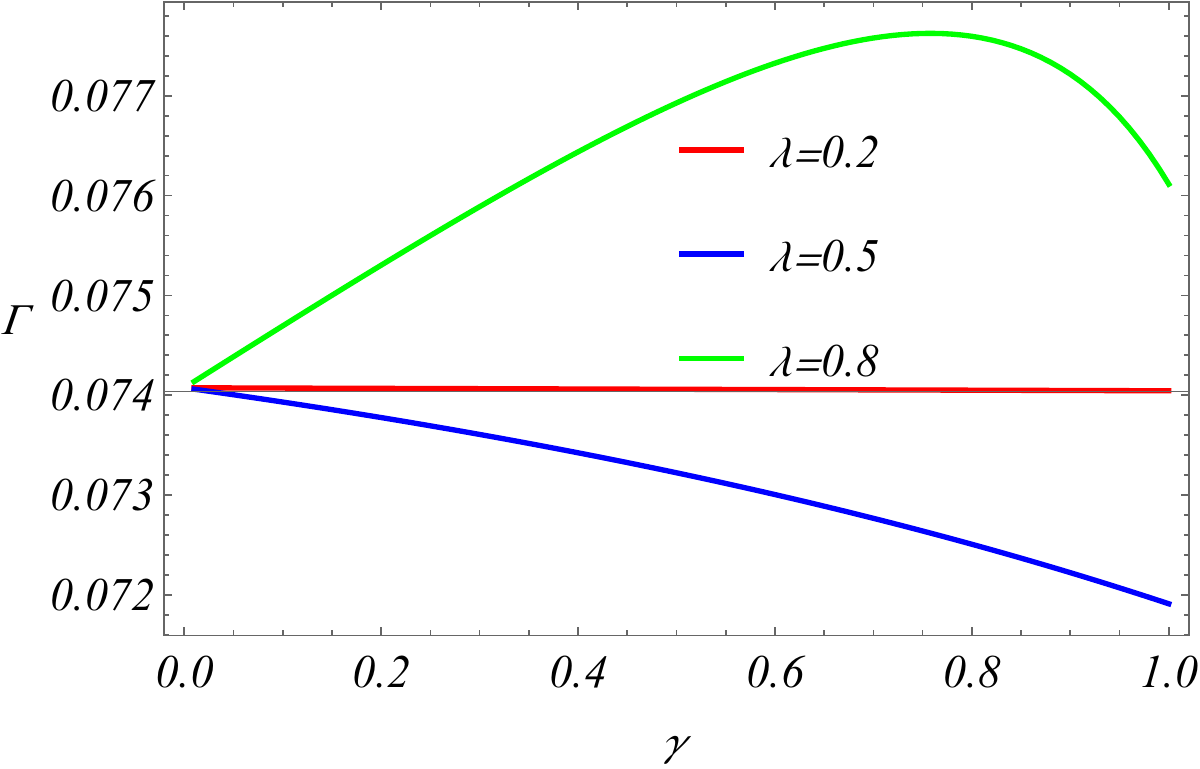}}
\caption{The behavior of angular velocity and Lyapunov exponent for different values of $\lambda$ and $\gamma$.} \label{QNM}
\end{figure}
The technique used to plot Fig. \ref{RSH} is again used here to plot the behavior of the angular velocity $\Omega$ and Lyapunov exponent $\Gamma$ vs $\lambda$ and $\gamma$ in Fig. \ref{QNM}. The upper panel corresponds to the plots for angular velocity $\Omega$ and the lower panel shows the plots for Lyapunov exponent $\Gamma$. From the left plot in the upper panel, it is found that the angular velocity remains approximately constant for the values up to $\lambda\approx0.2$ and then increases with increase in $\lambda$. The rate of increment in $\Omega$ grows as $\gamma$ increases. In the right plot, for $\lambda=2$, the horizontal curve shows that the angular velocity remains constant which can also be seen in the other plot for the initial values of $\lambda$. However, $\Omega$ increases with increase in $\gamma$ and the increment rate also grows with increase in $\lambda$. In the lower panel, the left plot shows that $\Gamma$ is constant for the values up to $\lambda\approx0.2$. However, $\Gamma$ starts decreasing for some values of $\lambda$ and then increases rapidly for the remaining values of $\lambda$. It can be seen that for smaller values of $\gamma$, the decrease and increase in $\Gamma$ is less as compared to the larger values of $\gamma$. While for the right plot, Lyapunov exponent is constant for $\lambda=0.2$ and decreases for $\lambda=0.5$ as $\gamma$ increases. However, $\Gamma$ increases for $\lambda=0.8$ up to certain value of $\gamma$ and then decreases. This fact is not obvious from the left plot because it does not contain curves for the values of $\gamma$ greater than 0.8.

\section{Rotating Black Hole in Kalb-Ramond Gravity}
As the simplest generalization of a static BH, one can obtain its rotating counterpart. The rotating BHs with an additional parameter corresponding to spin behave slightly different as compared to the static BHs in terms of capturing photons in orbits and generating the shadows. The rotating BHs are significant for obtaining a comparison of the shadows with the EHT data. As we know that the supermassive BHs are rotating in nature, therefore in order to accomplish a feasible and a rigorous comparison of the shadows it is important to consider rotating BHs. The Newman-Janis algorithm is one of the most commonly used methods to derive a rotating BH metric from a static BH metric. The static BH metric (\ref{7}) can also be converted into its rotating counterpart using the Newman-Janis algorithm. Since, Kumar et al. \cite{55} has already developed the rotating counterpart of the static BH solution (\ref{7}), therefore we do not mention the algorithm again. The rotating BH metric in KR gravity reads
\begin{eqnarray}
ds^{2}&=&-\bigg(\frac{\Delta(r)-a^2\sin^2{\theta}}{\rho^2}\bigg)dt^{2}+\frac{\rho^2}{\Delta(r)}dr^2+\rho^2d\theta^{2}+\frac{\sin^{2}\theta}{\rho^2}\bigg(\big(r^2+a^2\big)^2-\Delta(r) a^2\sin^{2}\theta\bigg)d\phi^{2} \nonumber\\
&&+\frac{2a\sin^{2}\theta}{\rho^2}\bigg(\Delta(r)-a^2-r^2\bigg)dtd\phi, \label{29}
\end{eqnarray}
whereas the metric functions of the above metric are identified as
\begin{eqnarray}
\Delta(r)&=&a^2+r^2f(r)=r^2+a^2-2Mr+\gamma r^{\frac{2(\lambda-1)}{\lambda}}, \label{30}\\
\rho^2&=&r^2+a^2\cos^2\theta \label{31}
\end{eqnarray}
with $a$ as the spin parameter. It may be noted that the rotating BH metric (\ref{29}) reduces to Kerr and Kerr-Newman BHs when $\lambda=0$ and $\lambda=1$, respectively. Moreover, when $a=0$, the rotating BH metric (\ref{29}) reduces to the static metric (\ref{7}). It is also worth mentioning that the rotating BH (\ref{29}) in KR gravity has the isometries corresponding to the rotational and time-translational invariance. Therefore, ensuring the existence of Killing vector fields $\big(\frac{\partial}{\partial t}\big)^\mu$ and $\big(\frac{\partial}{\partial \phi}\big)^\mu$. One of the major aim in this study is to constrain the parameters of BH in KR gravity for which we can regard the BH as one of either M87* or Sgr A* BHs. This is accomplished by comparing the shadows of BH in KR gravity with the observational shadow results obtained by EHT collaborations. As we have mentioned that the astrophysical BHs are rotating in nature, therefore we further proceed with the rotating BH metric (\ref{29}) in order to make a feasible comparison.

\begin{figure}[t]
\begin{center}
\subfigure{
\includegraphics[width=0.38\textwidth]{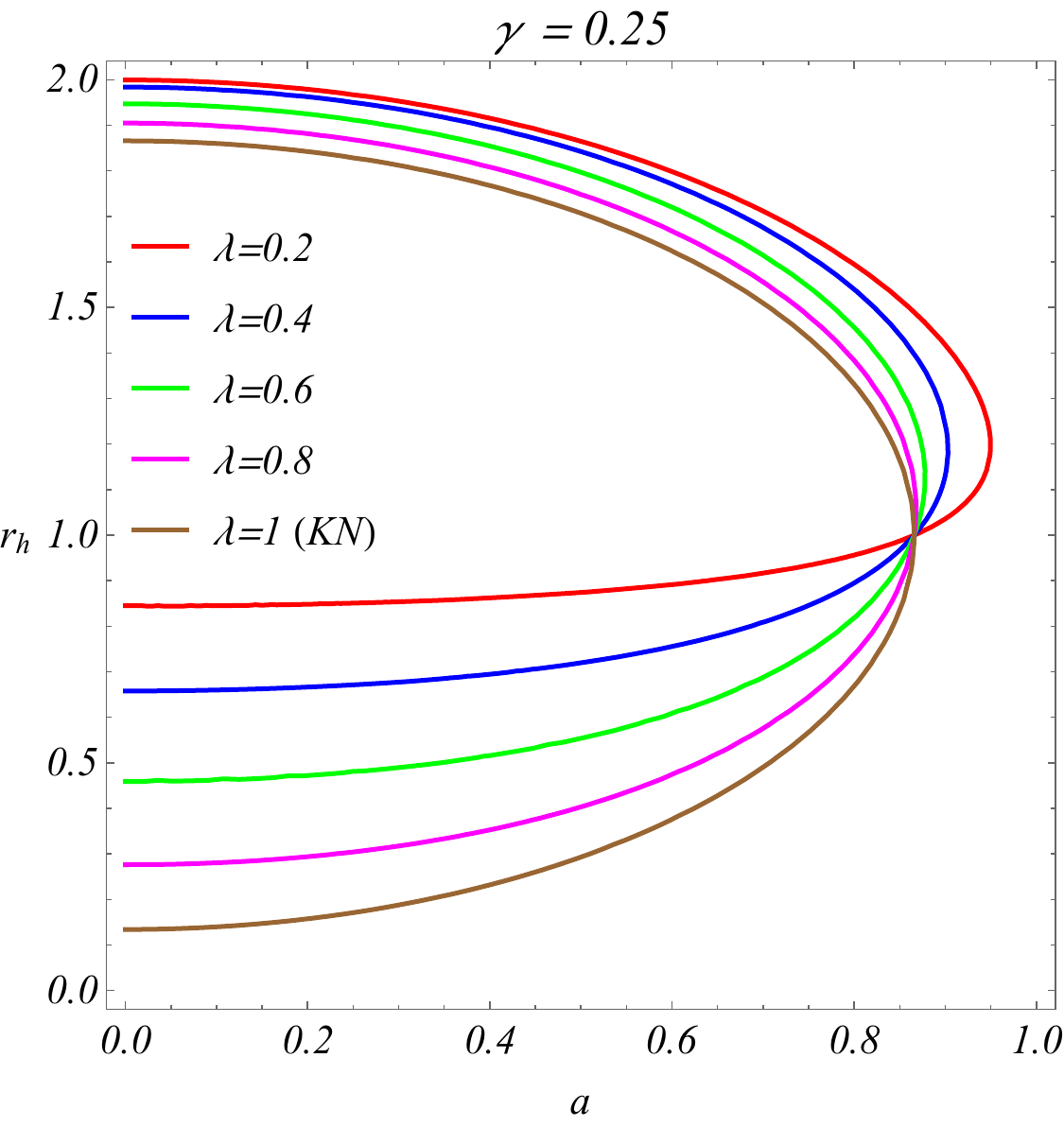}}~~~~~~~~
\subfigure{
\includegraphics[width=0.38\textwidth]{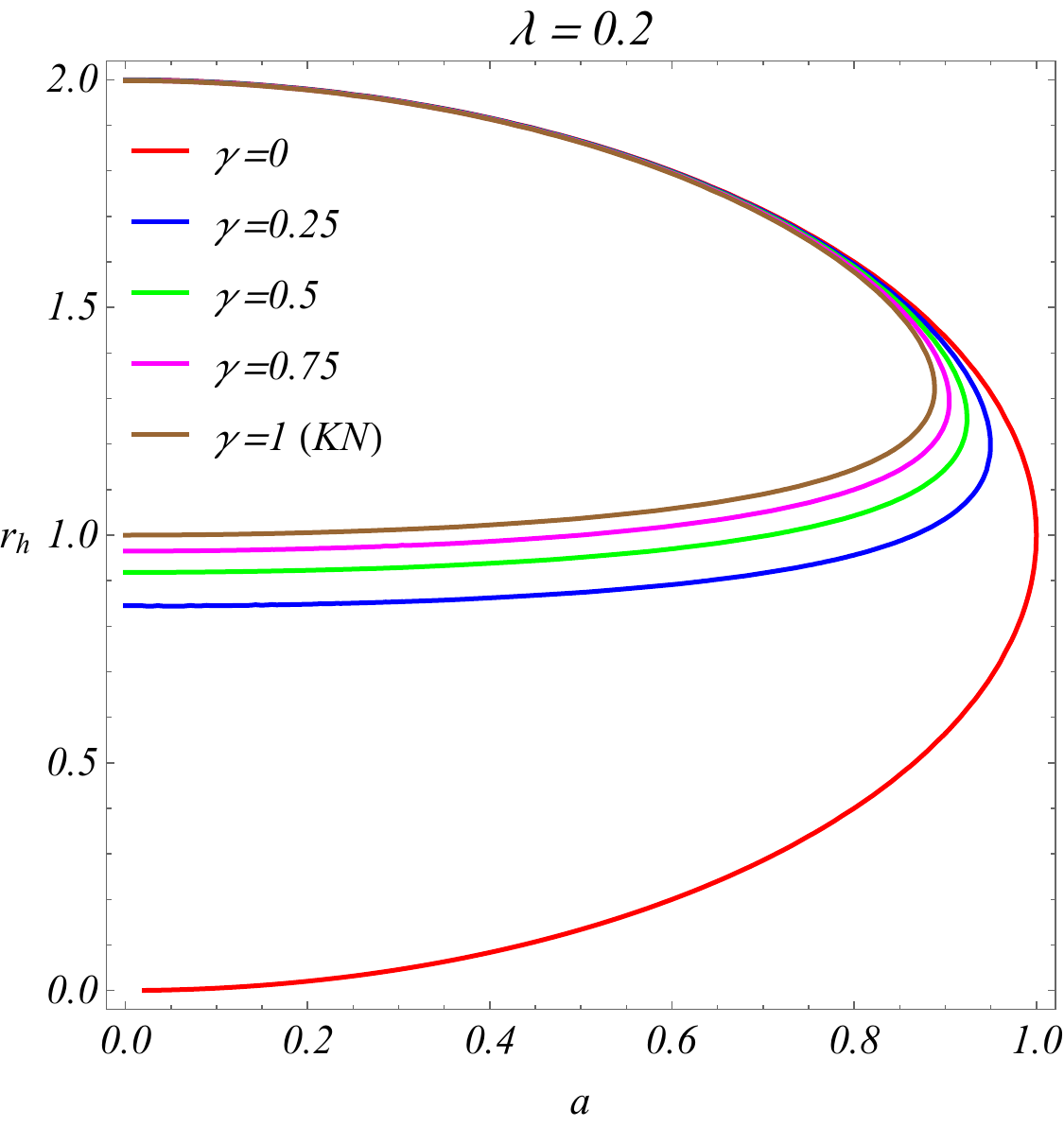}}\\
\subfigure{
\includegraphics[width=0.38\textwidth]{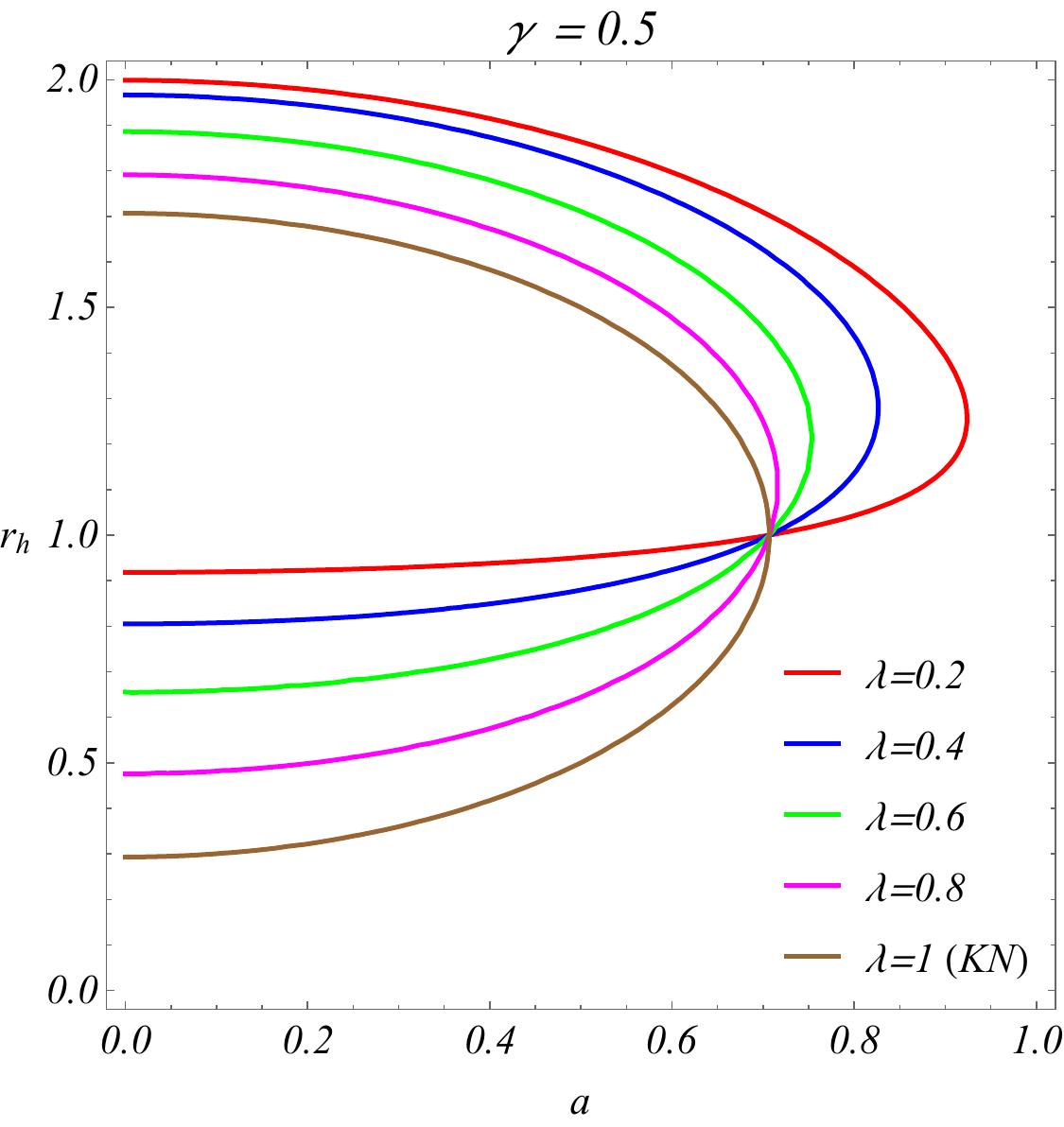}}~~~~~~~~
\subfigure{
\includegraphics[width=0.38\textwidth]{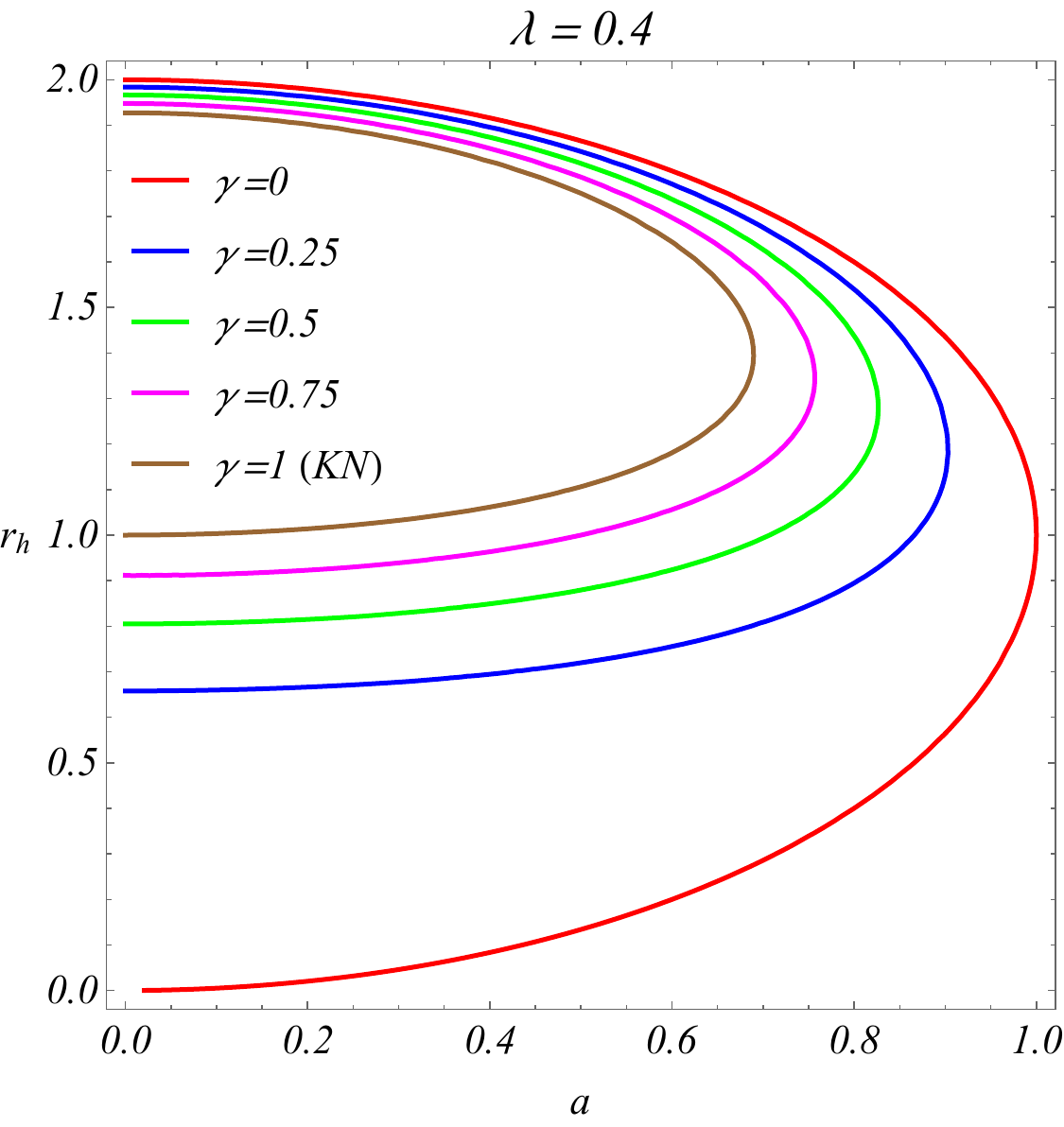}}\\
\subfigure{
\includegraphics[width=0.38\textwidth]{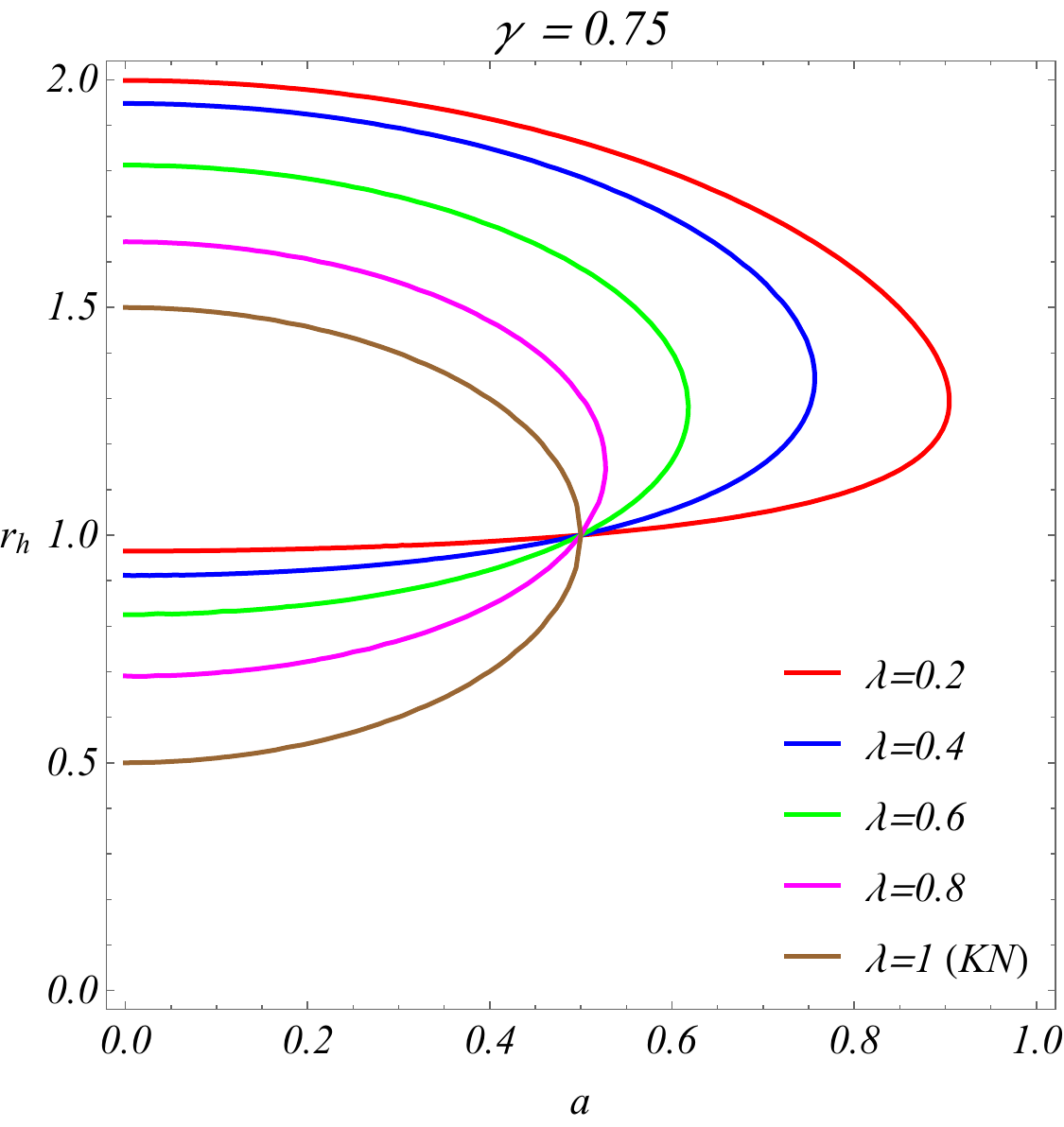}}~~~~~~~~
\subfigure{
\includegraphics[width=0.38\textwidth]{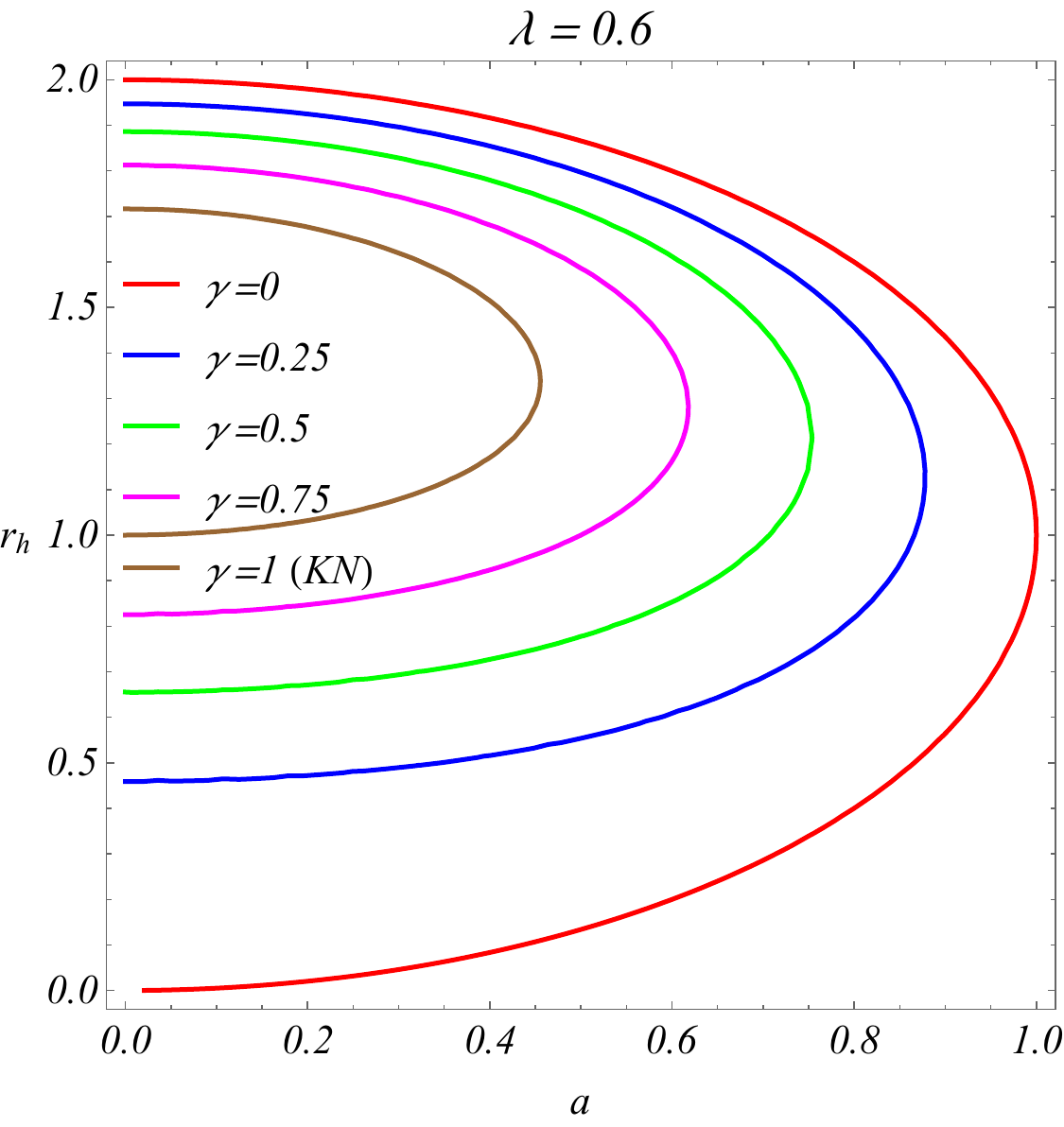}}
\end{center}
\caption{Plots showing the effects of $\gamma$, $\lambda$ and $a$ on horizon radius for the rotating BH in KR gravity.} \label{HR}
\end{figure}

The BH size depends upon the radius of the event horizon. Therefore, in order to study the influence of $\lambda$, $\gamma$ and $a$ on the BH size, we solve the equation $\Delta(r)=0$ using the numerical techniques and plot the radius of horizon versus the spin parameter $a$ as shown in Fig. \ref{HR}. Instead of plotting the metric function $\Delta(r)$ versus $r$, we preferred to plot the horizon radius $r_h$ versus $a$ for different values of $\lambda$ and $\gamma$. It is because in the metric function plots, we need to fix all of the BH parameters. Whereas, in the horizon radius plots, we can have $a$ as independent parameter and fix only $\lambda$ and $\gamma$. Therefore, a more general behavior of the horizons is obtained from the horizon radius plots. In all plots and for each curve, it is found that the event horizon decreases and Cauchy horizon increases as $a$ grows larger up to its extremal value. Moreover, for each plot, the extremal value of $a$ becomes smaller as the value of $\lambda$ and $\gamma$ rise for each individual curve. Furthermore, for each plot in the left panel, the value of both event and Cauchy horizons decreases for a constant spin $a$ as $\lambda$ grows larger for each individual curve. From top to bottom in the left panel, as $\gamma$ increases, the difference between event and Cauchy horizons reduces. In the right panel, the reduction in difference between event and Cauchy horizons with increase in $\gamma$ can be seen more clearly. As seen before, the photon sphere's radius is unaffected for all values of $\gamma$ when $\lambda\lessapprox0.2$. The same effect is also observed in the case of horizons. In the top right plot, we can clearly see that for $\lambda=0.2$ and regardless of the choice of $\gamma$, the change in event horizon is negligible. Whereas, the event horizon changes with increase in $\lambda$.

\subsection{Null Geodesics and Shadow} \label{nullgeod}
The emerging photons from a bright source or the accretion disk around the BH may get trapped in the locality of the outer horizon. Some photons disappear into the horizon, whereas the rest of them escape away to the infinity. This gives rise to the optical image of the BH, termed as shadow that is enclosed by a luminous null sphere \cite{64,65,66,67}. It is not quite strange that a captured photon in the photon emission ring orbits the BH multiple times before leaving it. It is because the photon resides in the unstable orbit for few revolutions. Whereas, a photon in the stable orbit does not leave its orbit. These orbits are characterized by the effective potential function. Next, we study the null geodesics and effective potential to draw an analysis for the unstable null orbits and to discover the effect of BH parameters on the orbits governed by the effective potential. However, to derive the effective potential, we need the null geodesic equations. For the rotating BH in KR gravity, the null geodesic equations can be obtained by employing the Hamilton-Jacobi formalism \cite{68}. The most common methods to study the motion of particles and objects are the Lagrangian and Hamiltonian formalisms. However, in relativity, these formalisms give only two constants of motion, the energy $E$ and the angular momentum $L$ along with mass of the particle under motion. To make the set of equations completely integrable, we need one more constant. In this way the Hamilton-Jacobi formalism is useful because from the separation of $r$ and $\theta$ coordinates in Hamilton-Jacobi equation, we get a constant known as the Carter constant \cite{15,68}. First, we begin with the Lagrangian formulation and derive the constants of motion $E$ and $L$. The Lagrangian is given by Eq. (\ref{8}) with the metric tensor $g_{\mu\nu}$ from the metric (\ref{29}). The generalized momenta can be written as
\begin{equation}
p_\mu=g_{\mu\nu}\dot{q}^{\nu}. \label{32}
\end{equation}
The constants of motion $E$ and $l$ are obtained by expanding the Lagrangian and using Eqs. (\ref{30}) and (\ref{32}) therein and therefore we get
\begin{eqnarray}
E&:=&-\frac{\partial\mathcal{L}}{\partial\dot{t}}=-g_{tt}\dot{t}-g_{\phi t}\dot{\phi}=p_t, \label{33} \\
L&:=&\frac{\partial\mathcal{L}}{\partial\dot{\phi}}=g_{\phi t}\dot{t}+g_{\phi \phi}\dot{\phi}=p_\phi. \label{34}
\end{eqnarray}
The Hamilton-Jacobi equation is given by
\begin{equation}
2\frac{\partial \mathcal{S_J}}{\partial\tau}+g^{\mu\nu}\frac{\partial \mathcal{S_J}}{\partial x^\mu}\frac{\partial \mathcal{S_J}}{\partial x^\nu}=0, \label{35}
\end{equation}
where $\mathcal{S_J}$ is the Jacobi action that can be supposed as
\begin{equation}
\mathcal{S_J}=\frac{1}{2}m_p\tau-Et+L\phi+\mathcal{A}_r(r)+\mathcal{A}_\theta(\theta), \label{36}
\end{equation}
where $m_p$ is the mass of particle in the orbit around the BH, $\mathcal{A}_r(r)$ and $\mathcal{A}_\theta(\theta)$ are functions that will be determined later. Now using the separation of variables in the Hamilton-Jacobi equation, we get the null geodesic equations corresponding to $m_p=0$ as
\begin{eqnarray}
\rho^2\frac{dt}{d\tau}&=&a\big(L-aE\sin^2{\theta}\big)+\frac{r^2+a^2}{\Delta(r)}\big(E\big(r^2+a^2\big)-aL\big), \label{37} \\
\rho^2\frac{dr}{d\tau}&=&\pm\sqrt{\mathcal{R}(r)}, \label{38} \\
\rho^2\frac{d\theta}{d\tau}&=&\pm\sqrt{\Theta(\theta)}, \label{39} \\
\rho^2\frac{d\phi}{d\tau}&=&\big(L\csc^2{\theta}-aE\big)-\frac{a}{\Delta(r)}\big(aL-E\big(r^2+a^2\big)\big), \label{40}
\end{eqnarray}
where
\begin{eqnarray}
\mathcal{R}(r)&=&\big(E(r^2+a^2)-aL\big)^2-\Delta(r)\big(\mathcal{Z}+(L-aE)^2\big), \label{41} \\
\Theta(\theta)&=&\mathcal{Z}+\cos^2{\theta}\big(a^2E^2-L^2\csc^2{\theta}\big), \label{42}
\end{eqnarray}
where $\mathcal{Z}$ is the Carter constant. The radial equation corresponding to the equatorial trajectories $\big(\theta=\frac{\pi}{2}\big)$ can also be written as $\dot{r}^2+V_{eff}(r)=0$, where $V_{eff}(r)$ is the effective potential given as
\begin{eqnarray}
V_{eff}(r)=-\frac{\mathcal{R}(r)}{2r^4}=-\frac{\big(E(r^2+a^2)-aL\big)^2-\Delta(r)\big(\mathcal{Z}+(L-aE)^2\big)}{2r^4}. \label{43}
\end{eqnarray}
\begin{figure}[t]
\begin{center}
\subfigure{
\includegraphics[width=0.48\textwidth]{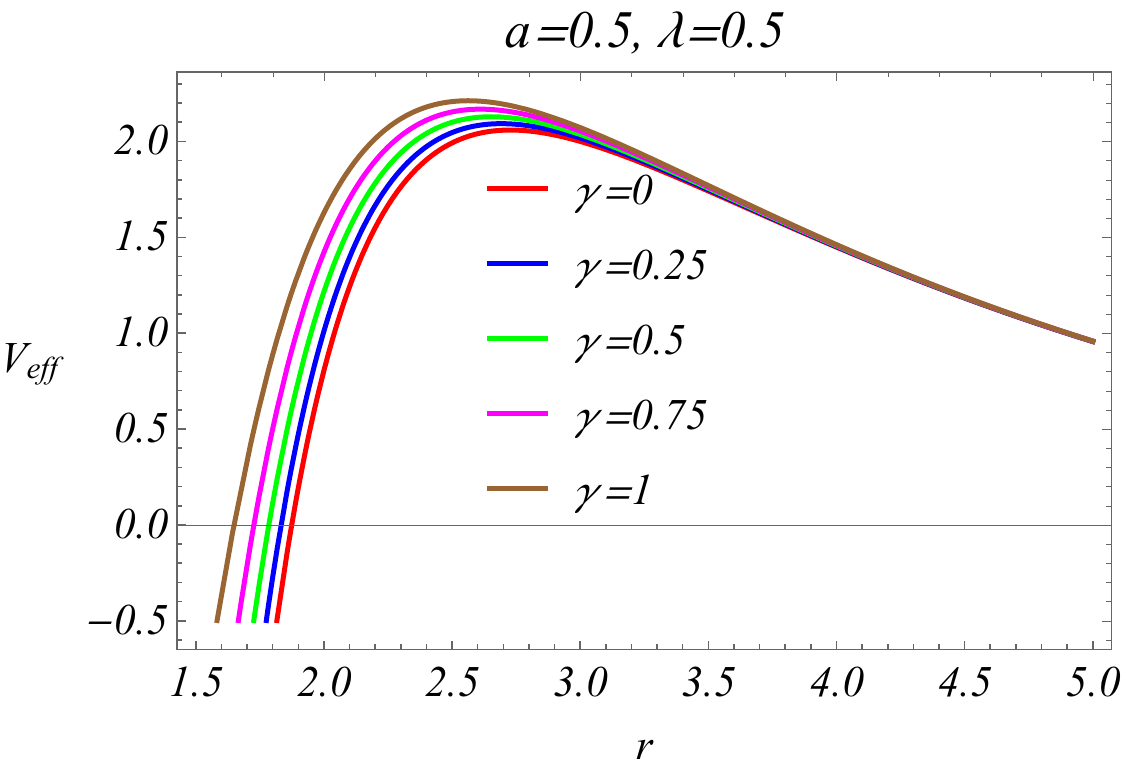}}~~~~
\subfigure{
\includegraphics[width=0.48\textwidth]{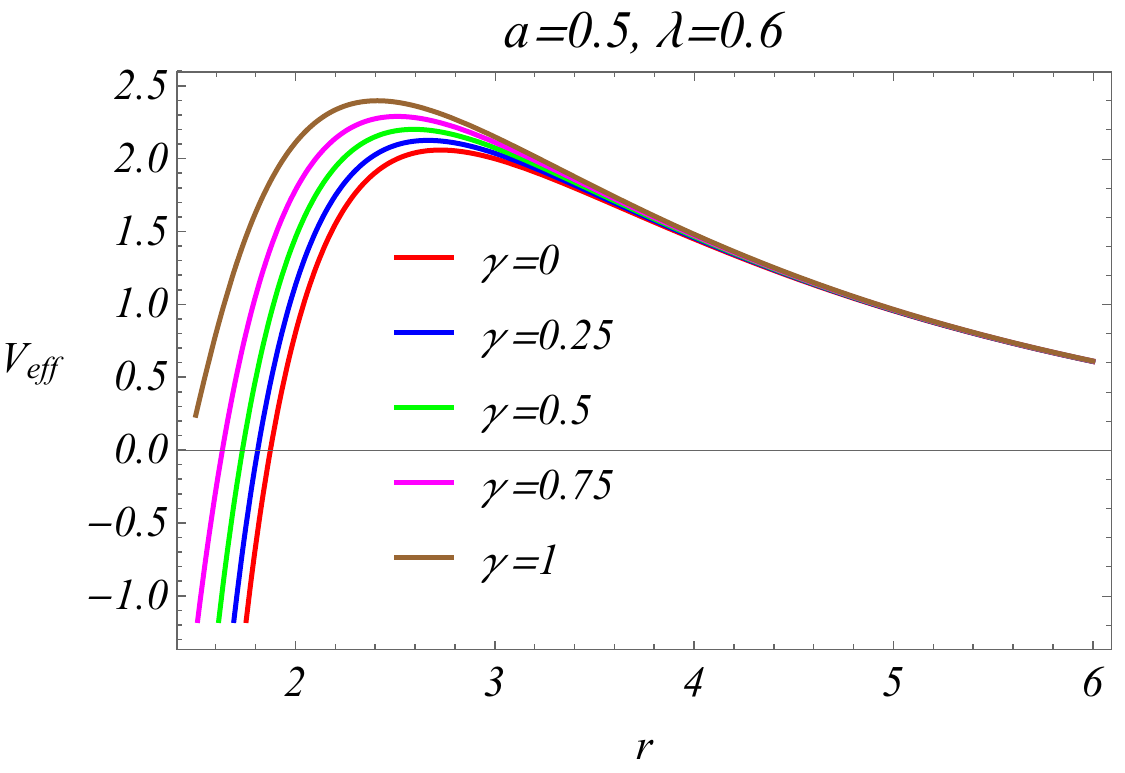}}\\
\subfigure{
\includegraphics[width=0.48\textwidth]{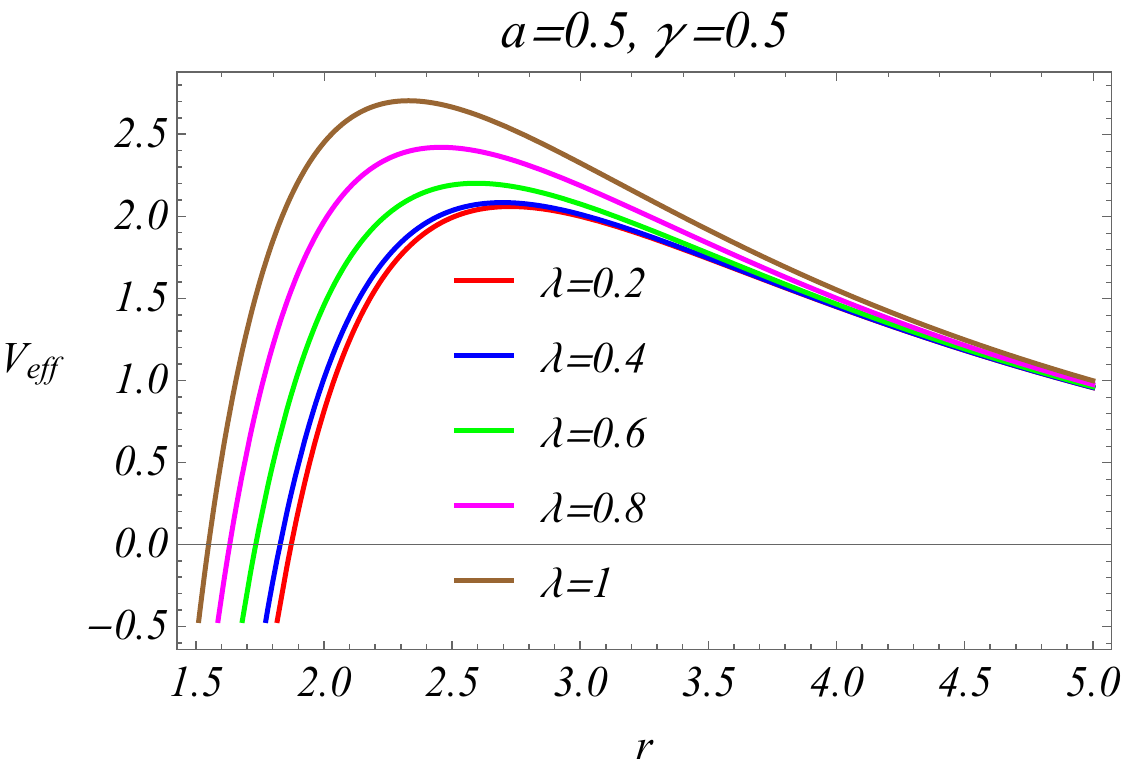}}~~~~
\subfigure{
\includegraphics[width=0.48\textwidth]{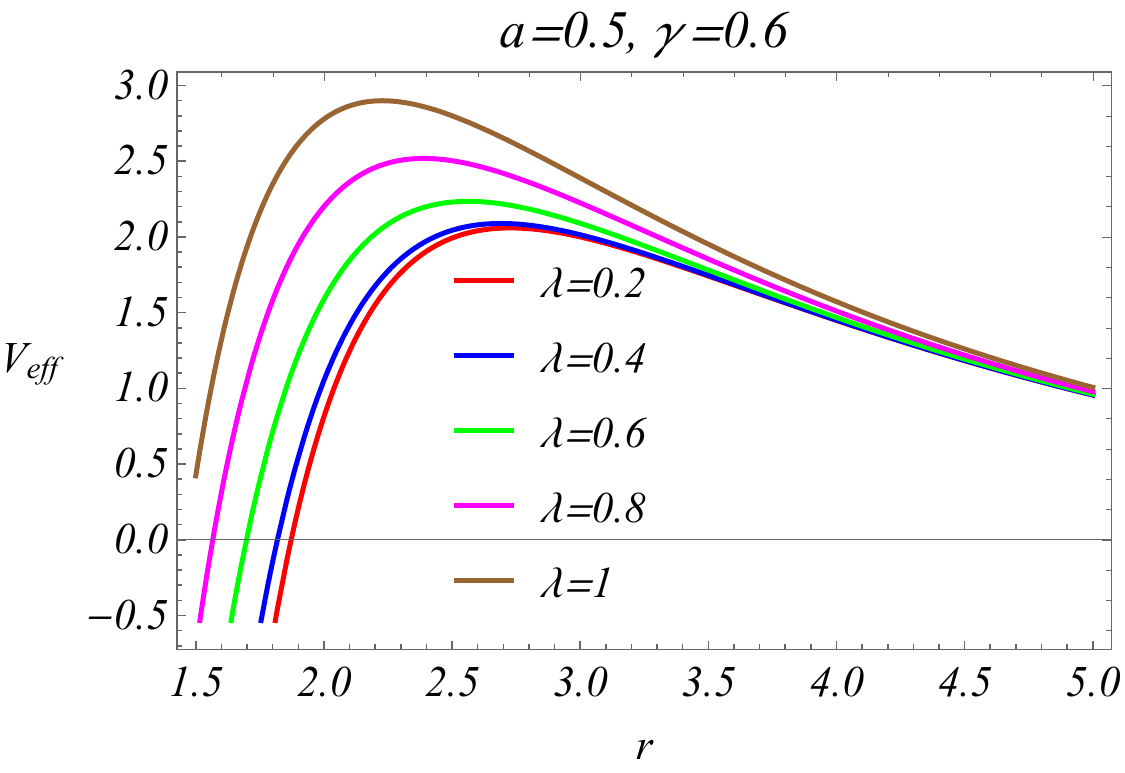}}\\
\subfigure{
\includegraphics[width=0.48\textwidth]{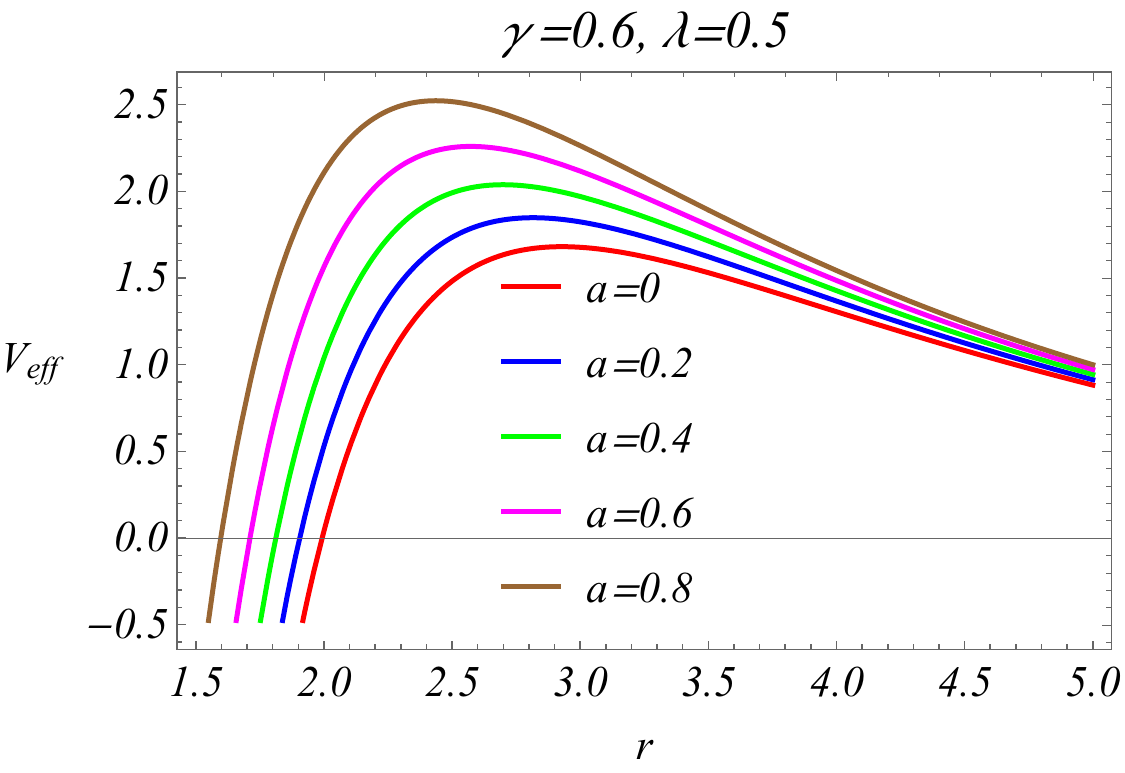}}~~~~
\subfigure{
\includegraphics[width=0.48\textwidth]{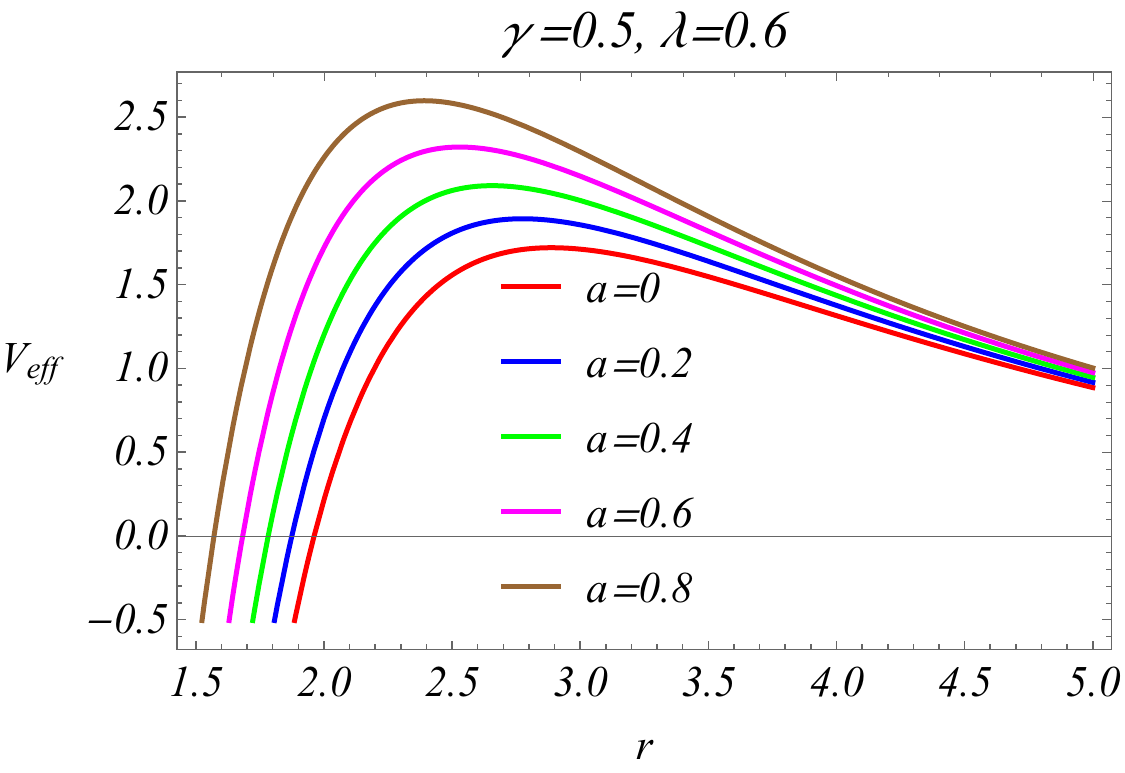}}
\end{center}
\caption{Plots showing the behaviour of effective potential $V_{eff}$ versus $r$.} \label{EP}
\end{figure}
As we are interested in the unstable null circular orbits in the vicinity of event horizon. Therefore, the photons must reside at the surface of 2-sphere with $r=constant$. These orbits are governed by the equations $\dot{r}=0$ and $\ddot{r}=0$, or equivalently, $\mathcal{R}(r)=0$ and $\partial_r\mathcal{R}(r)=0$. These conditions are connected to the effective potential for circular orbits as $V_{eff}(r)=0$ and $\partial_rV_{eff}(r)=0$. Whereas, for the unstable orbits, the local maxima corresponds to the condition $\partial^2_rV_{eff}(r)<0$. From Fig. \ref{EP}, we can see the plots for the behavior of effective potential $V_{eff}(r)$ versus $r$ for certain values of $\gamma$, $\lambda$ and $a$. We know that the local maxima in the effective potential curves describe the unstable circular orbits. Clearly, the peaks rise and are shifted towards the central singularity as we increase the values of $\gamma$, $a$ and $\lambda$ for all cases. It suggests that the unstable circular orbits are reduced in size with increase in $\gamma$, $a$ and $\lambda$. In other words, it means that the null sphere will become shorter with the increase in the values of $\gamma$, $a$ and $\lambda$.

Since, the unstable circular null orbits are responsible for the formation of BH shadow. Therefore, we calculate and plot the shadows for various parametric values for static and rotating BHs. As a reference, Kumar et al. \cite{55} have already calculated and plotted the shadows for some parametric values for an observer in the equatorial plane at infinity. However, to comprehend the study, we intend to calculate the shadows as viewed by the observer off the equatorial plane at infinity. For this, we mainly consider two cases, $\theta_0=\frac{\pi}{6}$ and $\theta_0=\frac{\pi}{3}$ as the observer's angular location. We derive the mathematical scheme for shadows by considering the conditions for circular null orbits and by letting $L_E=\frac{L}{E}$ and $K_E=\frac{\mathcal{Z}+(L-aE)^2}{E^2}$. Therefore, we get
\begin{eqnarray}
K_E(r_p)&=&16r^2\frac{\Delta(r)}{\Delta'(r)^2}\bigg|_{r=r_p}=\frac{4r^2\lambda^2\big(r^2+a^2-2Mr+\gamma r^{\frac{2(\lambda-1)}{\lambda}}\big)}{\big(\lambda(r-M)+\gamma(\lambda-1)r^{\frac{\lambda-2}{\lambda}}\big)^2}\bigg|_{r=r_p}, \label{44} \\
L_E(r_p)&=&\bigg(\frac{r^2+a^2}{a}-\frac{4r\Delta(r)}{a\Delta'(r)}\bigg)\bigg|_{r=r_p}=\bigg(\frac{r^2+a^2}{a}-\frac{2r\big(r^2+a^2-2Mr+\gamma r^{\frac{2(\lambda-1)}{\lambda}}\big)}{a\big(r-M+\gamma\big(1-\frac{1}{\lambda}\big)r^{1-\frac{2}{\lambda}}\big)}\bigg)\bigg|_{r=r_p}, \label{45}
\end{eqnarray}
The Eqs. (\ref{44}) and (\ref{45}) are essentially helpful to calculate the shadow of non-rotating BH. Furthermore, the impact parameters $\xi=\frac{L}{E}$ and $\eta=\frac{\mathcal{Z}}{E^2}$ turn out to be
\begin{eqnarray}
\xi(r_p)&=&L_E(r_p), \label{46}\\
\eta(r_p)&=&\frac{r^2}{a^2\Delta'(r)}\big(16\Delta(r)\big(a^2-\Delta(r)\big)-r^2\Delta'(r)^2+8r\Delta(r)\Delta'(r)\big)\bigg|_{r=r_p}, \nonumber\\
&=&-\frac{\lambda^2r^{3+\frac{4}{\lambda}}\big(r\big(r-3M\big)^2-4Ma^2\big)+r^6\gamma^2\big(\lambda+1\big)^2+2\gamma\lambda r^{4+\frac{2}{\lambda}}\big(2a^2+r(r-3M)(\lambda+1)\big)}{\big(ar\gamma(\lambda-1)+a\lambda r^{\frac{2}{\lambda}}(r-M)\big)^2}\bigg|_{r=r_p}. \label{47}
\end{eqnarray}
As we know that the shadow image of a BH is always two dimensional, therefore, the shadow is projected on the Cartesian celestial plane with coordinates $\alpha$ and $\beta$ given as
\begin{eqnarray}
\alpha&=&-\lim\limits_{r\rightarrow\infty}\bigg(r^2\sin\theta_0\bigg[\frac{d\phi}{dr}\bigg]_{\theta\rightarrow\theta_0}\bigg), \label{48}\\
\beta&=&\lim\limits_{r\rightarrow\infty}\bigg(r^2\bigg[\frac{d\theta}{dr}\bigg]_{\theta\rightarrow\theta_0}\bigg), \label{49}
\end{eqnarray}
such that the observer has been shifted to infinity at an inclination angle $\theta_0$. Solving the differentials in Eqs. (\ref{48}) and (\ref{49}) and by applying the limits, we obtain
\begin{eqnarray}
\alpha(r_p)&=&-\xi(r_p) \csc\theta_0, \label{50} \\
\beta(r_p)&=&\pm\sqrt{\eta(r_p)+a^2\cos^2\theta_0-\xi(r_p)^2\cot^2\theta_0}. \label{51}
\end{eqnarray}
\begin{figure}[t]
\begin{center}
\subfigure{
\includegraphics[width=0.32\textwidth]{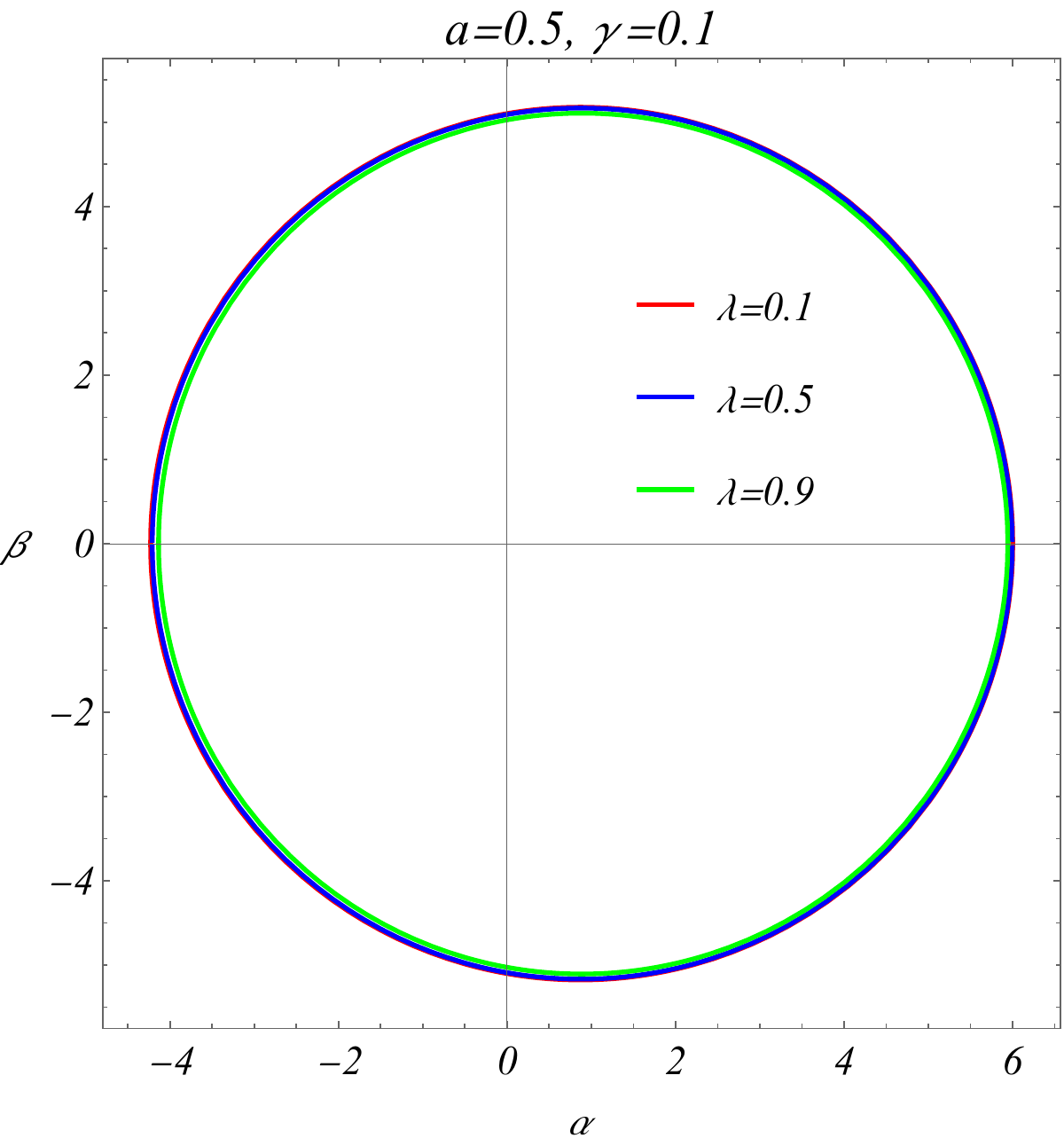}}~~
\subfigure{
\includegraphics[width=0.32\textwidth]{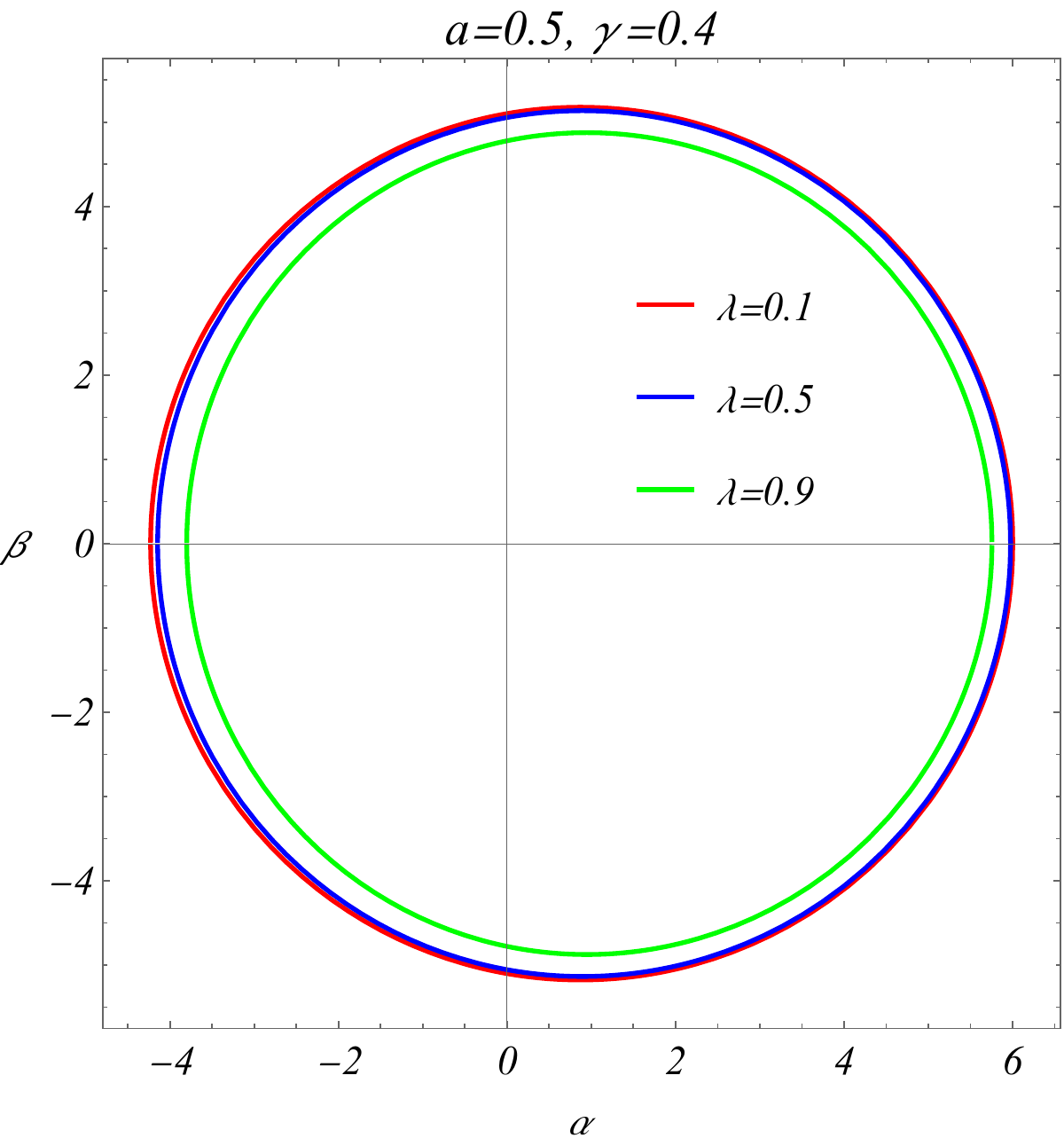}}~~
\subfigure{
\includegraphics[width=0.32\textwidth]{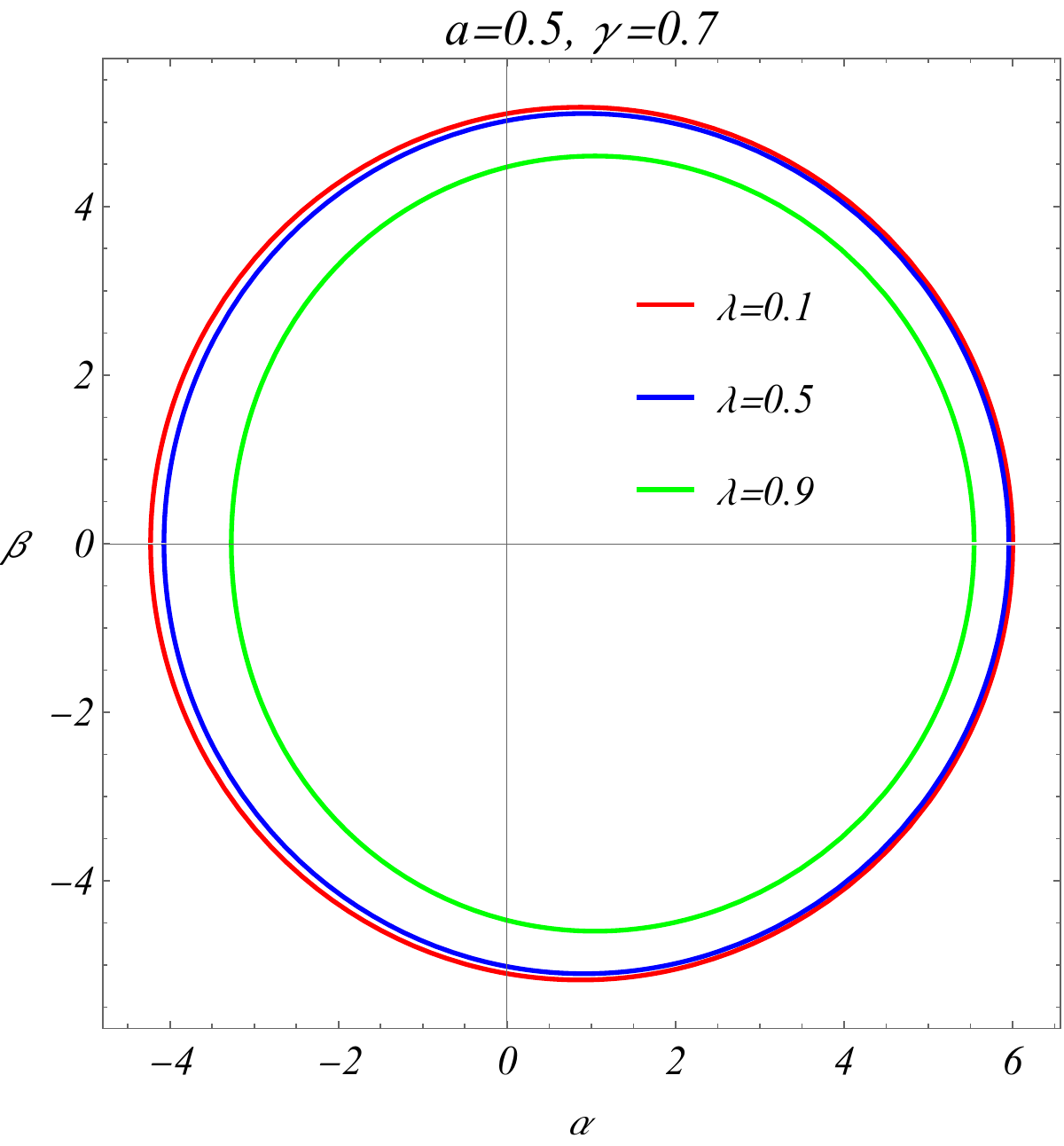}}\\
\subfigure{
\includegraphics[width=0.32\textwidth]{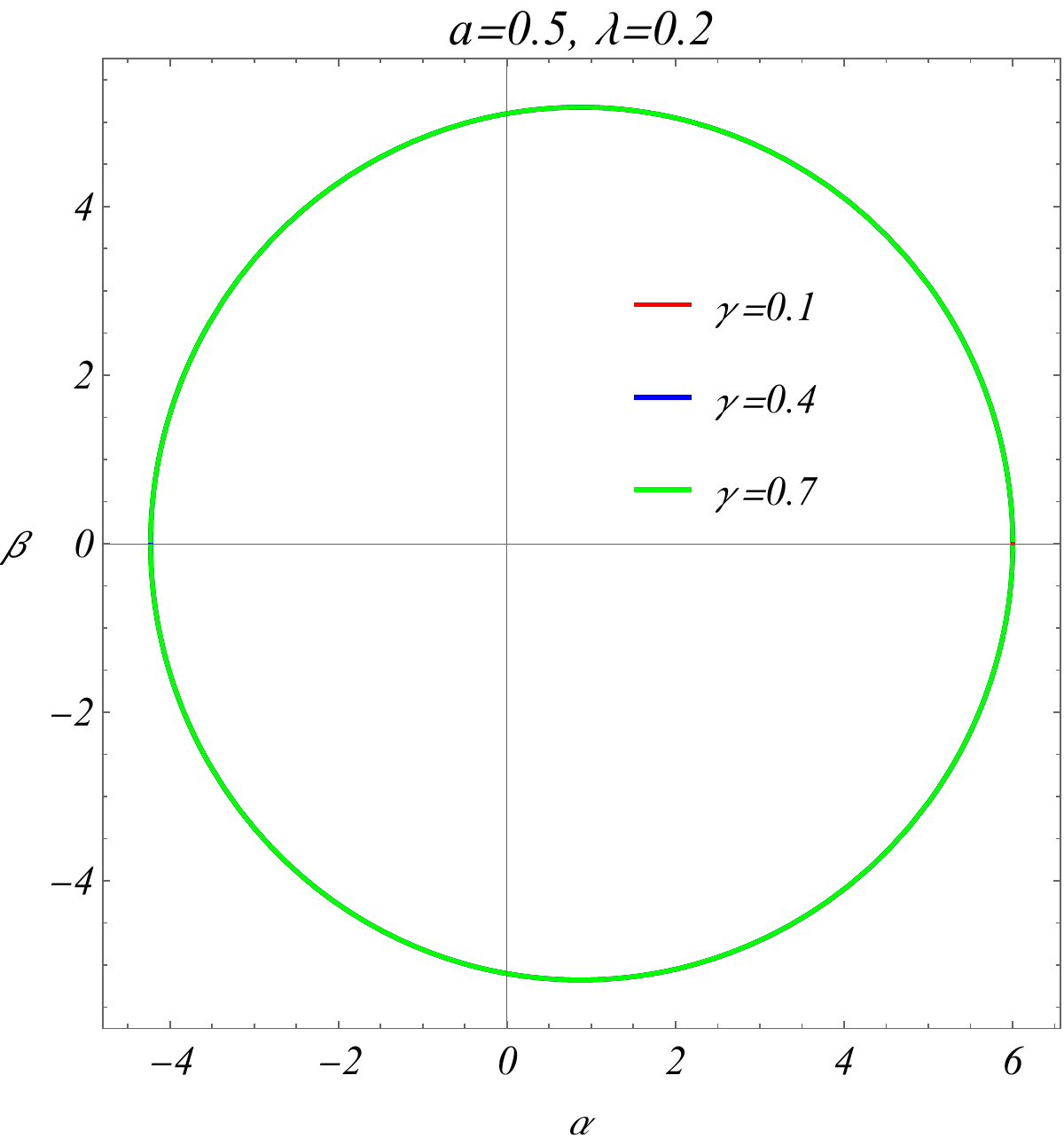}}~~
\subfigure{
\includegraphics[width=0.32\textwidth]{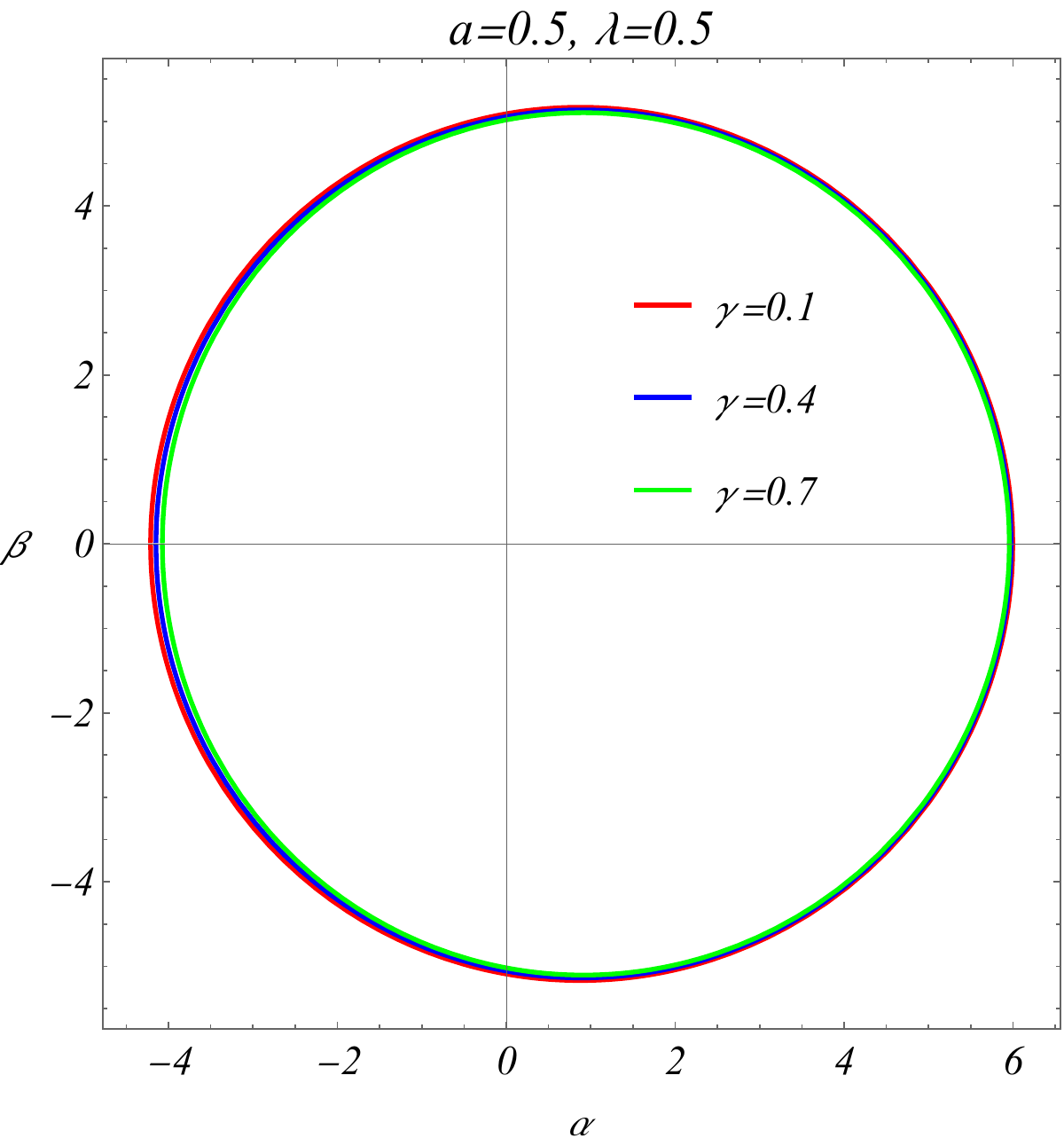}}~~
\subfigure{
\includegraphics[width=0.32\textwidth]{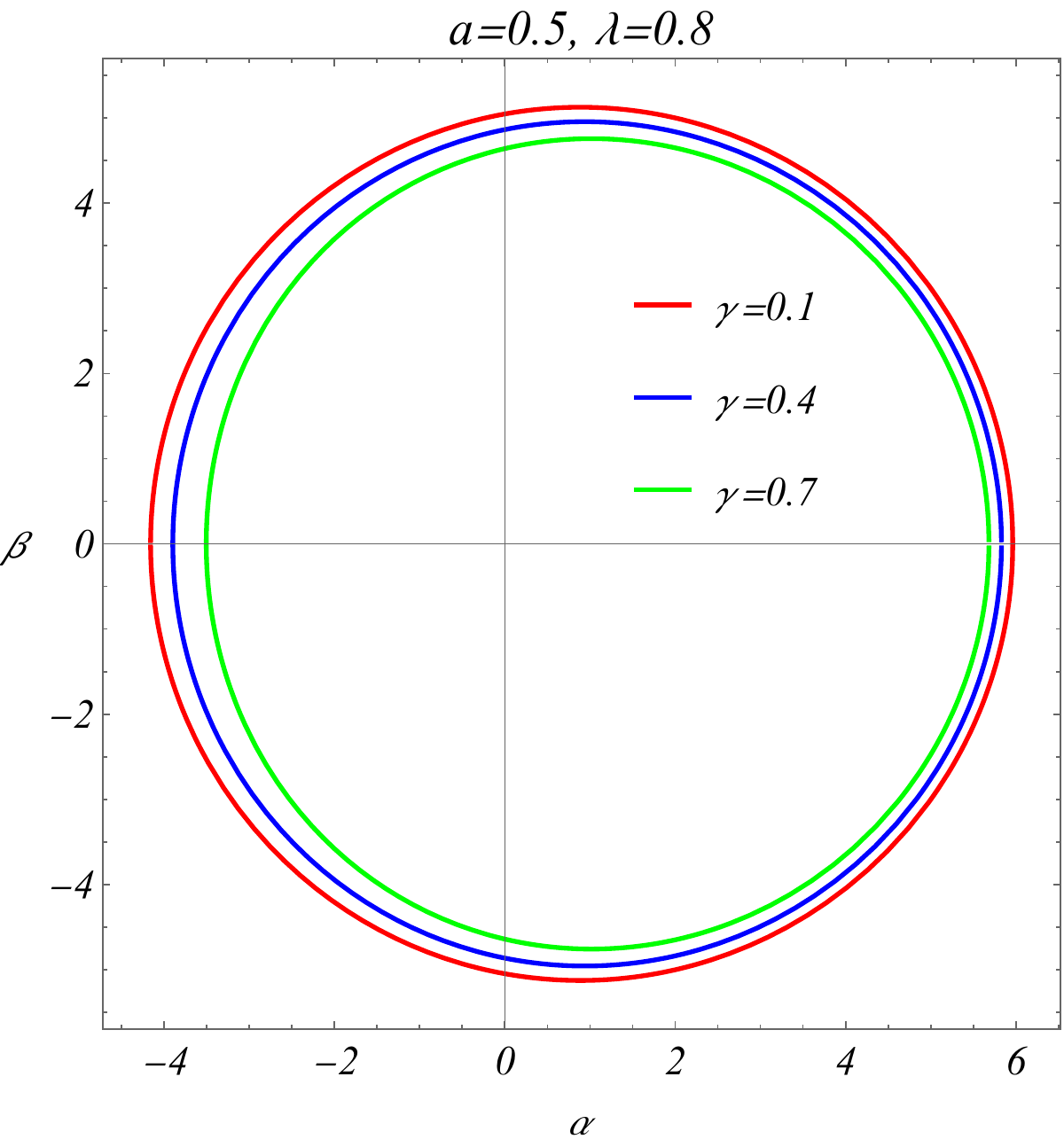}}\\
\subfigure{
\includegraphics[width=0.32\textwidth]{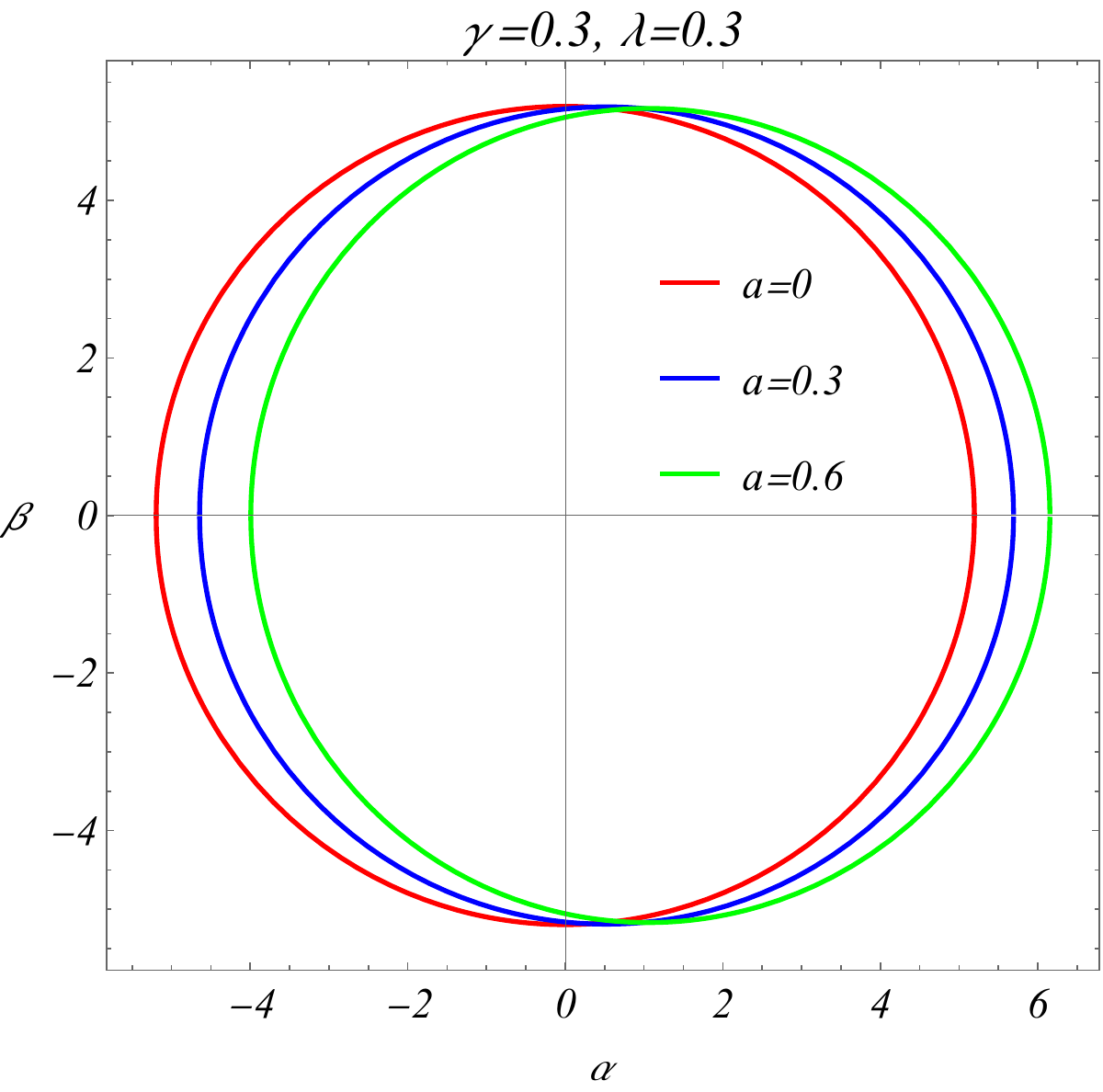}}~~
\subfigure{
\includegraphics[width=0.32\textwidth]{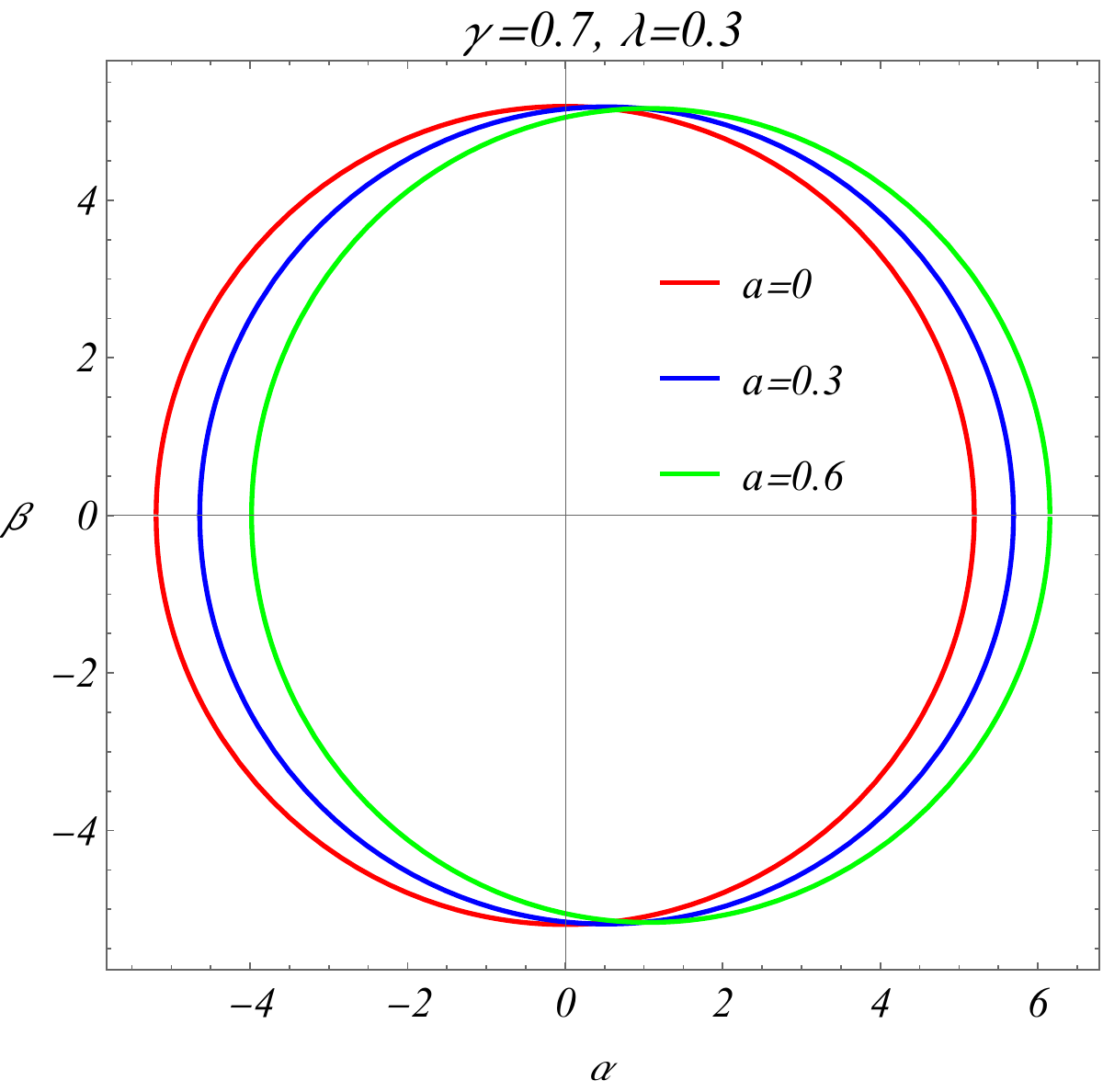}}~~
\subfigure{
\includegraphics[width=0.32\textwidth]{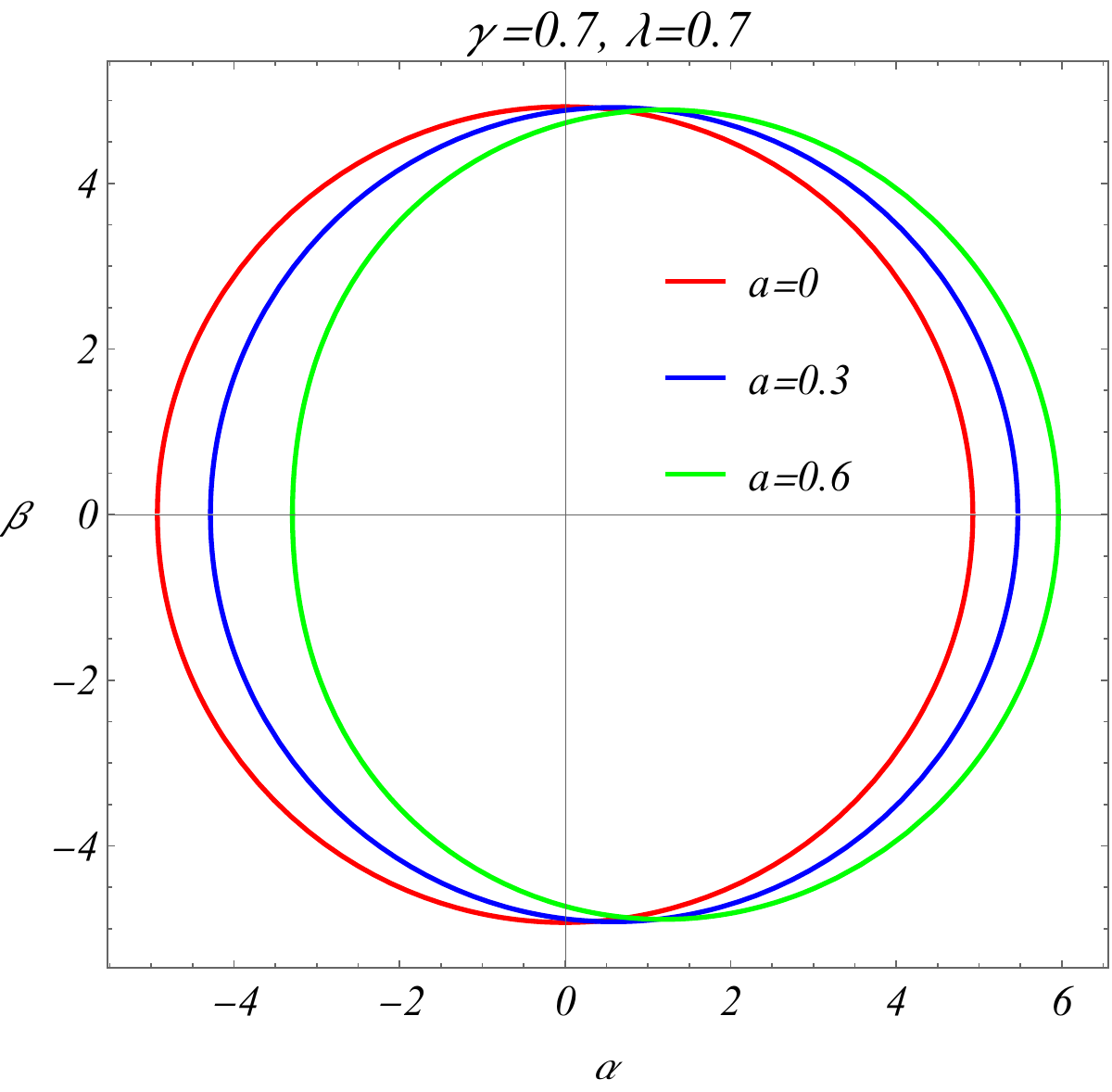}}
\end{center}
\caption{Shadow projections as observed by an observer at $r_0\rightarrow\infty$ and $\theta_0=\frac{\pi}{3}$.} \label{Sh6}
\end{figure}
\begin{figure}[t]
\begin{center}
\subfigure{
\includegraphics[width=0.32\textwidth]{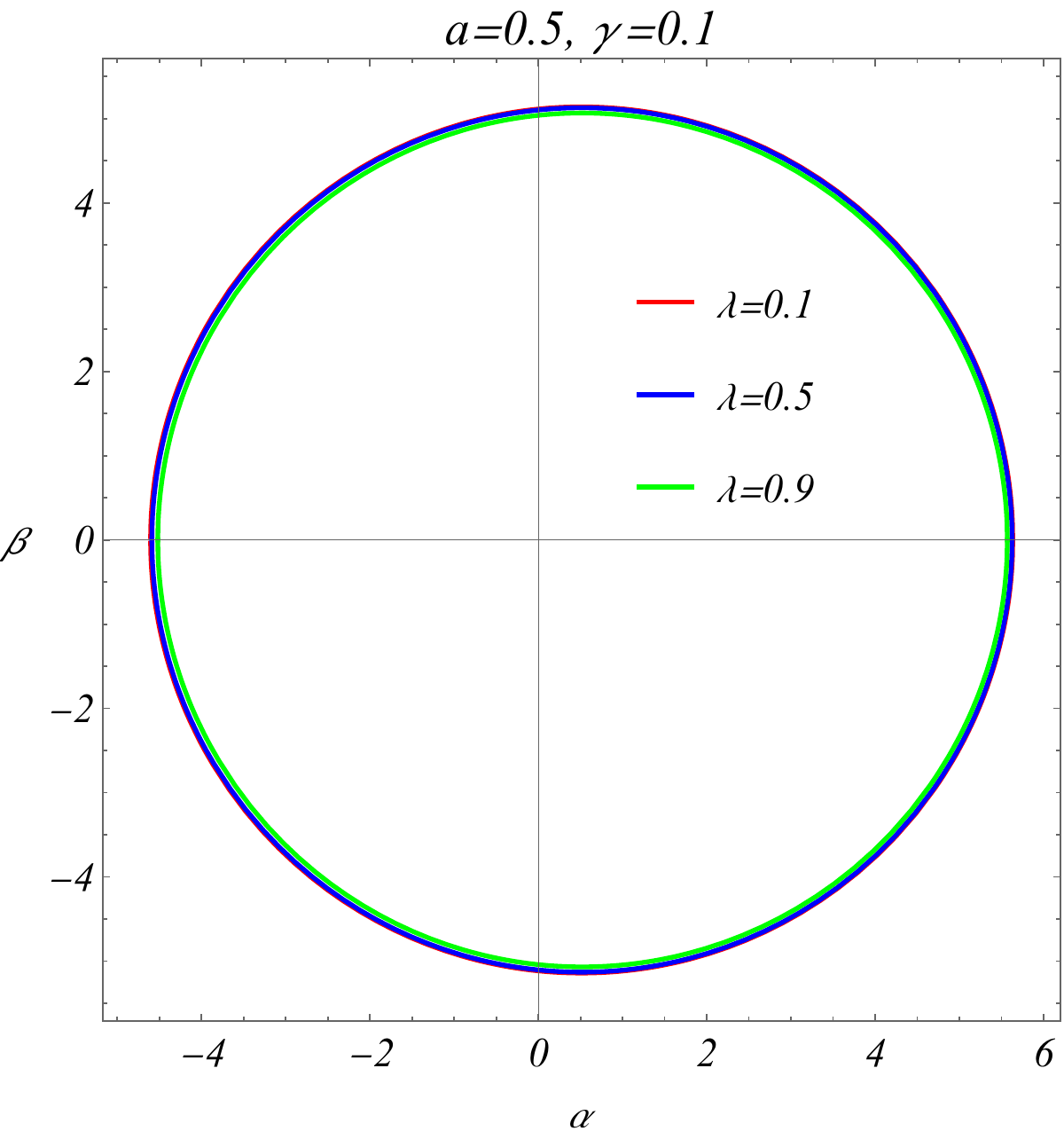}}~~
\subfigure{
\includegraphics[width=0.32\textwidth]{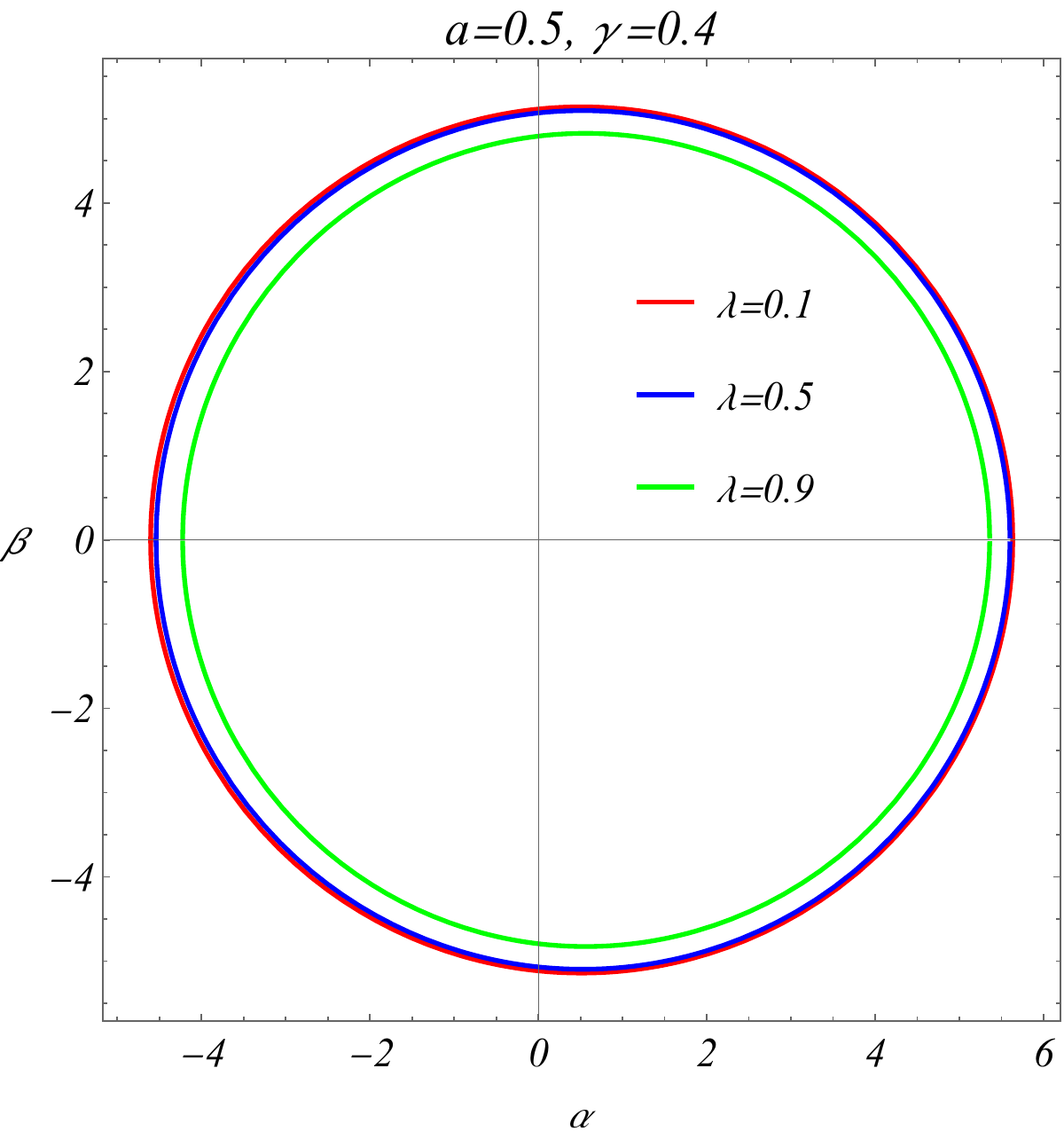}}~~
\subfigure{
\includegraphics[width=0.32\textwidth]{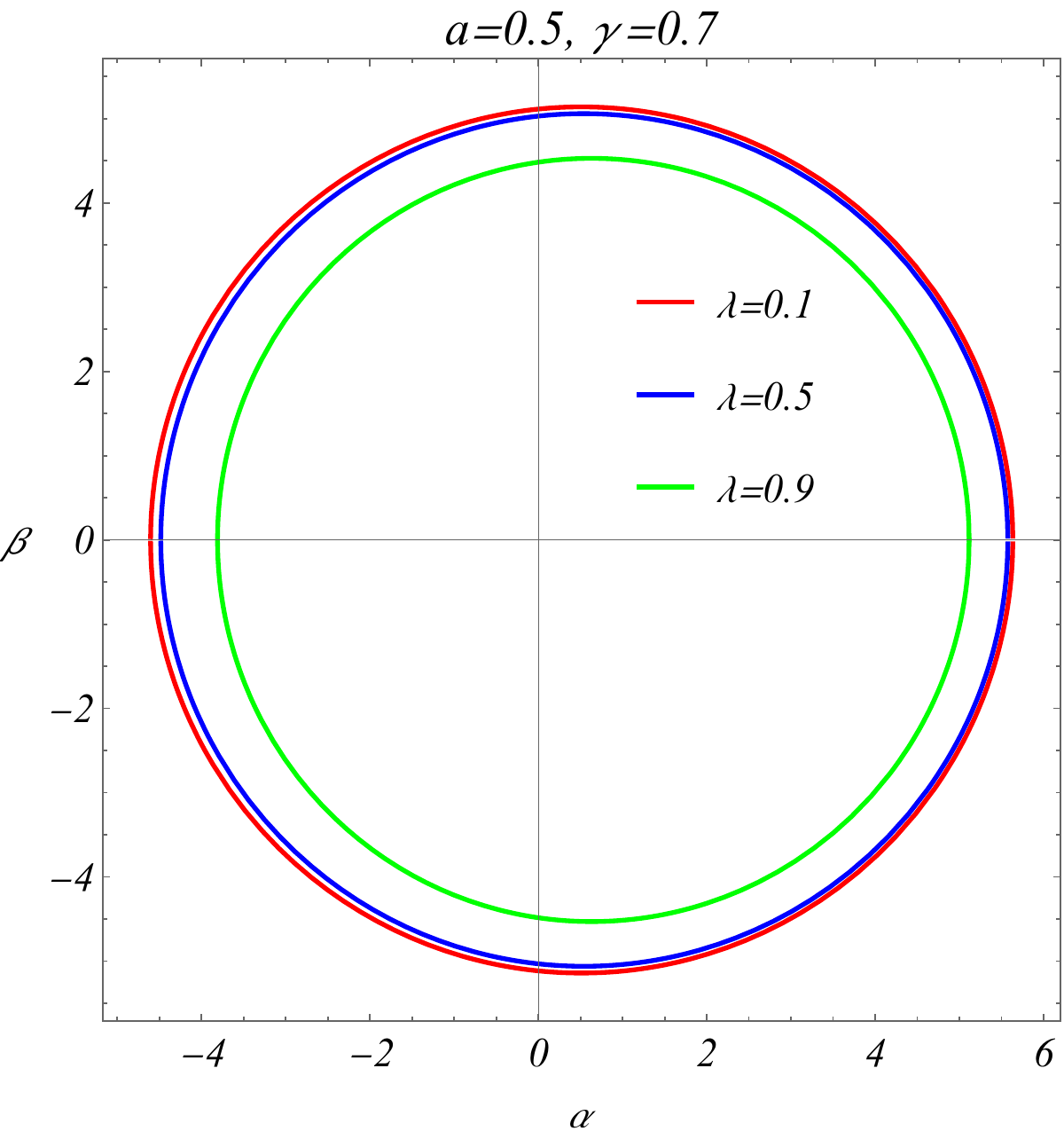}}\\
\subfigure{
\includegraphics[width=0.32\textwidth]{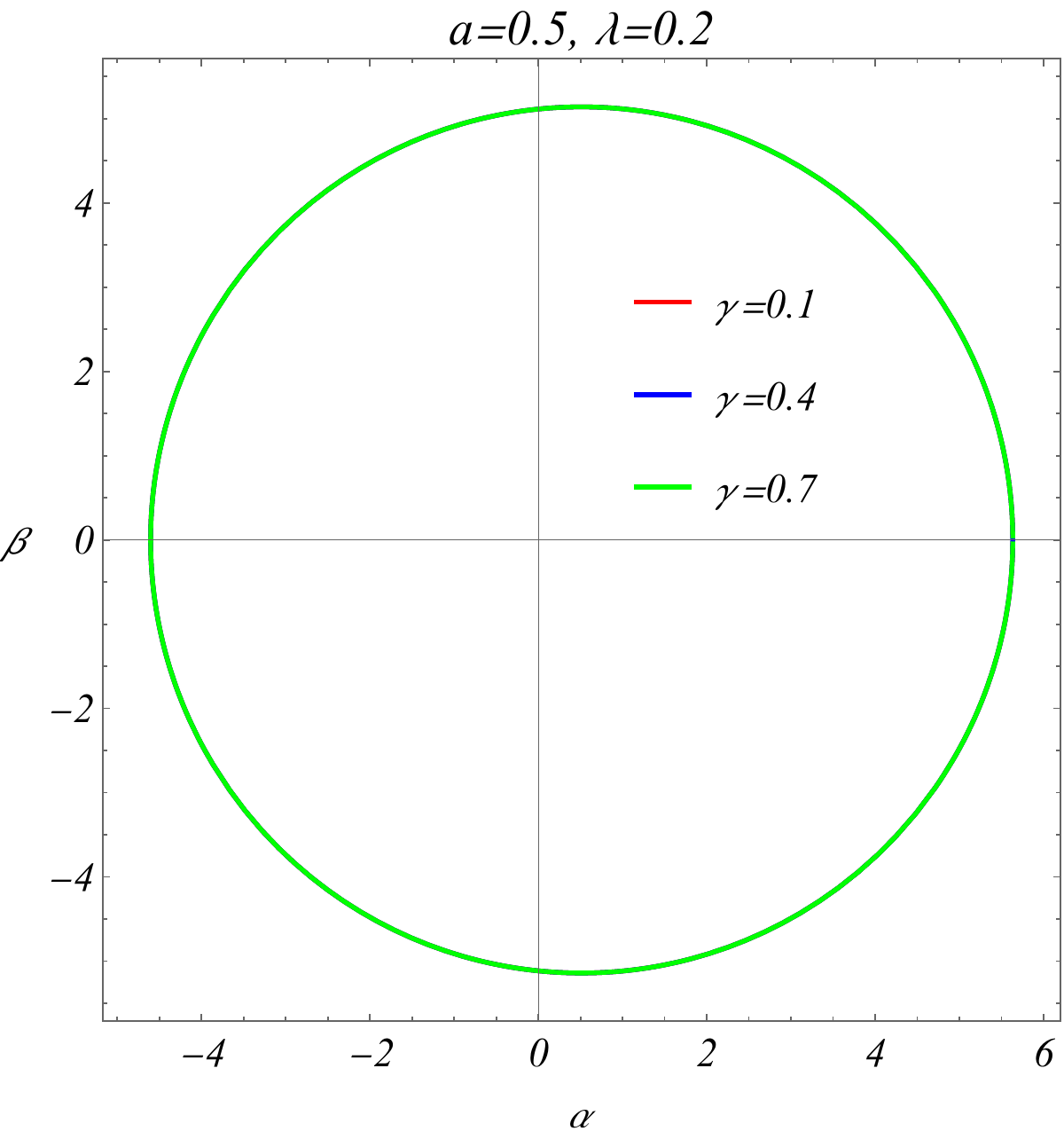}}~~
\subfigure{
\includegraphics[width=0.32\textwidth]{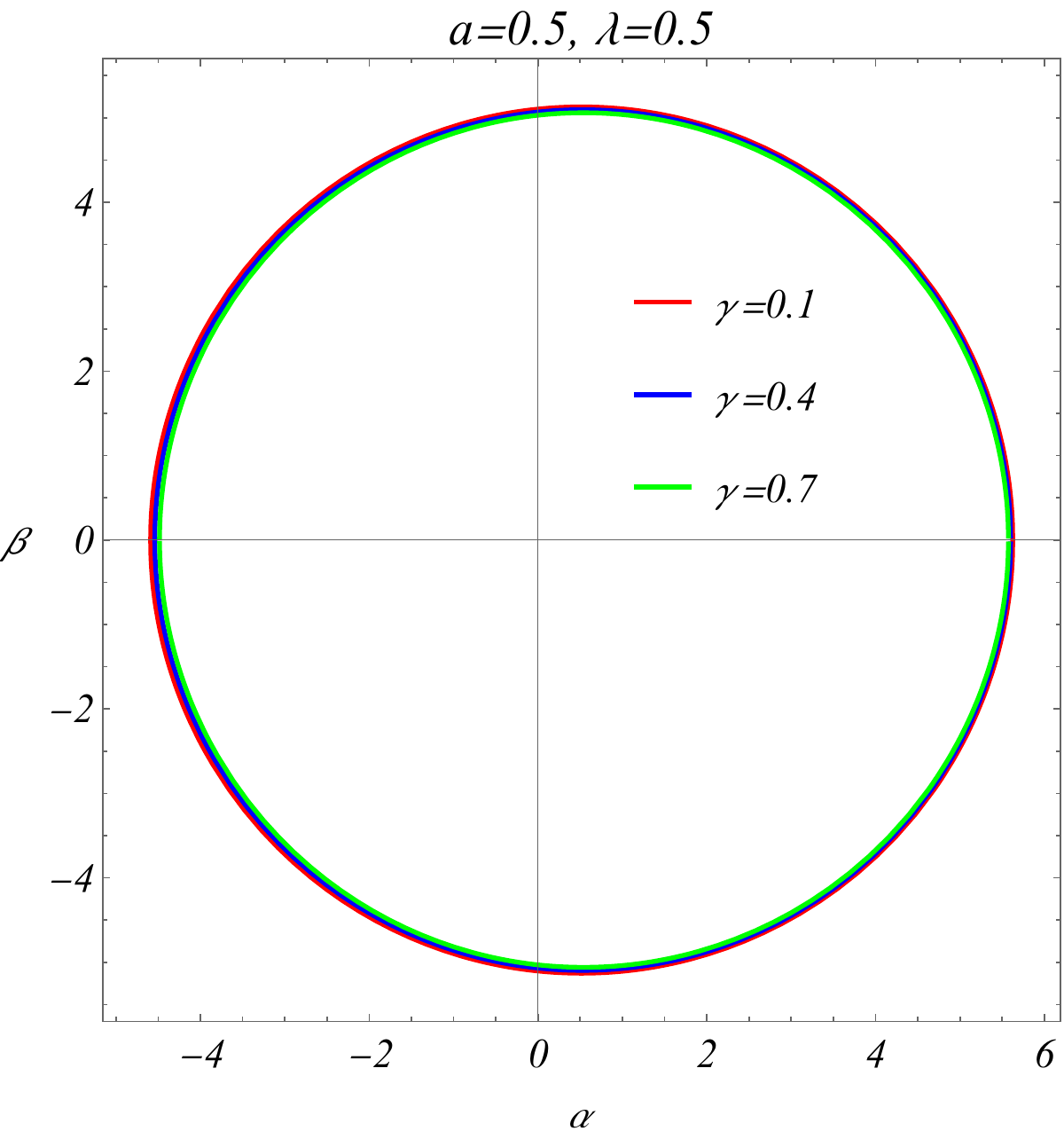}}~~
\subfigure{
\includegraphics[width=0.32\textwidth]{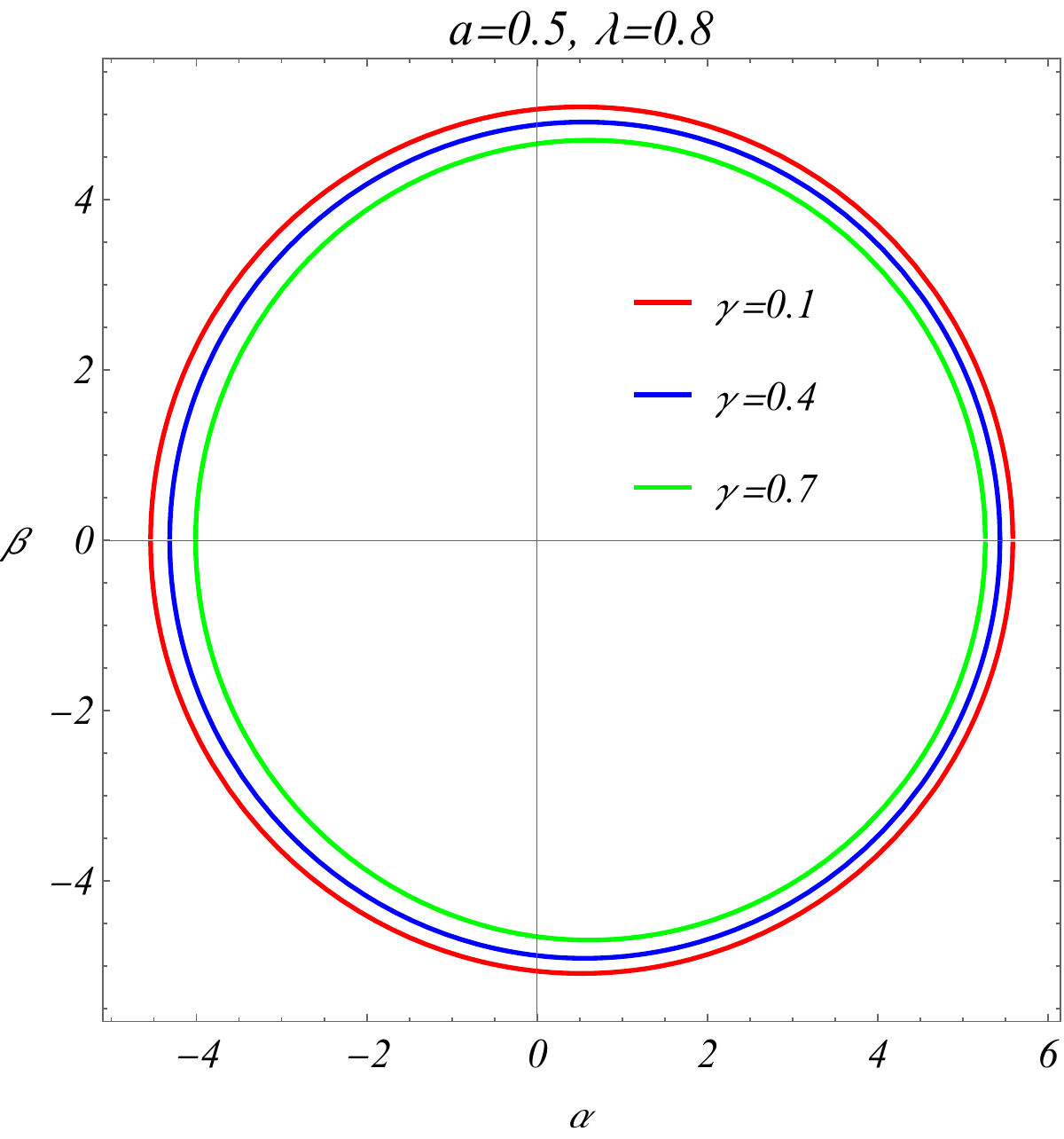}}\\
\subfigure{
\includegraphics[width=0.32\textwidth]{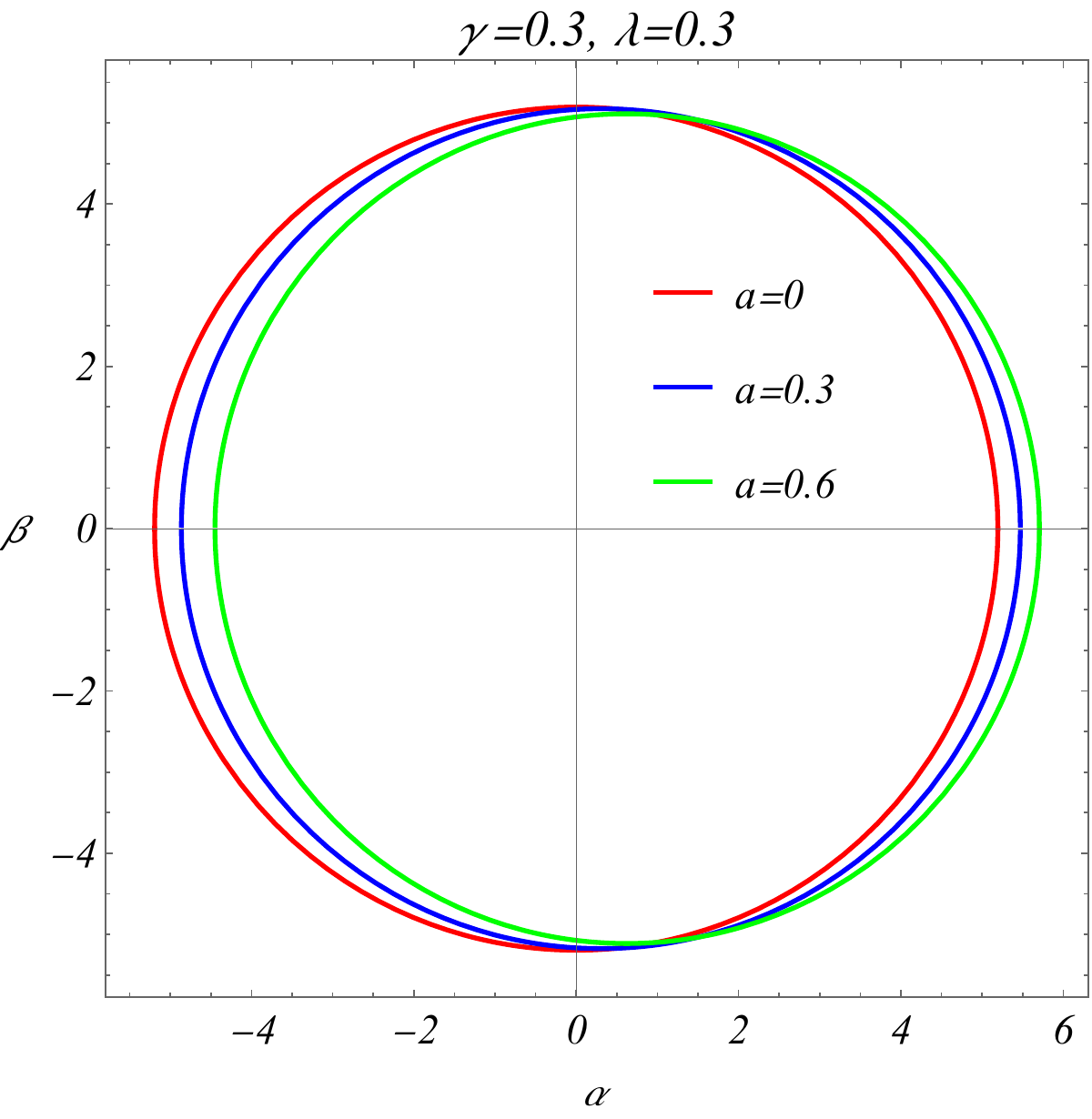}}~~
\subfigure{
\includegraphics[width=0.32\textwidth]{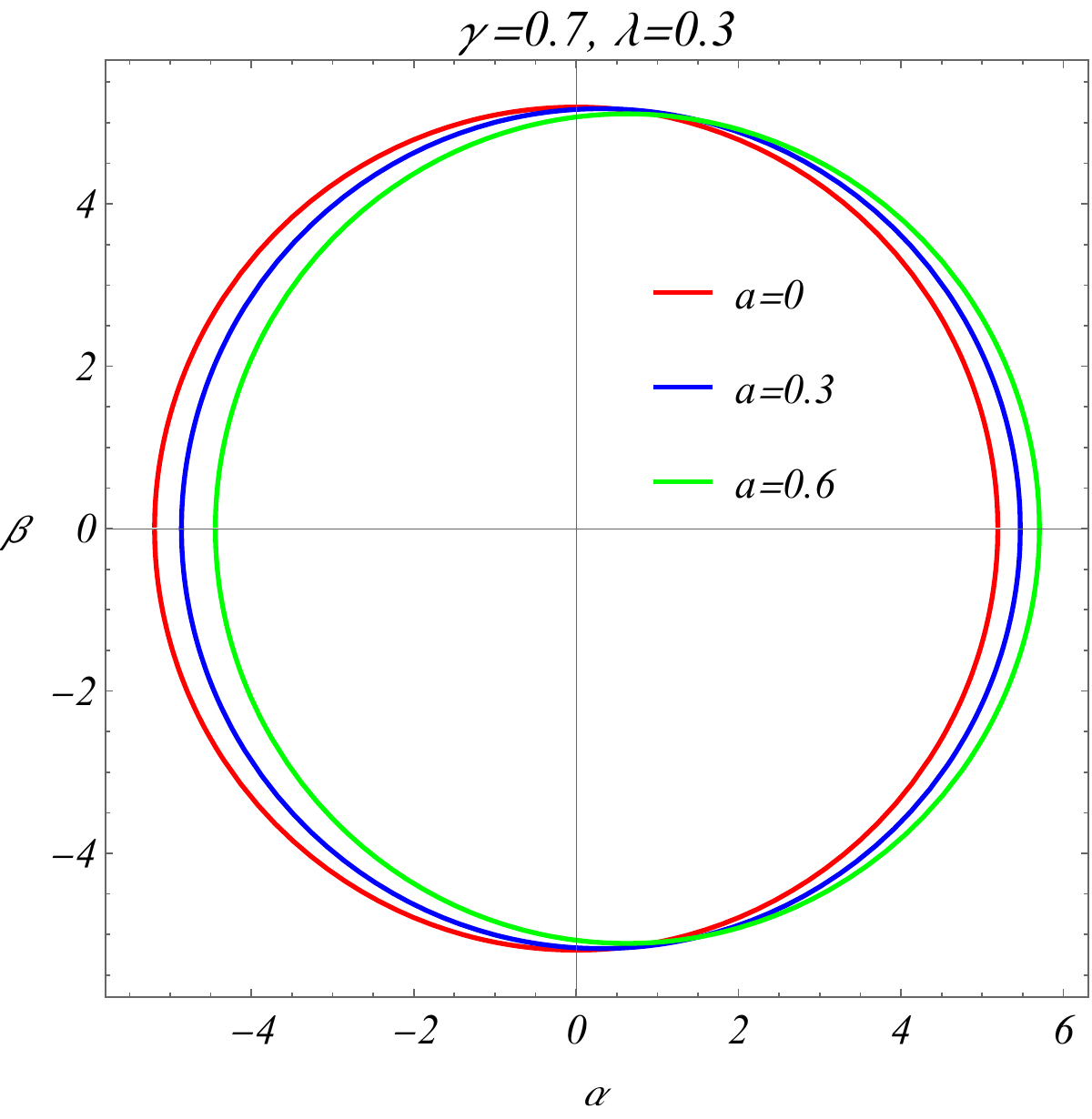}}~~
\subfigure{
\includegraphics[width=0.32\textwidth]{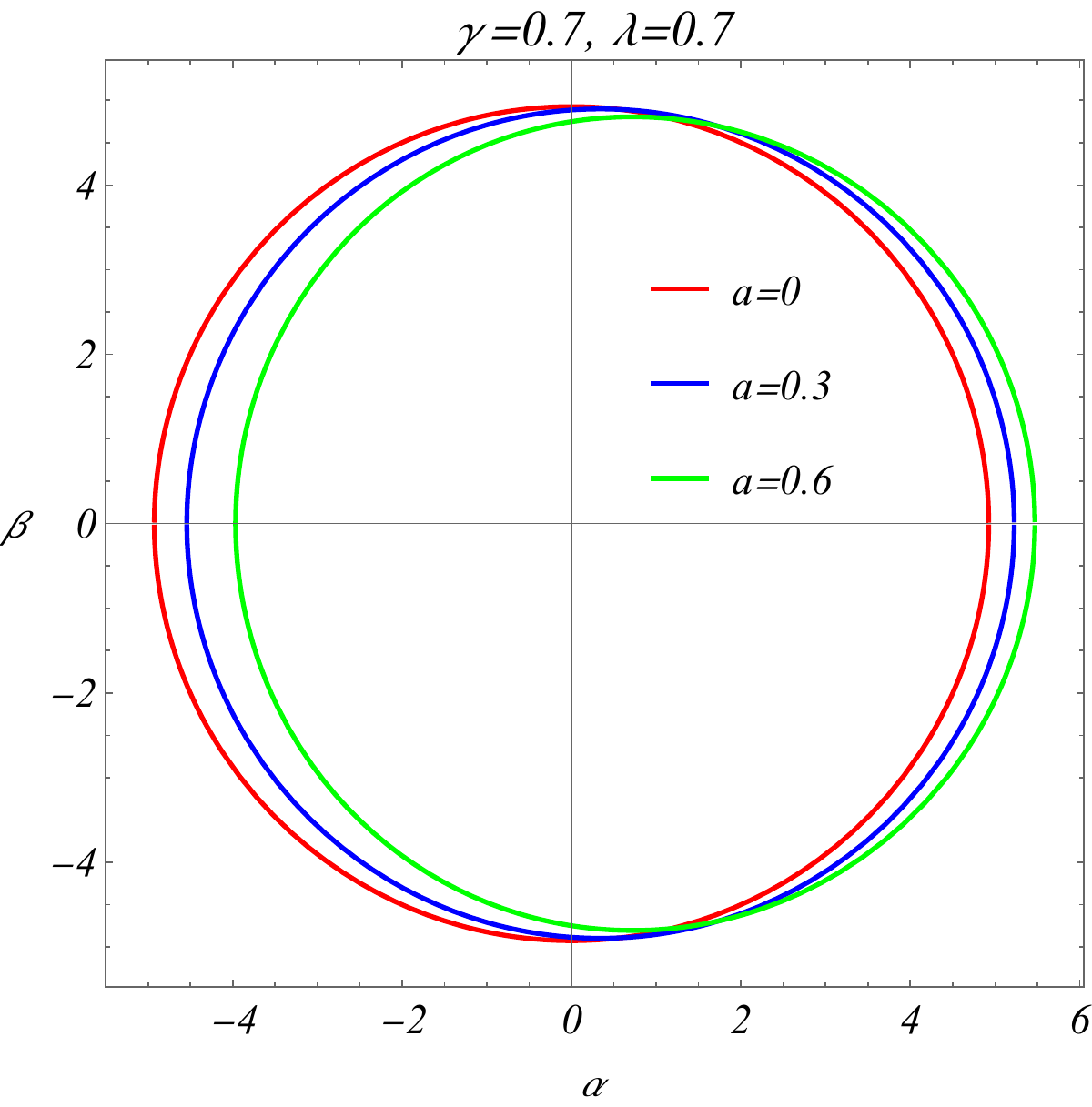}}
\end{center}
\caption{Shadow projections as observed by an observer at $r_0\rightarrow\infty$ and $\theta_0=\frac{\pi}{6}$.} \label{Sh7}
\end{figure}
In the neighborhood of rotating BH, there exists a null spherical shell of a definite thickness characterized by a parametric radial distance $r_p$. The photons travel asymptotically towards this null sphere in spiral trajectories. Therefore, the shadow equations comprise an extra parameter $r_p\in\big[r_{p,min},r_{p,max}\big]$ such that $r_{p,min}$ and $r_{p,max}$ are the smallest and largest radial values of null sphere. These extreme radial values of $r_p$ are obtained by solving the equation $\beta(r)=0$ for real and positive roots. The null sphere around a static BH is a spherical shell having a unit thickness. Therefore, $r_p$ cannot be considered as a parameter. This is because $K_E(r_p)$ has a unique value corresponding to a fixed $r_p$. However, we cannot determine the value of $L_E(r_p)$. Hence, $L_E(r_p)$ can be considered as the parameter. The end point limits of $L_E(r_p)$ lie in the interval that can be obtained by the zeros of $\Theta(\theta_0)$.

Considering the observer off the equatorial plane inclined at $\theta_0=\frac{\pi}{6}$ and $\theta_0=\frac{\pi}{3}$, the shadows have been calculated for some specific values of $a$, $\lambda$ and $\gamma$. The graphical projections are plotted in Figs. \ref{Sh6} and \ref{Sh7} corresponding to $\theta_0=\frac{\pi}{3}$ and $\theta_0=\frac{\pi}{6}$, respectively. The top panel in Fig. \ref{Sh6} has the shadow loops corresponding to the different values of $\lambda$ with fixed $a$, whereas $\gamma$ varies in each plot. It is observed that the shadow size is not much altered with increase in $\lambda$ for small value of $\gamma$. However, as the value of $\gamma$ increases in the panel, the variation of the shadow size is much more visible. Moreover, the size of the shadow reduces with increase in the value of $\lambda$. The flatness due to spin is also observed as $\gamma$ and $\lambda$ increase in the panel. Therefore we can deduce that the flatness due to spin is also dependent upon the other parameters. The curves in the middle panel correspond to the different values of $\gamma$, $\lambda$ varies for each plot in the panel and $a$ is again fixed. The shadow curves show that there is no significant variation in shadow size for small $\lambda$ as $\gamma$ varies in the left plot. However, some variation is observed on the flattened side of the shadow with increase in $\gamma$ for middle plot. Whereas, for higher value of $\lambda$ corresponding to the right plot, a significant variation in the shadow size is observed with a slight increase in the flatness on one side. Following the result from the upper panel, it is again deduced that the flatness of the shadow is also dependent upon $\lambda$ and $\gamma$ as well. In the lower panel, the shadow loops are corresponding to both static and rotating BHs. For a fixed value of $\lambda$ and $\gamma$ in the left plot, the shadow is seen to shift towards right with increase in $a$. For $a=0$, we get exactly circular shadow in all plots. A similar kind of behavior is seen in the middle plot for which $\gamma$ is increased keeping $\lambda$ fixed. However, in the right plot when $\lambda$ is also increased, the shifting of the shadow towards the right is enhanced with increased flatness with increase in spin $a$. The overall shadow size is also altered within the panel.

In order to observe the variation in the size and shape of the shadow with the change in the observer's angle of location, we calculated the shadows for an observer located at $\theta_0=\frac{\pi}{6}$ and are plotted in Fig. \ref{Sh7}. Note that for this case, we kept the same parametric values in order to make a rigorous comparison between the shadows in Figs. \ref{Sh6} and \ref{Sh7}. Generally, the maximum flatness in the shadow is observed at the equatorial plane without affecting the size of the shadow and as we move away from the equatorial plane, the shadows become less flattened and eventually becomes circular at the poles. The same behavior is observed in this case, the shadow size has no significant change. However, the shape of the shadow is altered, i.e., the shadows have become less flattened on either side for all plots.

\subsection{Distortion}
The shadow of the static BH is pure circular image, whereas the rotating BH has a vertically elongated shadow that is shifted on either side with a flattened opposite side. It means that if a shadow shifts rightwards, then the left side of the shadow loop is flattened. This difference of the shadow as compared to the circular shadow is measured by a quantity known as distortion. Furthermore, to measure the distortion, we need an observable called the linear radius \cite{20,69} of the shadow, defined as
\begin{equation}
R_{sh}=\frac{(\alpha_t-\alpha_r)^2+\beta_t^2}{2|\alpha_t-\alpha_r|}. \label{52}
\end{equation}
The linear radius is the radius of an imaginary circle that meets the shadow curve at three distinct points having coordinates $(\alpha_t,\beta_t)$, $(\alpha_b,\beta_b)$ and $(\alpha_r,0)$ residing at the top, bottom and right most point on the shadow loop, respectively. The coordinates of a random point on the shadow loop are given as $(\alpha,\beta)$ such that the top, bottom and right most points are distinguished by the indices $t$, $b$ and $r$, respectively. For more details, these points can be seen on the shadow loop as given in Fig. $\textbf{9}$ in \cite{69}. The relation (\ref{52}) is useful for only rotating BHs as in the case of non-rotating BHs, the imaginary circle will definitely meet the left most point located with coordinates $(-\alpha_r,0)$. Therefore, the curve becomes a pure circle lying at the origin. Hence, the shadow and hypothetical circle will get a same radius. The Eq. (\ref{52}) require the relation for linear radius to further calculate the distortion. As mentioned earlier, our one of the major aim in this manuscript is to analyze the behavior and effect of the BH parameters on the distortion. Therefore, it was needed to calculate the shadows explicitly to obtain the numerical values of the shadow radii. The relation for distortion can be written as
\begin{equation}
\delta_s=\frac{|\bar{\alpha}_l-\alpha_l|}{R_{sh}}, \label{53}
\end{equation}
\begin{figure}[t]
\begin{center}
\subfigure{
\includegraphics[width=0.45\textwidth]{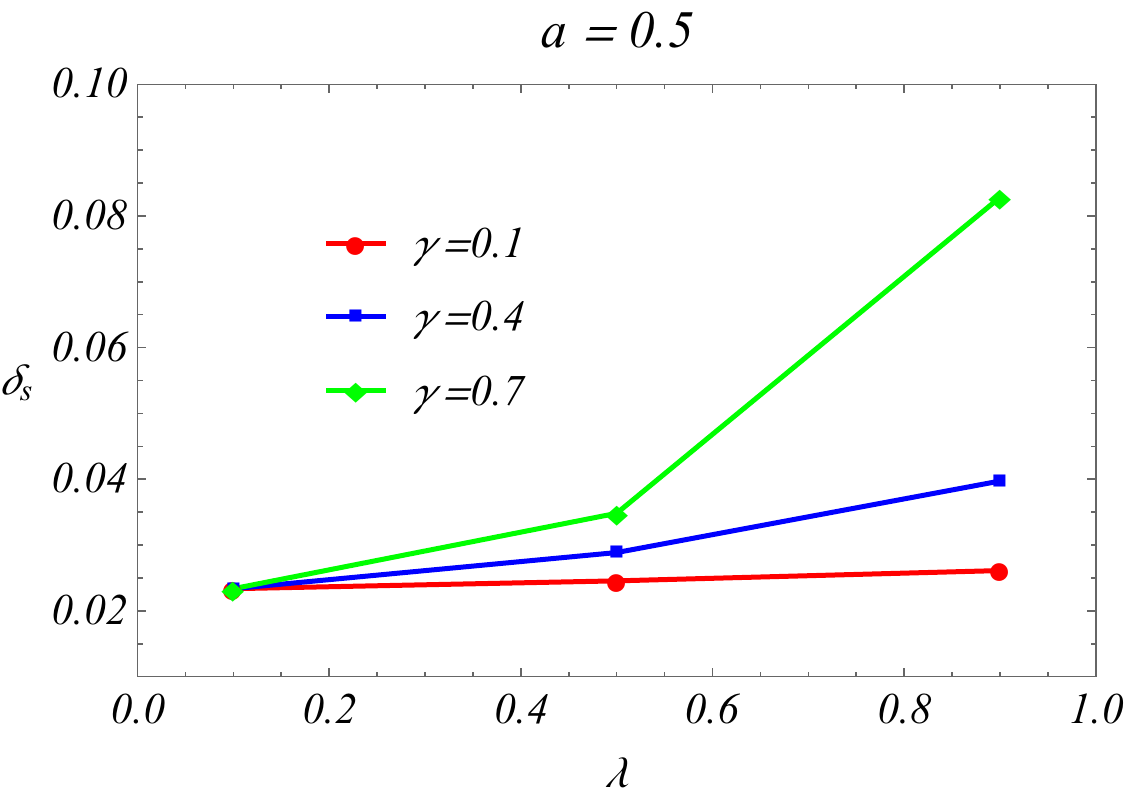}}~~~
\subfigure{
\includegraphics[width=0.45\textwidth]{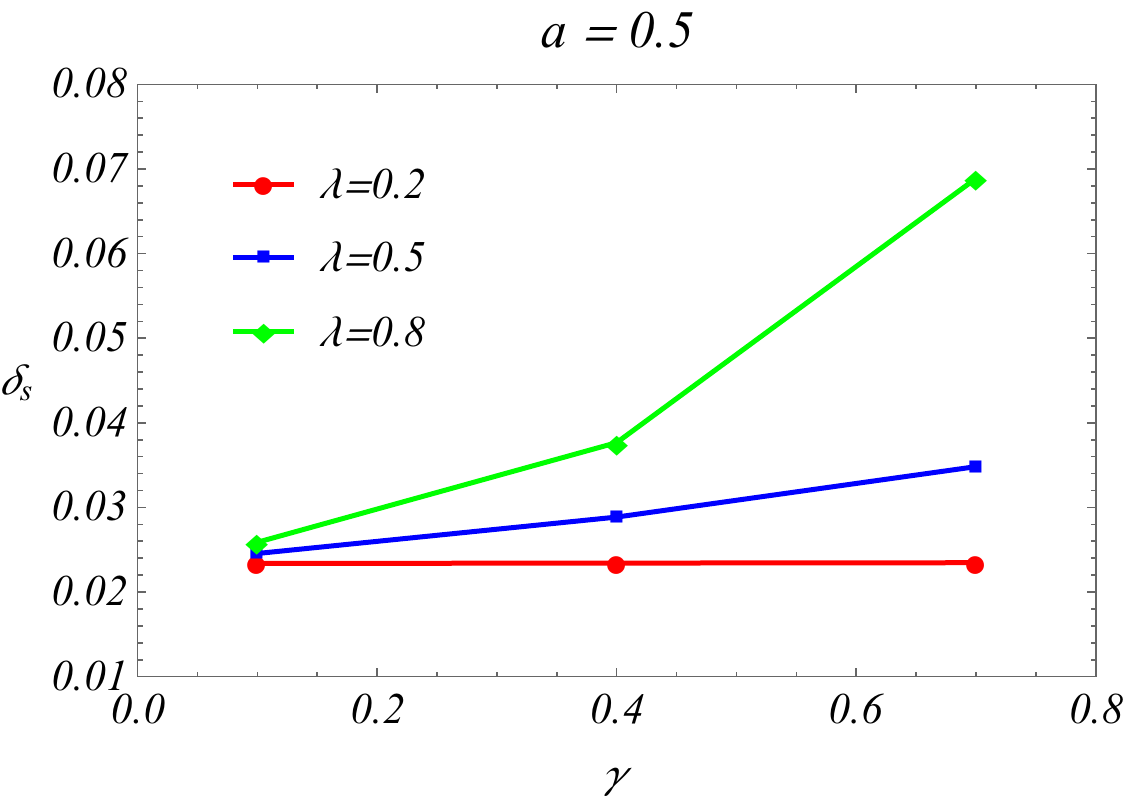}}
\subfigure{
\includegraphics[width=0.45\textwidth]{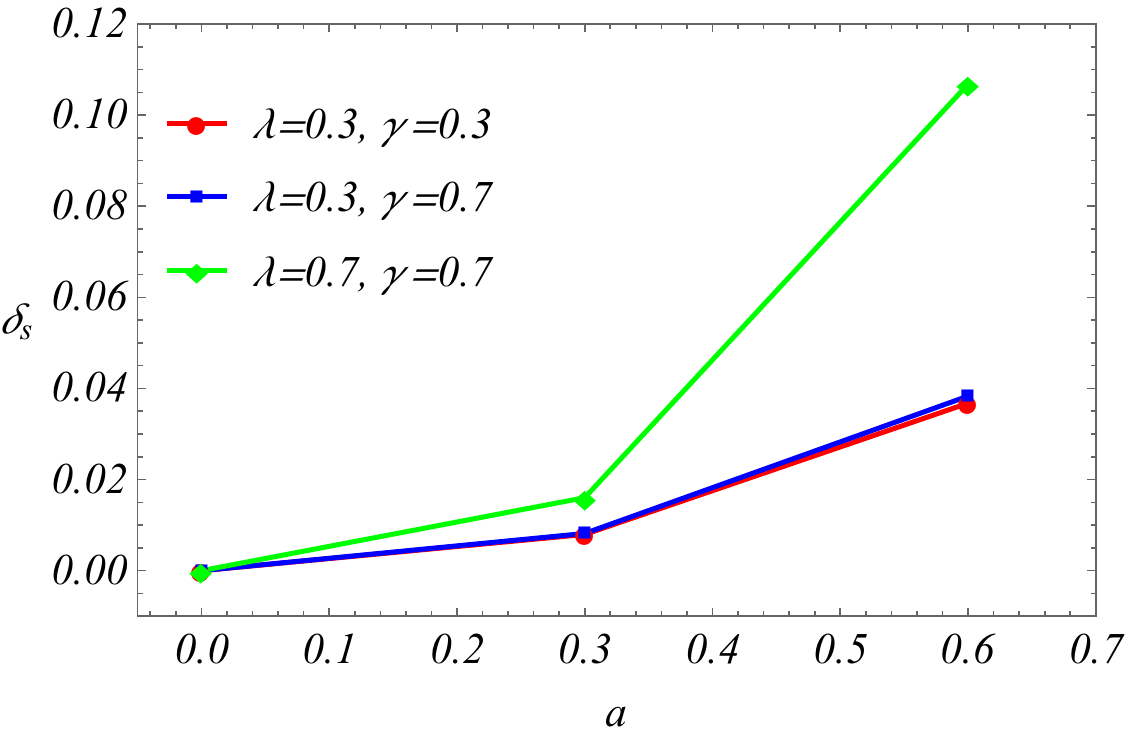}}
\end{center}
\caption{Plots showing the behavior of distortion corresponding to the shadows in Fig. \ref{Sh6}.} \label{Dis8}
\end{figure}
\begin{figure}[t]
\begin{center}
\subfigure{
\includegraphics[width=0.45\textwidth]{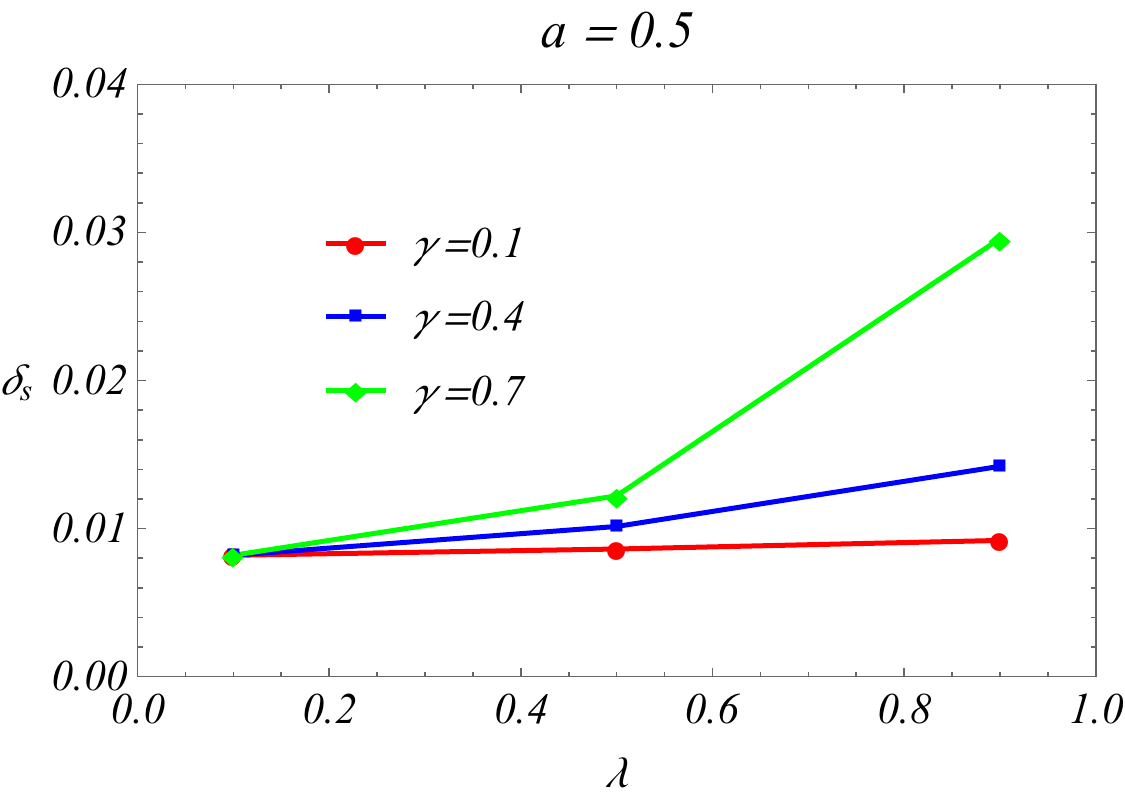}}~~~
\subfigure{
\includegraphics[width=0.45\textwidth]{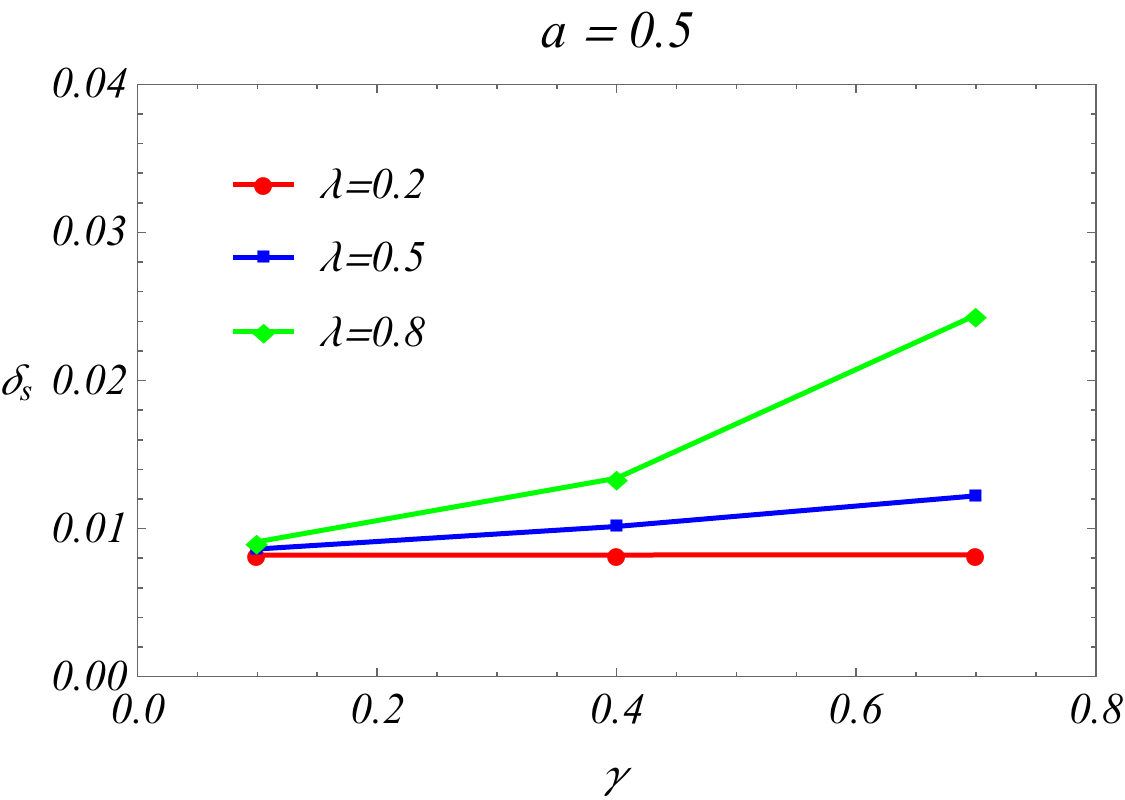}}
\subfigure{
\includegraphics[width=0.45\textwidth]{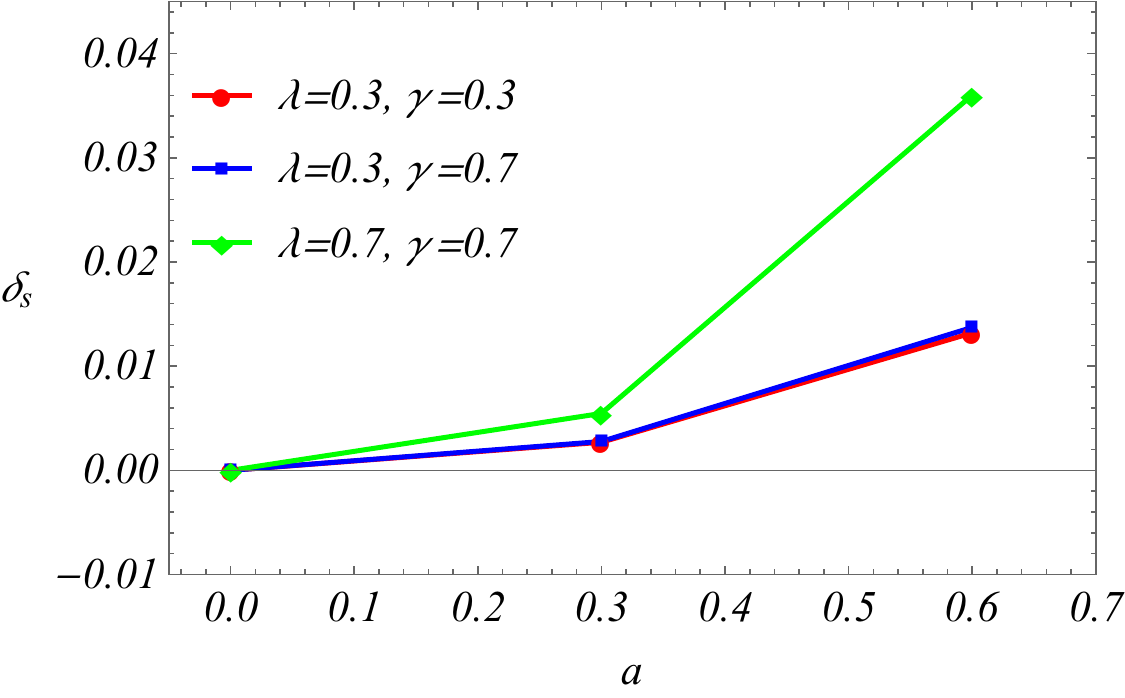}}
\end{center}
\caption{Plots showing the behavior of distortion corresponding to the shadows in Fig. \ref{Sh7}.} \label{Dis9}
\end{figure}
whereas the points $(\alpha_l,0)$ and $(\bar{\alpha}_l,0)$ are located on the shadow loop and hypothetical circle, respectively, crossing the $-\alpha$-axis. The index $l$ represents the shadow loop on the negative $\alpha$-axis, whereas the points on the imaginary circle are represented by $~\bar{}~$. The impact of $\gamma$, $\lambda$ and $a$ can be observed from the graphical sketches of distortion given in the Figs. \ref{Dis8} and \ref{Dis9}.

In Fig. \ref{Dis8}, the upper left plot describes the distortion with increasing value of $\lambda$ for different values of $\gamma$ for each curve and with fixed value of $a$. It is found that the distortion grows with increasing $\lambda$ as the value of $\gamma$ is increased for each curve. For smaller value of $\gamma$, the variation in distortion is negligible. Whereas, the variation in distortion increases as $\gamma$ increases for each curve. Therefore, it can be deduced that the effect of $\lambda$ on distortion is also dependent on $\gamma$ as well. Moreover, for smaller values of $\lambda$, the distortion is almost same for all values of $\gamma$. The rate of increment in distortion increases with increase in both $\lambda$ and $\gamma$. The upper right plot corresponds to the behavior of distortion with increase in $\gamma$ for different values of $\lambda$ for each curve and fixed spin. It is seen that with increasing $\gamma$, there is almost no distortion for small value of $\lambda$ given in red curve. An equivalent result is also observed in the left plot. As the value of $\lambda$ is raised, the distortion also grows w.r.t increasing $\gamma$. Now, as the value of $\lambda$ is further increased, the distortion grows more rapidly as $\gamma$ increases. The lower plot describes the distortion behavior with increase in $a$. There is no distortion for $a=0$ and increases more rapidly as $a$ grows larger. Therefore, it is obvious from these results that all of these parameters have similar kind of effect on the distortion. In general, we know that the spin causes the distortion, however, it is seen that $\gamma$ and $\lambda$ also cause the distortion to increase.

In Fig. \ref{Dis9}, the distortion is depicted corresponding to the shadows in Fig. \ref{Sh7}. Since, the shadows in Fig. \ref{Sh7} showed that there was no significant change in the shape and size of the shadows with the variation of the BH parameters $\lambda$, $\gamma$ and $a$. However, only the change in observer's location angle $\theta_0$ affected the shape of the shadows. This is also obvious from the distortion plots in Fig. \ref{Dis9}, i.e., the variation in the distortion with increase in all parameters in all cases is decreased in comparison with the distortion plots in Fig. \ref{Dis8}. Moreover, the distortion is decreased in comparison with the distortion in Fig. \ref{Dis8}. This is obviously because of the fact that the shadow becomes circular as the observer approaches the poles, off the equatorial plane. The behavior of the curves and the variation in the distortion with the increase in $\lambda$, $\gamma$ and $a$ in each curve is same as that in the Fig. \ref{Dis8}.

\subsection{Energy Emission Rate}
As we know that a BH is the region of extremely strong gravitational attraction. From the classical point of view, an object is lost forever if it is swallowed by a BH. However, under the Quantum Mechanical perspective, a BH does emit. Inside the event horizon of a BH, particles are created and annihilated due to the influence of quantum fluctuations. Due to extremely high density and pressure, a huge amount of energy is caused. Therefore, quantum tunneling process causes these positive energy particles to cross the event horizon and escape away to infinity. Thus, the associated energy with these particles is also released that enables the BH to evaporate. The measurement of an absorption process such as the absorption of photons by a molecule, is accomplished in terms of the probability known as absorption cross section. Away from a BH, the absorption cross section of high energy particles possess a relation with BH shadow that can be determined by a fixed value $\sigma_{lim}$. The geometric area of null sphere is found to be approximately equal to $\sigma_{lim}$. It is known that the innermost null orbit is the covering layer of the shadow and therefore it possesses a relation with the absorption cross section \cite{70,71,72} as
\begin{equation}
\sigma_{lim}\approx\pi R_{sh}^2. \label{54}
\end{equation}
The energy emission rate takes the form as
\begin{equation}
\mathcal{E}_{\omega t}:=\frac{d^2\mathcal{E}(\omega)}{dtd\omega} =\frac{2\pi^2\omega^3\sigma_{lim}}{e^{\frac{\omega}{T_H}}-1}\approx\frac{\pi^3\omega^3}{2\big(e^{\frac{\omega}{T_H}}-1\big)}\frac{\big((\alpha_t-\alpha_r)^2+\beta_t^2\big)^2}{|\alpha_t-\alpha_r|^2}, \label{55}
\end{equation}
where the angular frequency can be indicated by $\omega$ and $T_H=\frac{\kappa}{2\pi}$ is the Hawking temperature. The relation
\begin{equation}
\kappa(r_h)=\lim\limits_{\theta=0, r\rightarrow r_h}\frac{\partial_r\sqrt{g_{tt}}}{\sqrt{g_{rr}}} \label{56}
\end{equation}
determines the surface gravity of a rotating BH at the event horizon $r_h$ \cite{73} that reduces into
\begin{equation}
\kappa(r_h)=\frac{\Delta'(r_h)}{2(r_h^2+a^2)} \label{57}
\end{equation}
for the rotating metric. The surface gravity at the outer horizon for the static BH can be derived by considering $a=0$ in Eq. (\ref{57}) and can be written as
\begin{equation}
\kappa(r_h)=\frac{1}{2}f'(r_h). \label{58}
\end{equation}
\begin{figure}[t]
\begin{center}
\subfigure{
\includegraphics[width=0.32\textwidth]{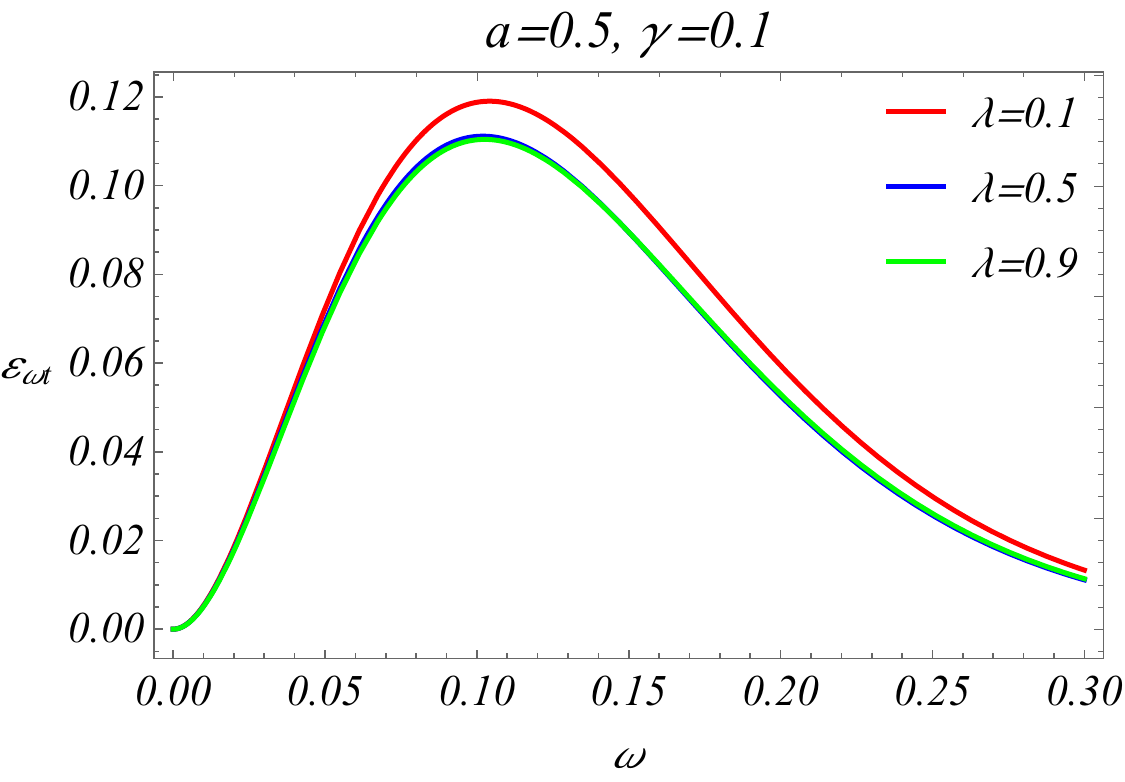}}~
\subfigure{
\includegraphics[width=0.32\textwidth]{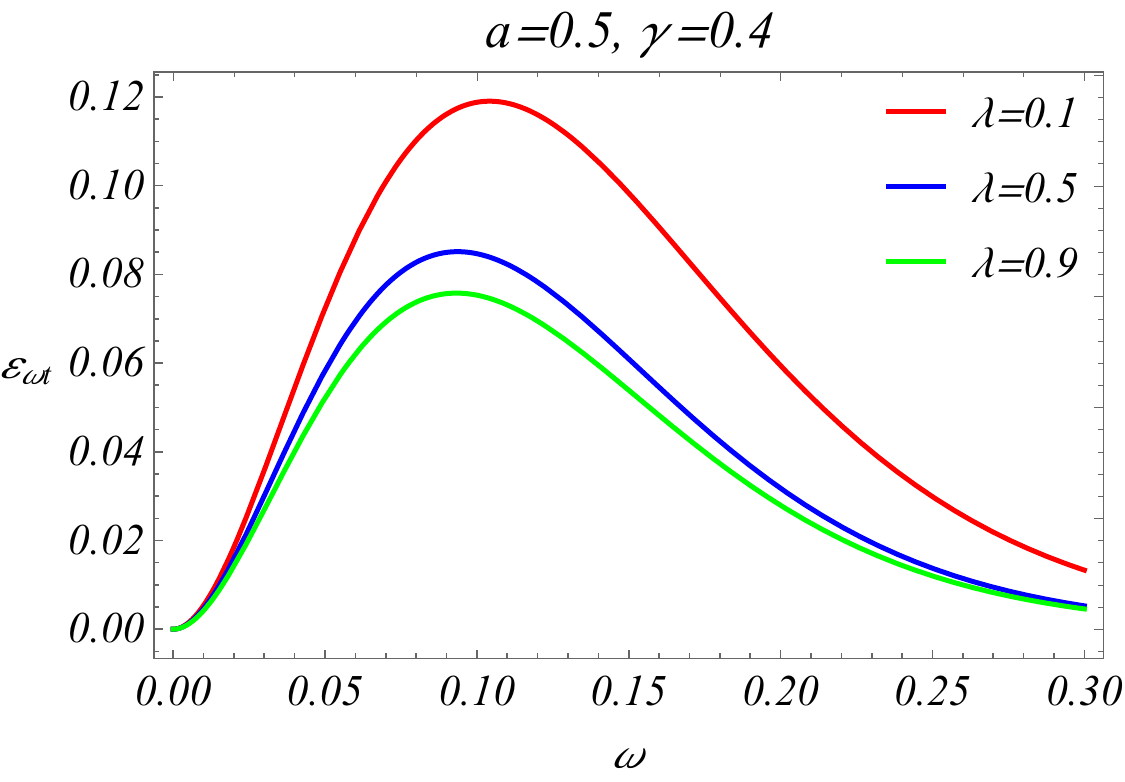}}~
\subfigure{
\includegraphics[width=0.32\textwidth]{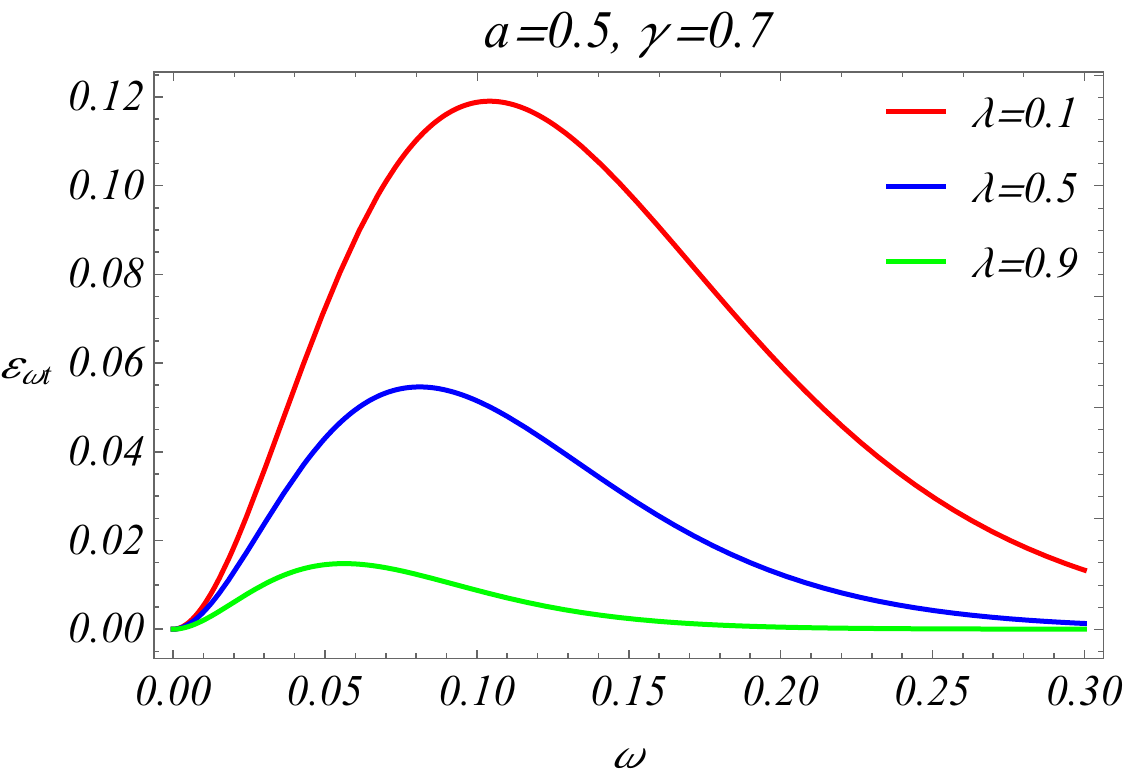}}
\subfigure{
\includegraphics[width=0.32\textwidth]{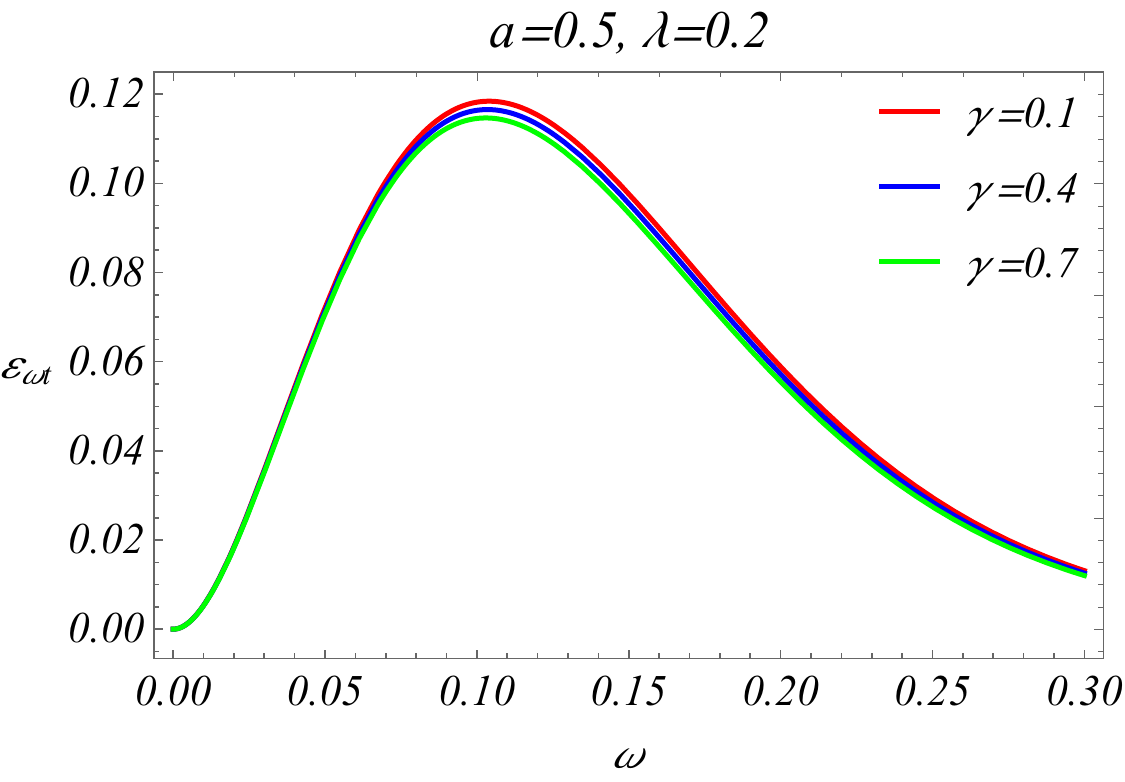}}~
\subfigure{
\includegraphics[width=0.32\textwidth]{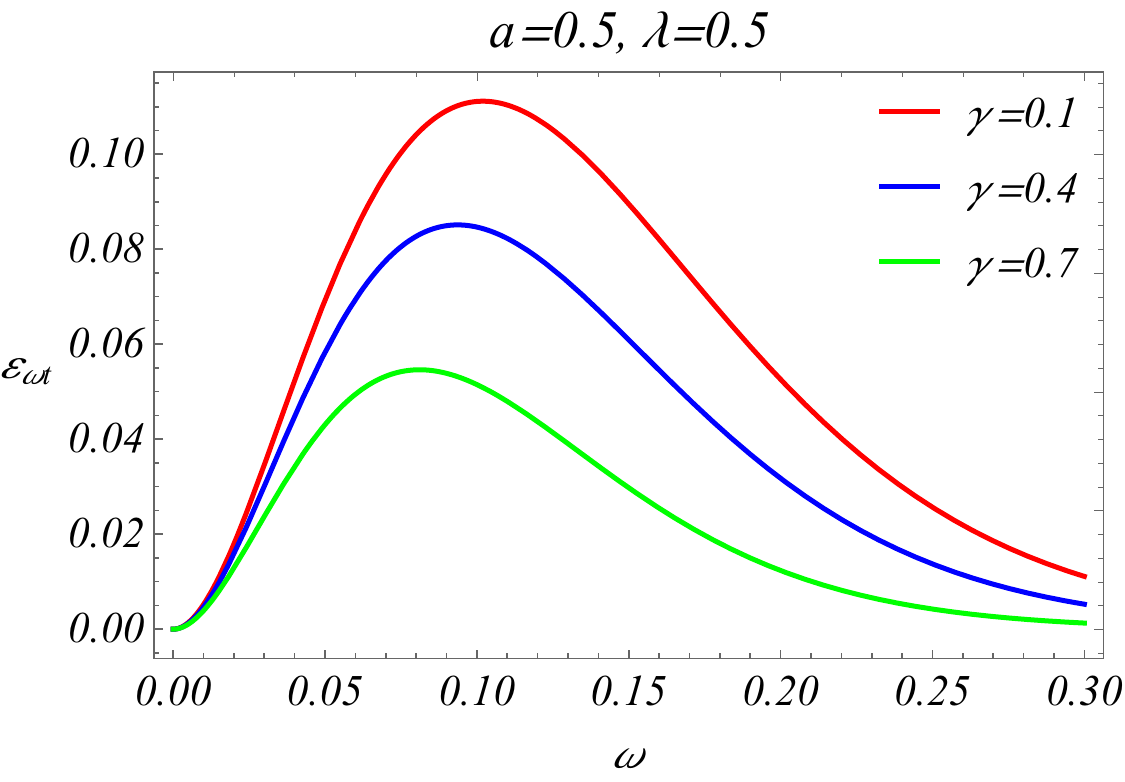}}~
\subfigure{
\includegraphics[width=0.32\textwidth]{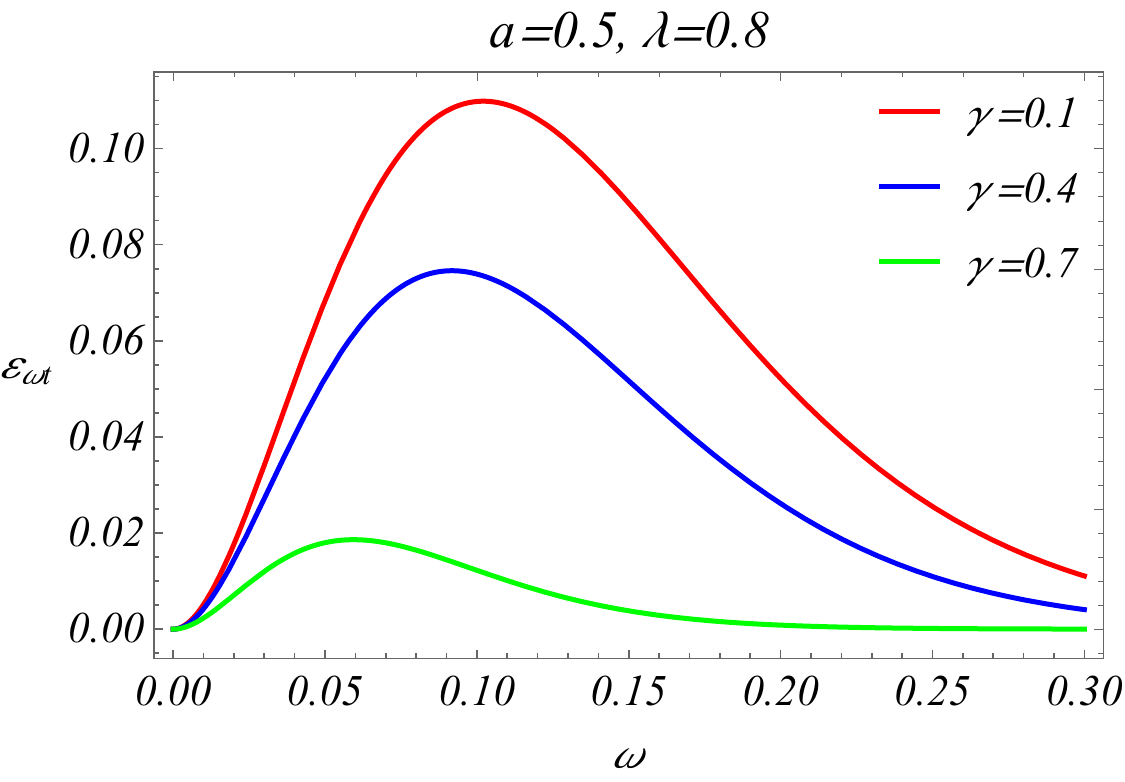}}
\subfigure{
\includegraphics[width=0.32\textwidth]{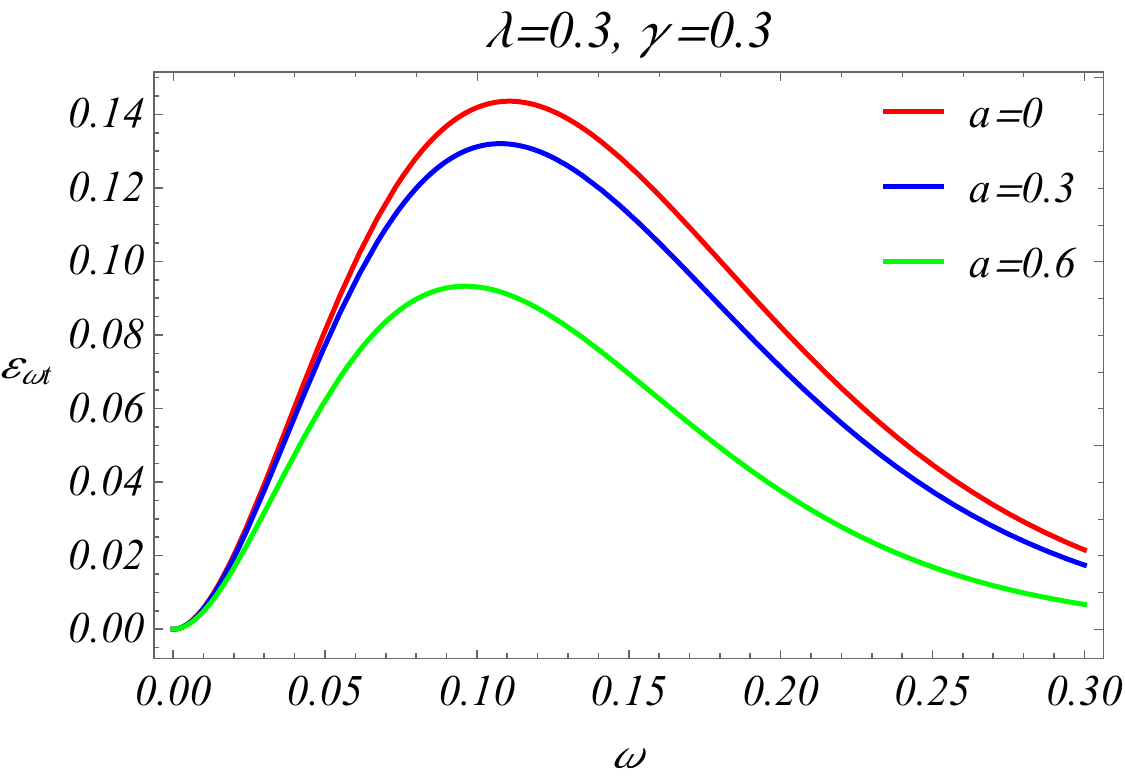}}~
\subfigure{
\includegraphics[width=0.32\textwidth]{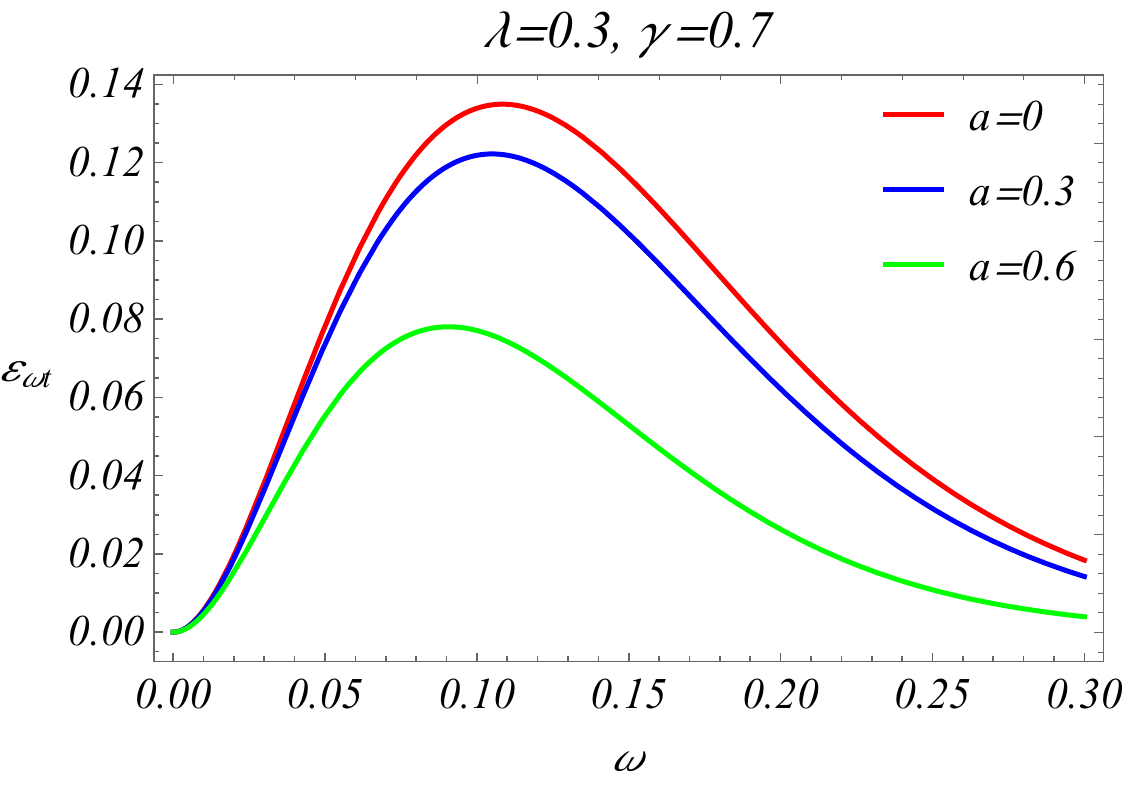}}~
\subfigure{
\includegraphics[width=0.32\textwidth]{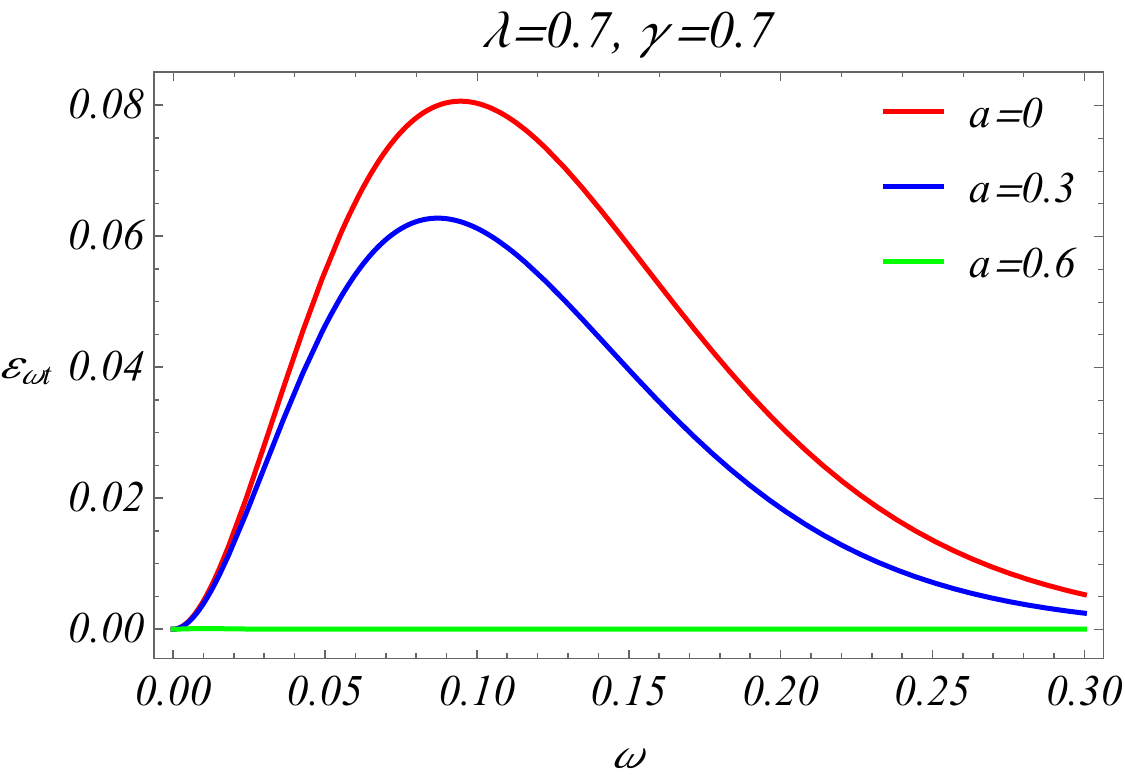}}
\end{center}
\caption{Plots for the energy emission rate corresponding to the shadows in Fig. \ref{Sh6}.} \label{EER10}
\end{figure}
\begin{figure}[t]
\begin{center}
\subfigure{
\includegraphics[width=0.32\textwidth]{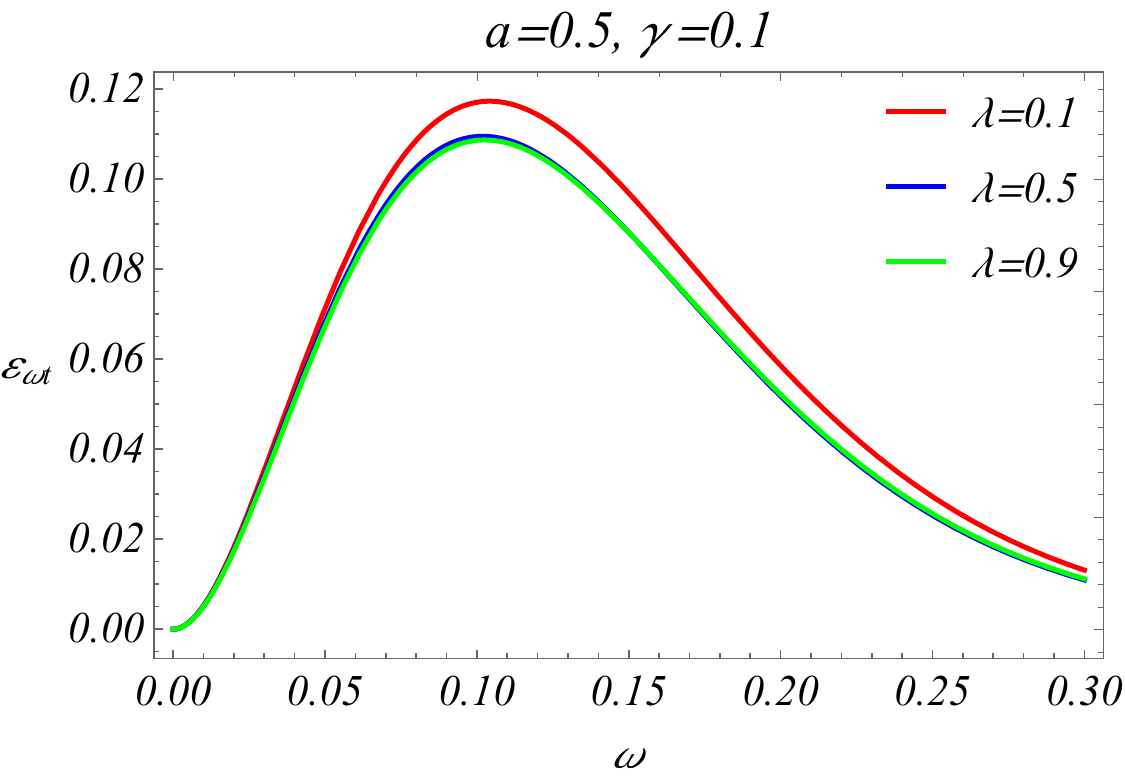}}~
\subfigure{
\includegraphics[width=0.32\textwidth]{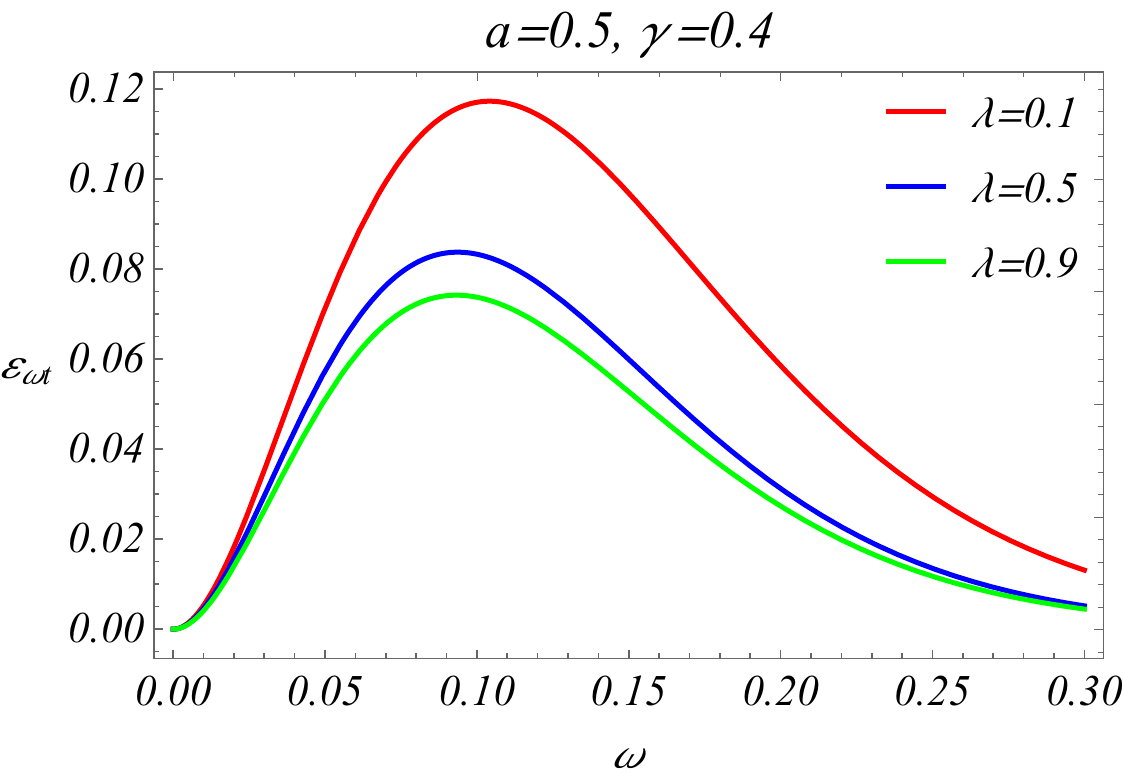}}~
\subfigure{
\includegraphics[width=0.32\textwidth]{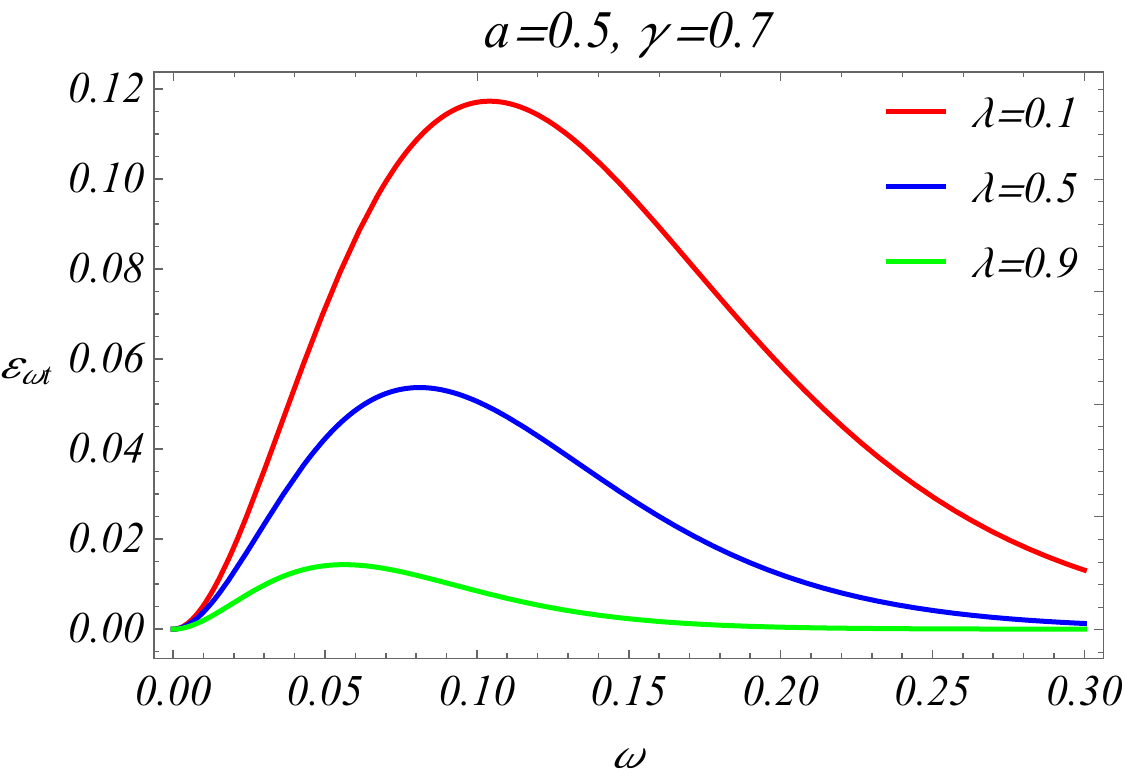}}
\subfigure{
\includegraphics[width=0.32\textwidth]{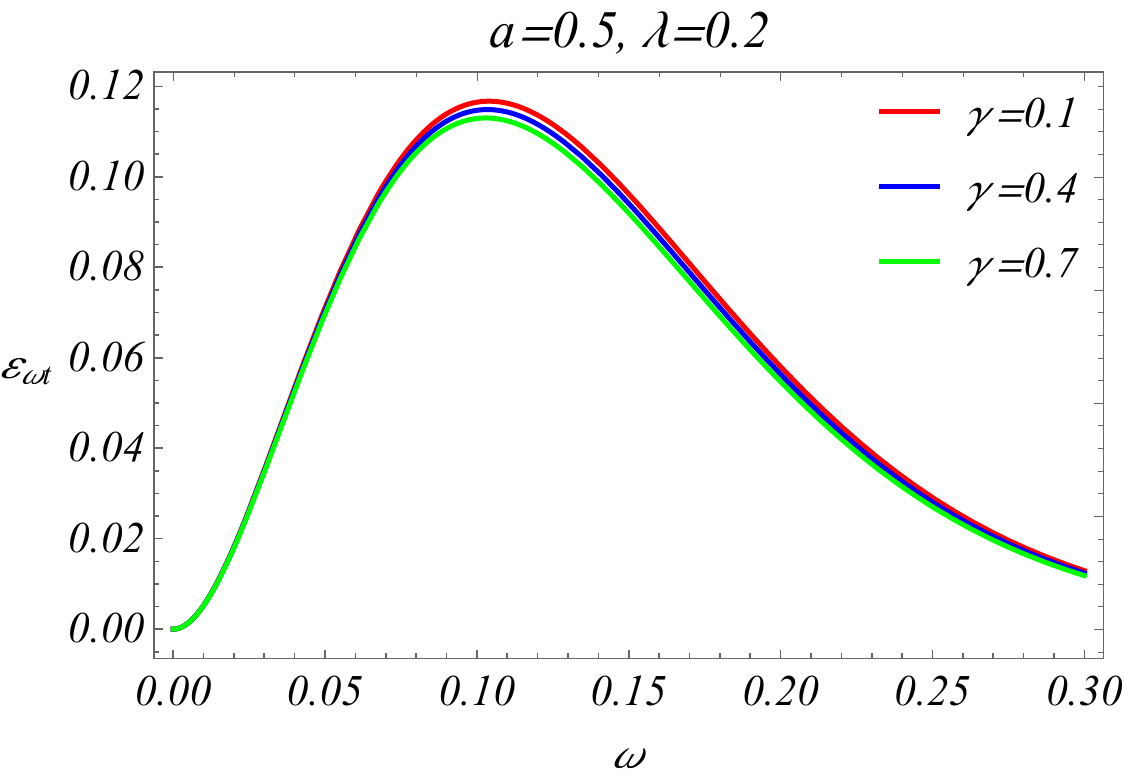}}~
\subfigure{
\includegraphics[width=0.32\textwidth]{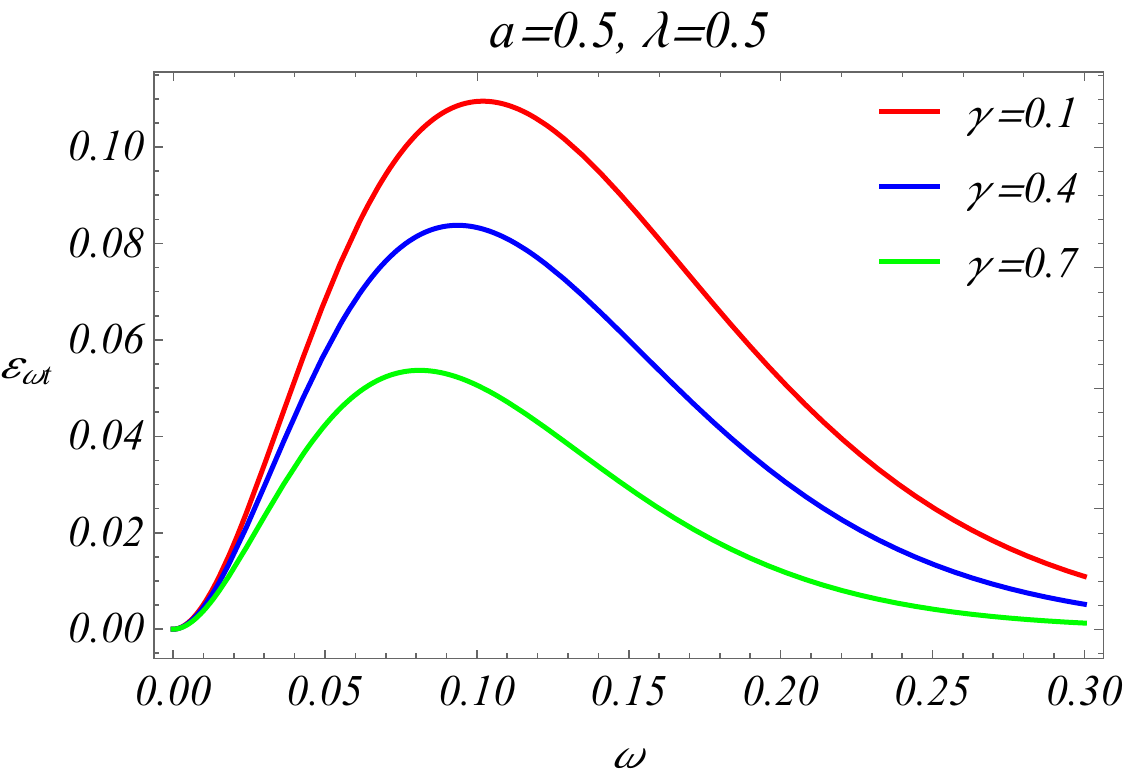}}~
\subfigure{
\includegraphics[width=0.32\textwidth]{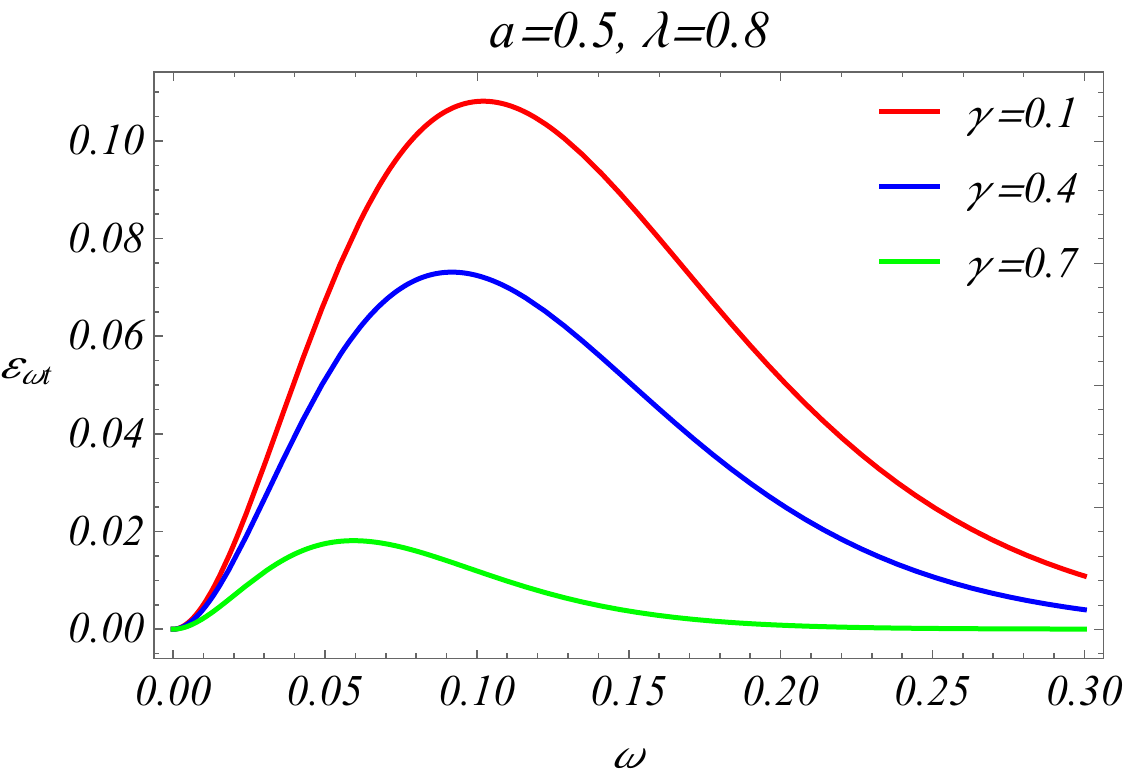}}
\subfigure{
\includegraphics[width=0.32\textwidth]{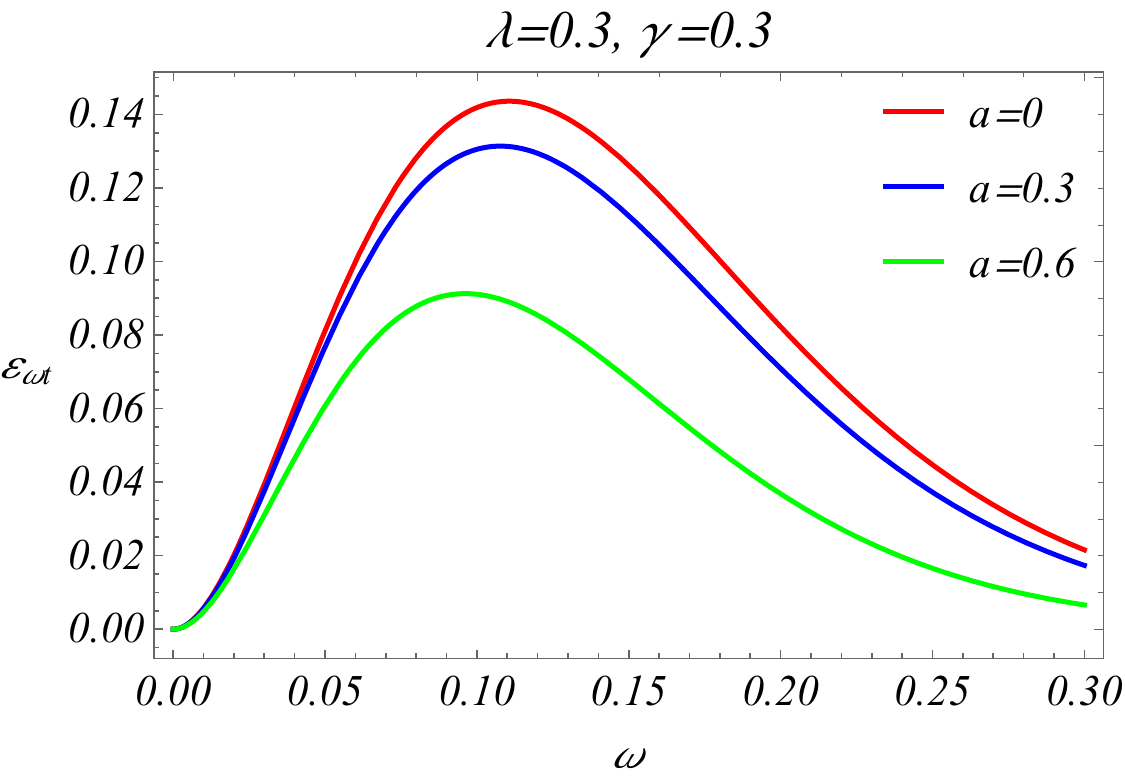}}~
\subfigure{
\includegraphics[width=0.32\textwidth]{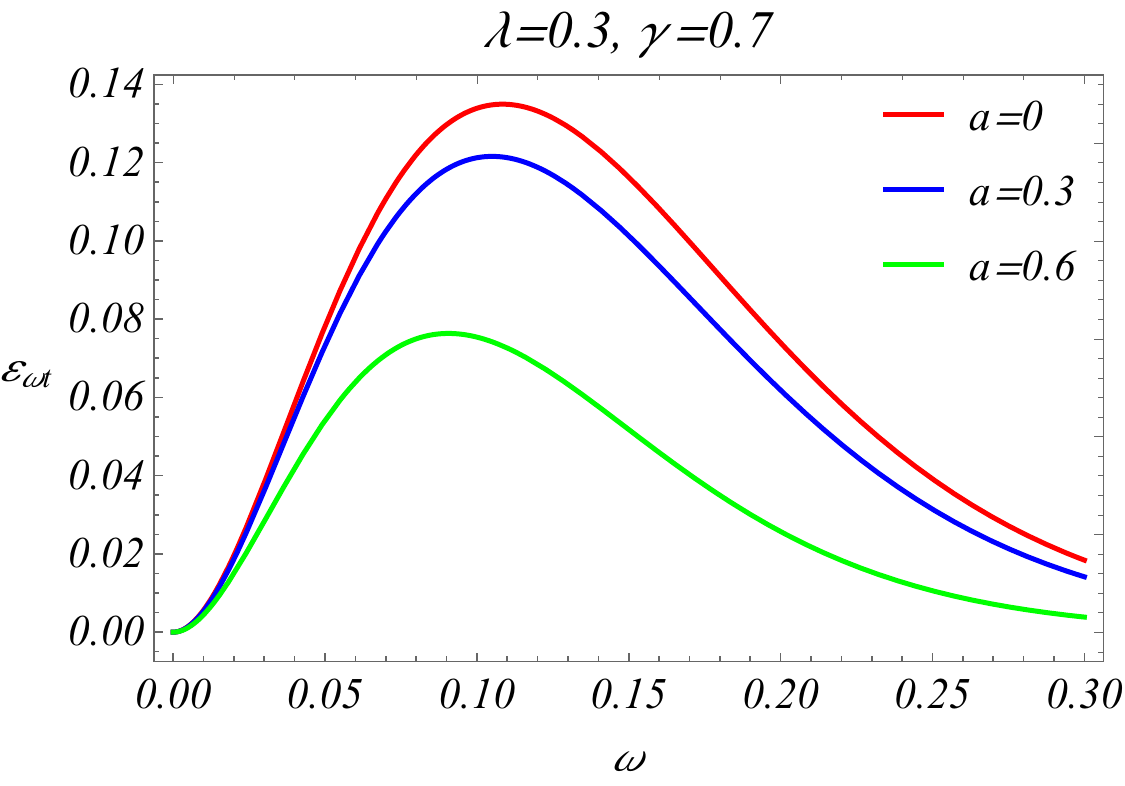}}~
\subfigure{
\includegraphics[width=0.32\textwidth]{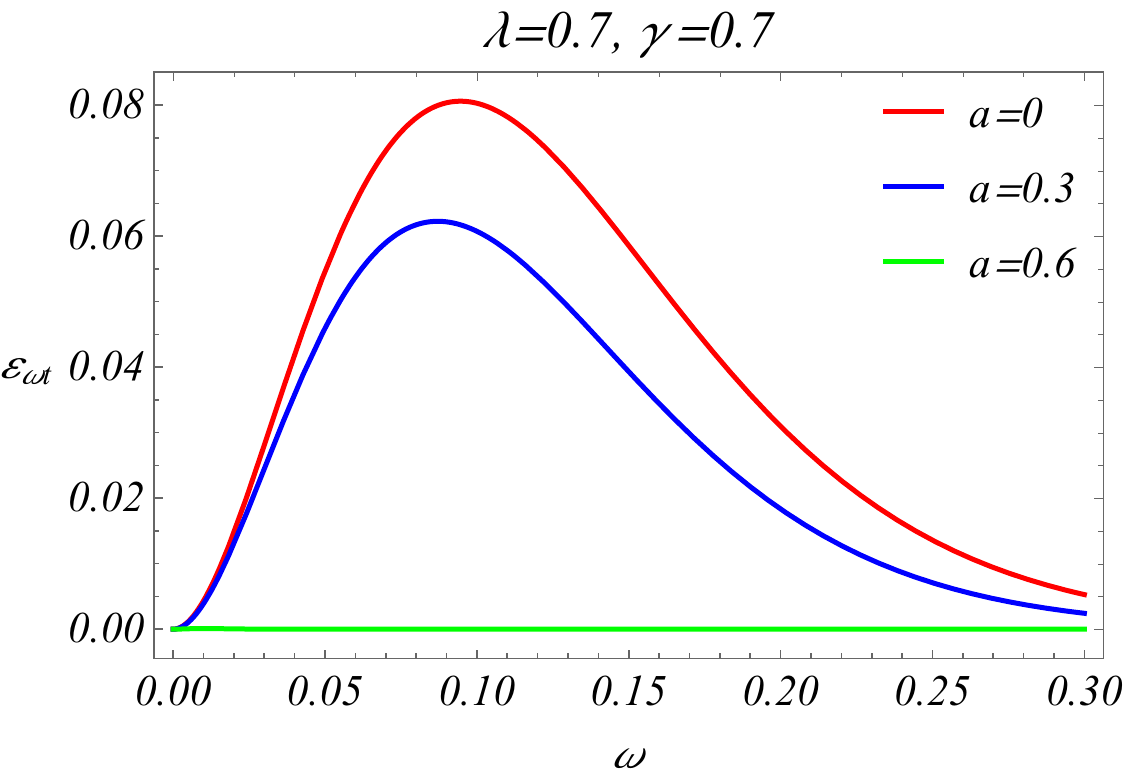}}
\end{center}
\caption{Plots for the energy emission rate corresponding to the shadows in Fig. \ref{Sh7}.} \label{EER11}
\end{figure}
Using the above formalism, we can calculate the BH evaporation rate. Since, the KR field being a quantum field, is associated with the string theory. Therefore, a BH in KR gravity will also be influenced by such a quantum field and will experience the underlying quantum effects. The study of energy emission rate is one way to deal with such effects. As said before, quantum tunneling is responsible for the particle emission from a BH. These particles carry energy that is released that we will measure in this work. When the particles are created and annihilated inside a BH, many quantum processes take place. This is how a BH in KR gravity is very much suitable for the study of quantum theoretic aspects. Moreover, the study of spacetime structure around a BH in KR gravity might be fascinating for scientists who work in the fields such as loop quantum gravity, spinfoam models, Unruh effect and Hawking radiation. We have calculated the rate of emitted energy for all those cases for which we calculated the shadows. The rate of emitted energy has been plotted in Figs. \ref{EER10} and \ref{EER11}. The top panel in Fig. \ref{EER10} corresponds to the curves for different values of $\lambda$ with fixed values of $a$ and $\gamma$ varies for each plot. In the left plot, we figure out that the emission peak drops down and hence causes the evaporation of BH to become slow as we increase the value of $\lambda$ for each curve. However, for higher values of $\lambda$, the variation in BH evaporation rate becomes less. The same effect of $\lambda$ is observed in the middle plot, however, slightly lower emission peaks are seen for higher values of $\lambda$ as compared to the peaks in left plot. It shows that with increase in $\gamma$, the BH evaporation is delayed as well. Again, the right plot shows a similar kind of behavior of emission peaks and the effect of $\lambda$ on the BH evaporation. However, the height of emission peaks is reduced more as compared to the middle plot for higher values of $\lambda$. We can deduce that with increase in $\lambda$ and $\gamma$, BH evaporation is delayed. However, for any $\gamma$, the BH evaporation is approximately constant for small values of $\lambda$ as seen in red curves. These effects can be seen equivalently and alternatively in the middle panel. In the left plot, we can see that for small $\lambda$, the emission peaks are almost same for all curves corresponding to different values of $\gamma$. This effect is observed in the top panel with red curves. Now, with the increase in $\lambda$ in the middle plot, the evaporation of BH becomes slow with increase in $\gamma$. Again, the same effect is observed in the right plot, however, the BH evaporation rate is further slowed down with increase in both $\lambda$ and $\gamma$. The bottom panel shows that with the increased spin, the peaks drop down and thus causing a slow BH evaporation. The horizontal curve in the right plot corresponds to the extremal BH and thus shows that there is no energy emission. Therefore, we can summarize that all of the BH parameters cause a slow BH evaporation. The plots for BH evaporation rate in the Fig. \ref{EER11} correspond to the shadows in the Fig. \ref{Sh7}. It is found that the BH evaporation rate is not affected significantly by changing the observer's angular location. A fractionally small variation in BH evaporation rate is observed in all plots. The peak in each curve is shifted down by a small amount as compared to the corresponding curve in Fig. \ref{EER10}. There is no change in the behavior of the curves with respect to the BH parameters.

\section{Comparison with EHT Data}
Our aim in this section is to obtain the constraints on the BH parameters $\lambda$, $\gamma$ and $a$ by using the EHT data for supermassive BHs Sgr A* and M87*. To accomplish this, we compare the shadow size of rotating BH in KR gravity with the shadow size of Sgr A* and M87* BHs. Those parametric values for which the BH shadow size will be within 1-$\sigma$ or 2-$\sigma$ error ranges, will be the constraints on the BH parameters. For these parametric values, our rotating BH in KR gravity can be termed as the corresponding supermassive BH. Since the supermassive BHs are rotating in nature, therefore to accomplish a feasible comparison, we only consider the rotating BH. For the analysis, we follow a coordinate-independent method in which we use the observables shadow area $A$ and the oblateness $D$ \cite{74,75} defined as
\begin{eqnarray}
A&=&2\int_{r_-}^{r_+}\bigg(\beta(r)\frac{d\alpha(r)}{dr}\bigg)dr, \label{59} \\
D&=&\frac{\Delta\alpha}{\Delta\beta}, \label{60}
\end{eqnarray}
where $r_+$ and $r_-$ are the radii of the retrograde and prograde stable circular orbits, respectively. Using the numerical techniques, we calculate these observables.

\subsection{Constraints on Parameters from M87*}
Using the above mentioned area and oblateness, we calculate the bounds on the parameters of M87* BH considering it as a 4D rotating BH in KR gravity. If the distance between the observer and the BH is $d$, then the angular radius or diameter of the shadow is measured as follows \cite{44j,76}
\begin{eqnarray}
\theta_d=\frac{2R_A}{d}, \qquad R_A^2=\frac{A}{\pi}, \label{61}
\end{eqnarray}
where $R_A$ is the areal shadow radius. Using Eq. (\ref{59}) in (\ref{61}), the BH shadow angular diameter can be defined in terms of the BH parameters and the observer's angle of location. It also depends on the mass of BH implicitly. The mass of M87* and its distance from the Earth are $M=6.5\times10^9M_\odot$ and $d=16.8Mpc$, respectively \cite{33,34}, where $M_\odot$ is the mass of Sun. Note that, the errors in the mass and distance measurements of the BHs have not been considered for simplicity. The angular diameter of M87* BH image is $\theta_d=42\pm3\mu as$ \cite{29}. Finally, by using the Eq. (\ref{61}), we can determine the angular diameter of the BH shadow for the rotating BH in KR gravity. This angular diameter can be compared with the angular diameter of M87* for BH parameters estimation.
\begin{figure}[t]
\begin{center}
\subfigure{
\includegraphics[width=0.41\textwidth]{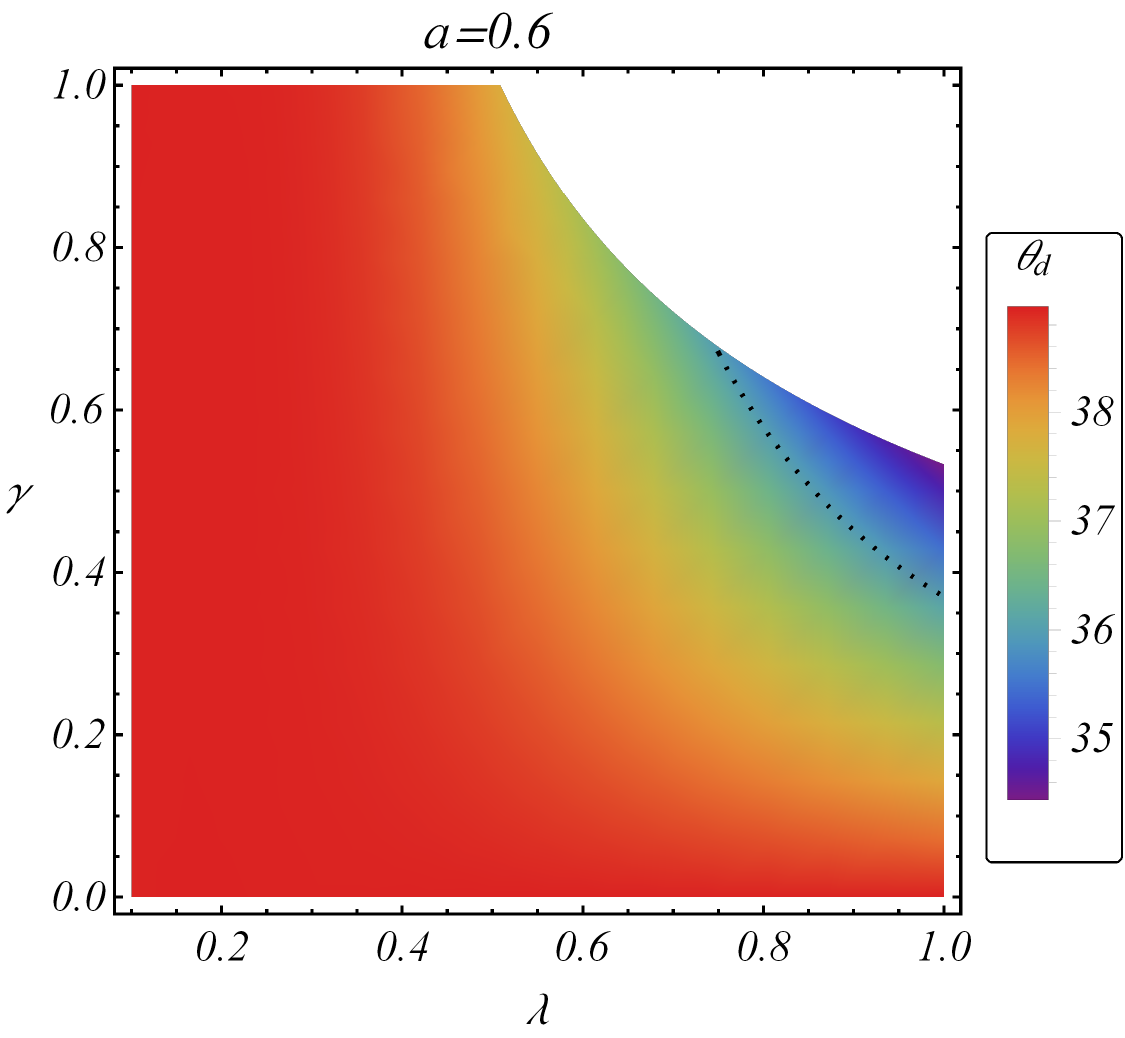}}~~~~~
\subfigure{
\includegraphics[width=0.41\textwidth]{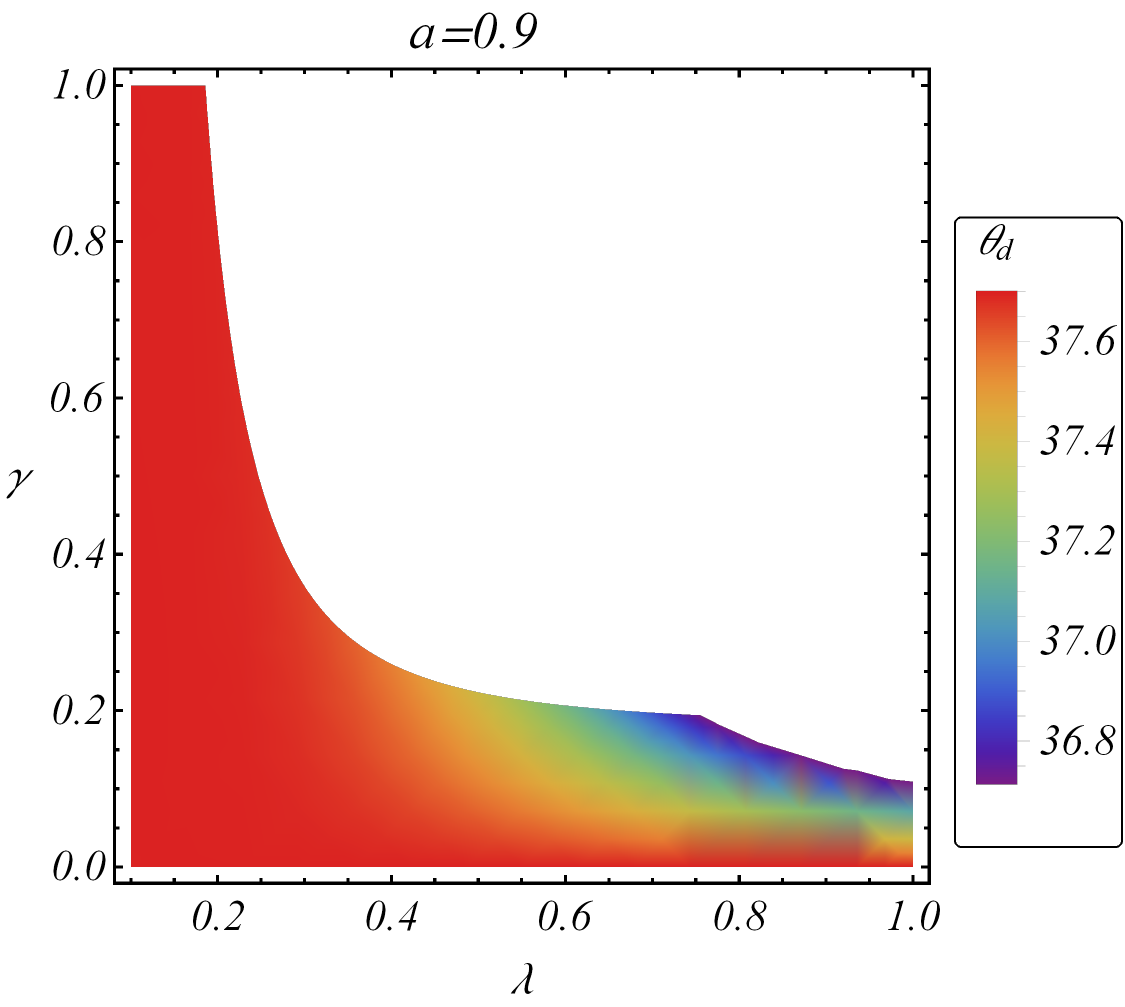}}\\
\subfigure{
\includegraphics[width=0.41\textwidth]{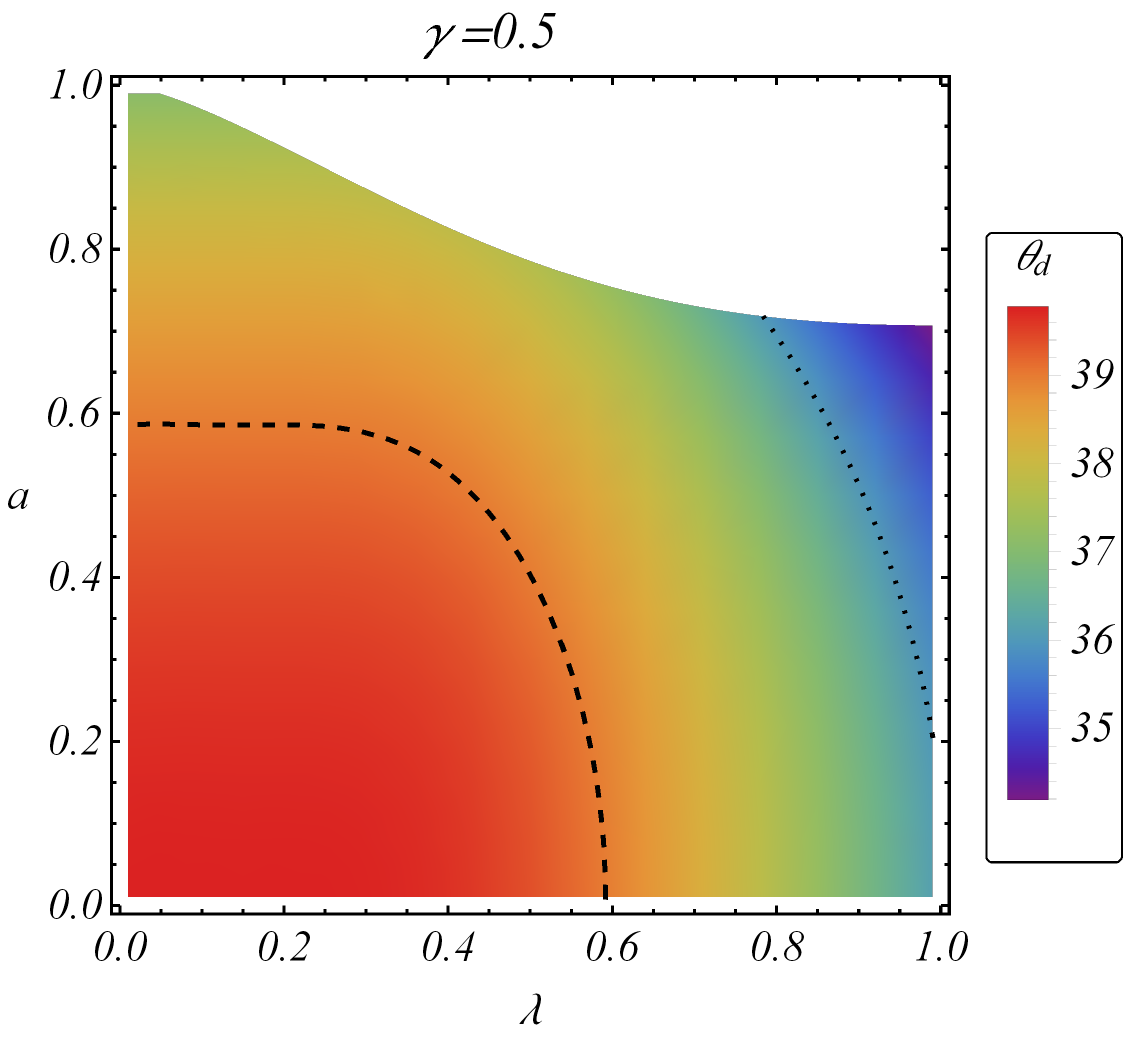}}~~~~~
\subfigure{
\includegraphics[width=0.41\textwidth]{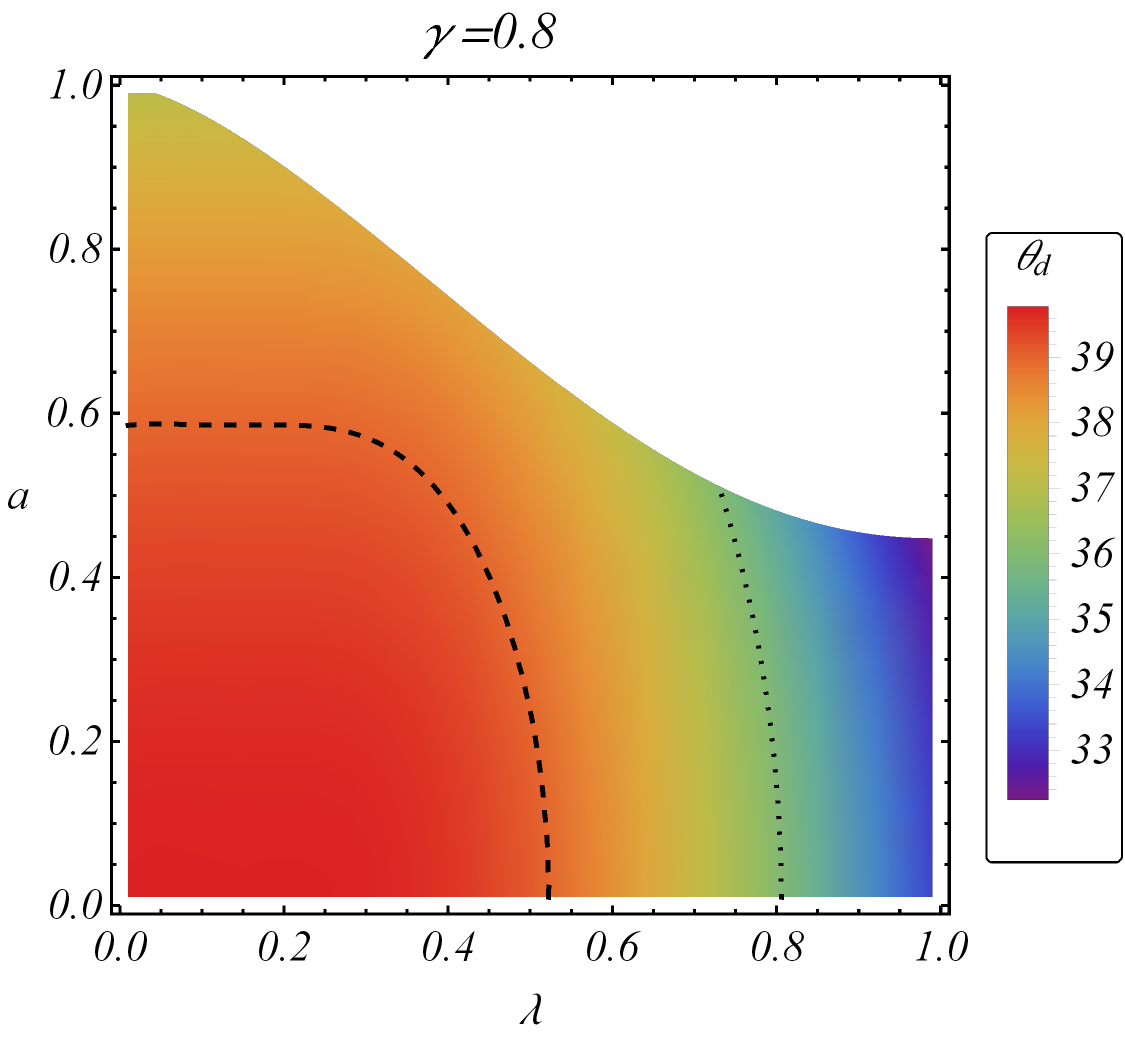}}\\
\subfigure{
\includegraphics[width=0.41\textwidth]{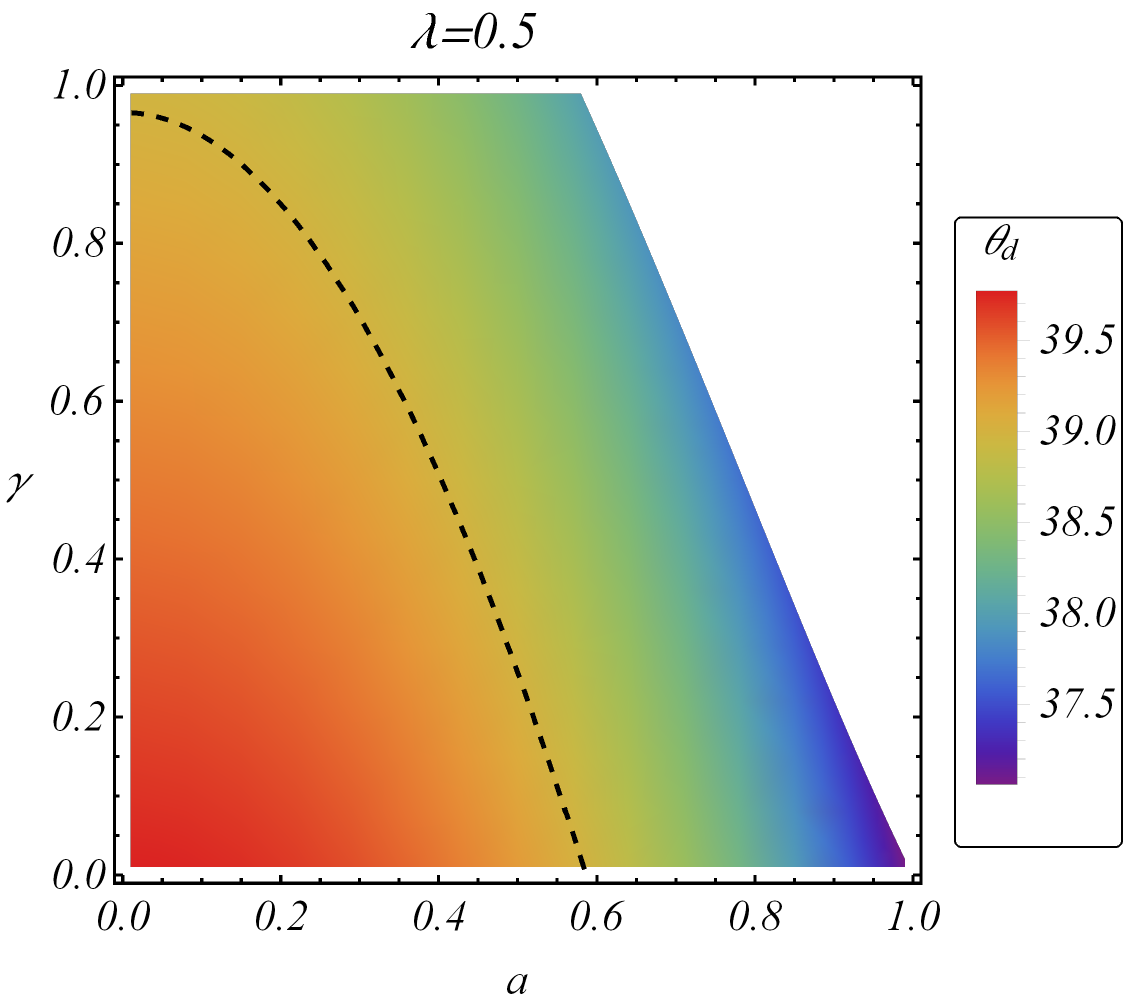}}~~~~~
\subfigure{
\includegraphics[width=0.41\textwidth]{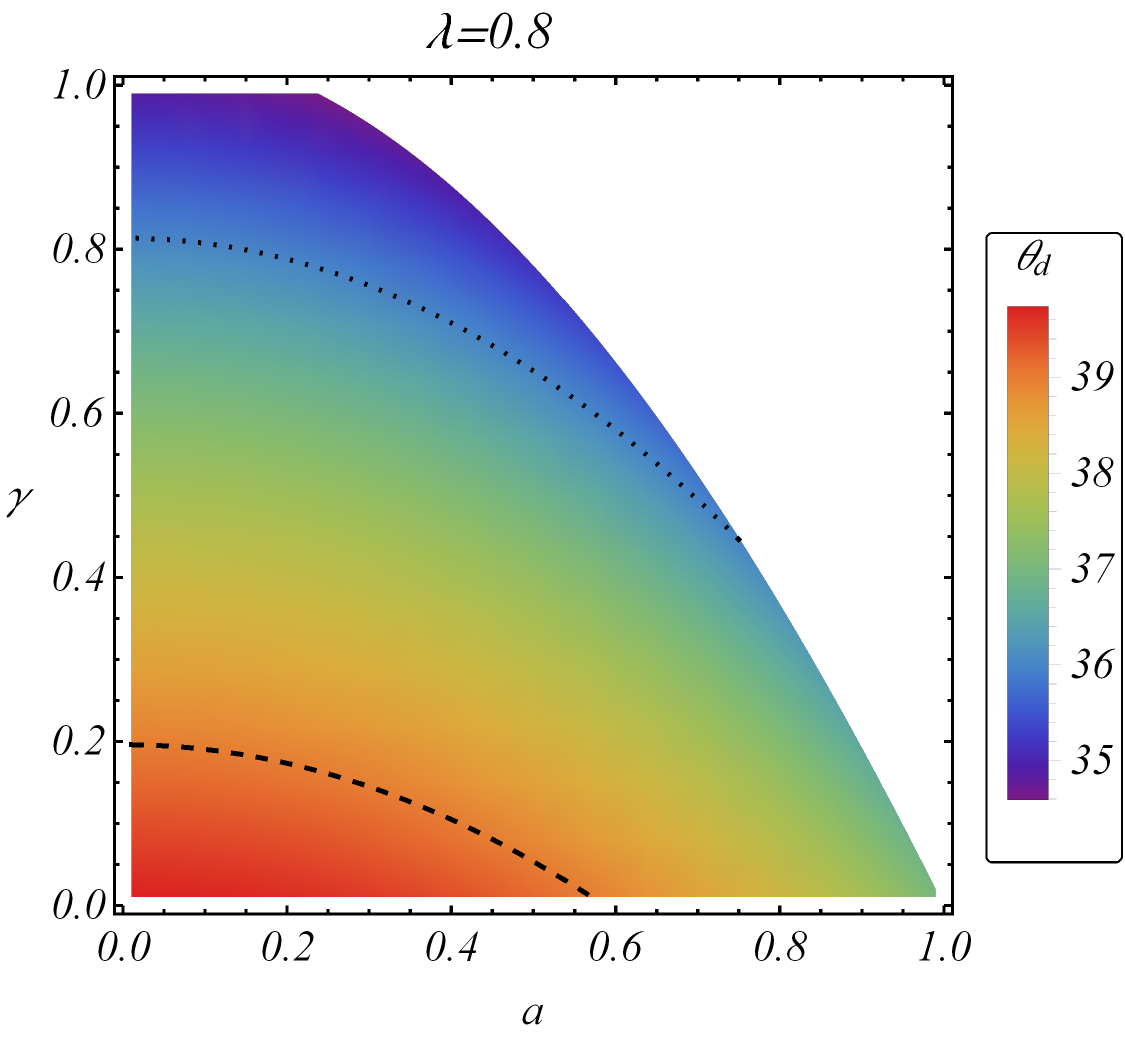}}
\end{center}
\caption{Plots for the angular diameter of rotating BH in KR gravity in terms of parametric spaces in comparison with M87* BH. For the calculations, $\theta_0=17^\circ$ observer's angle has been considered.} \label{M87}
\end{figure}
\begin{figure}[t]
\begin{center}
\subfigure{
\includegraphics[width=0.41\textwidth]{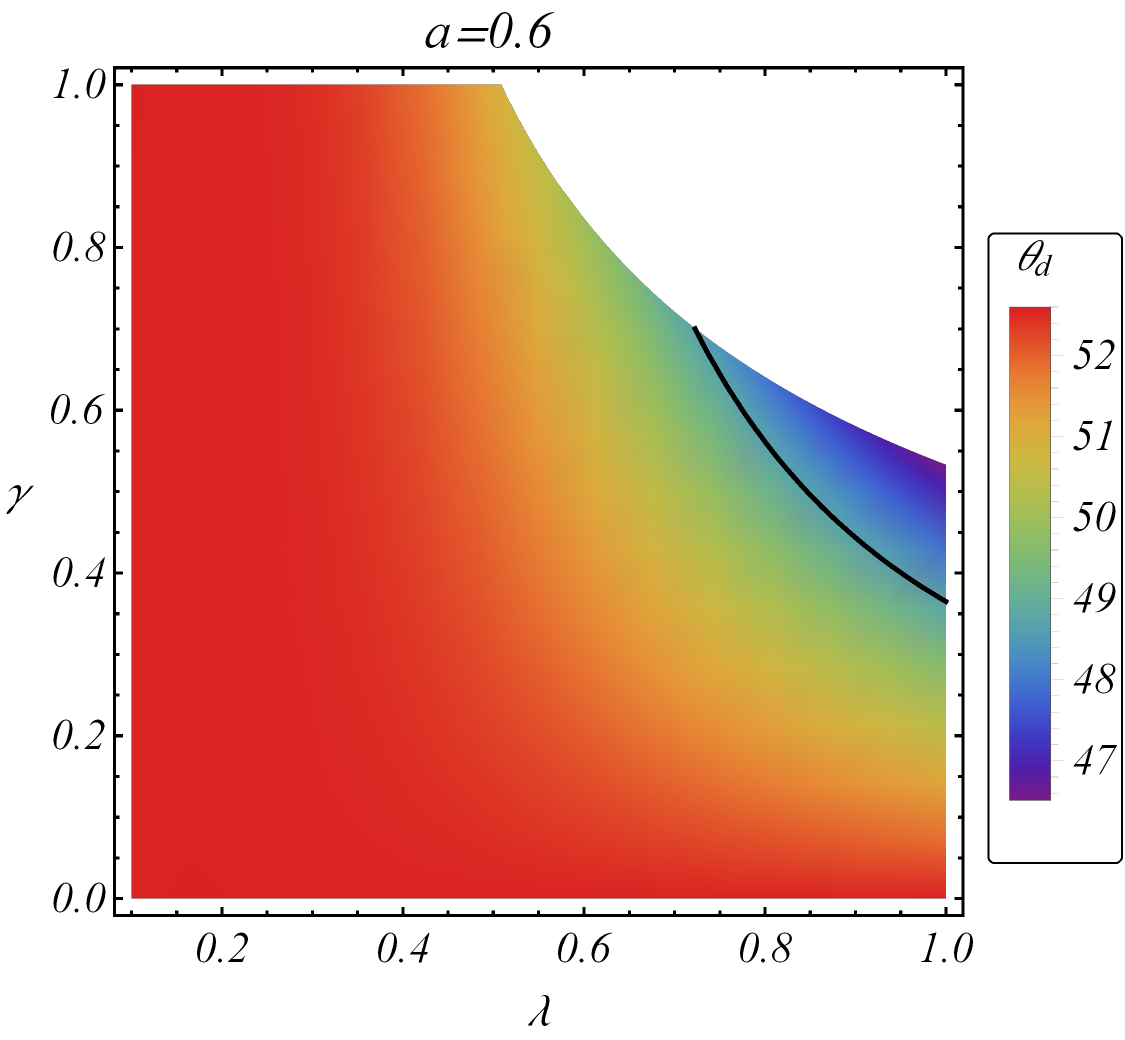}}~~~~~
\subfigure{
\includegraphics[width=0.41\textwidth]{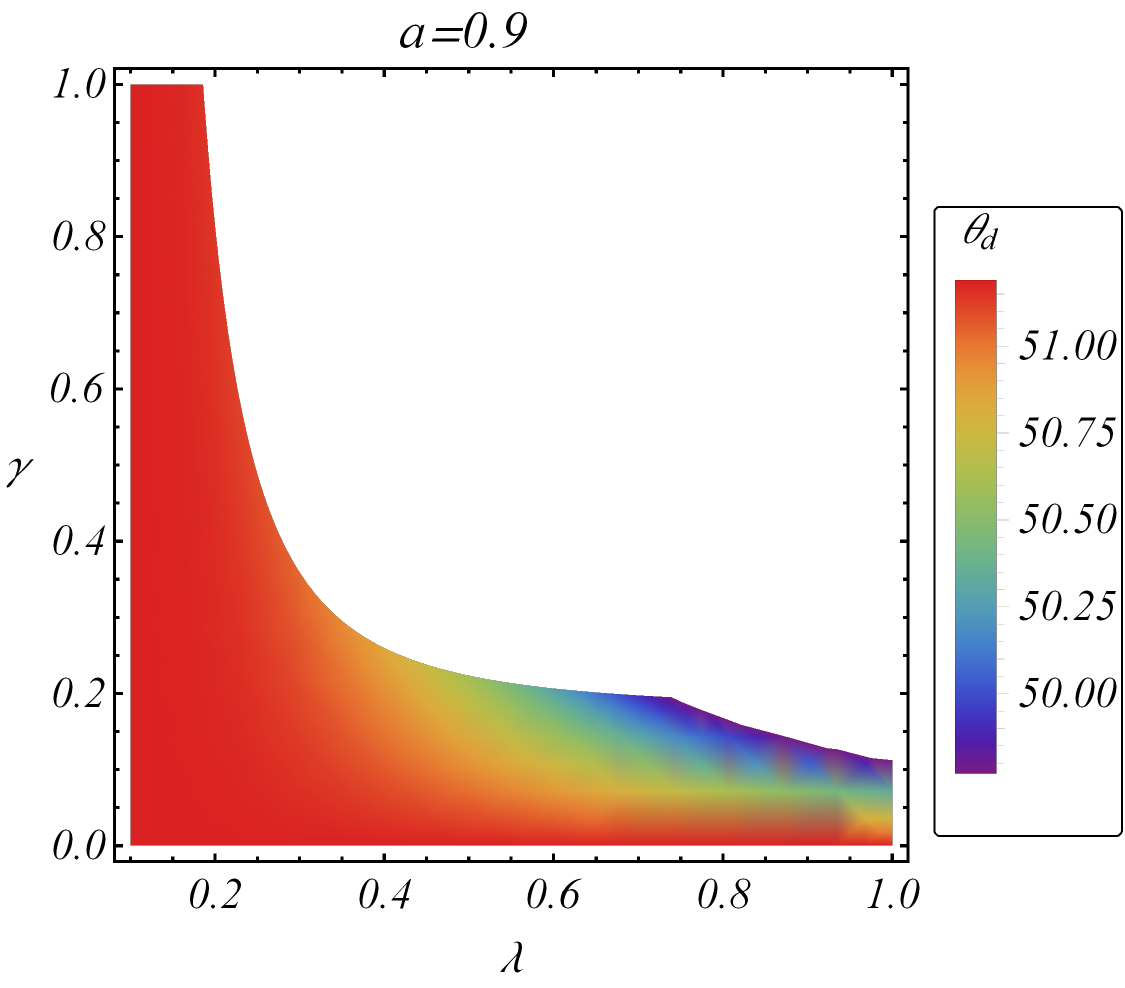}}\\
\subfigure{
\includegraphics[width=0.41\textwidth]{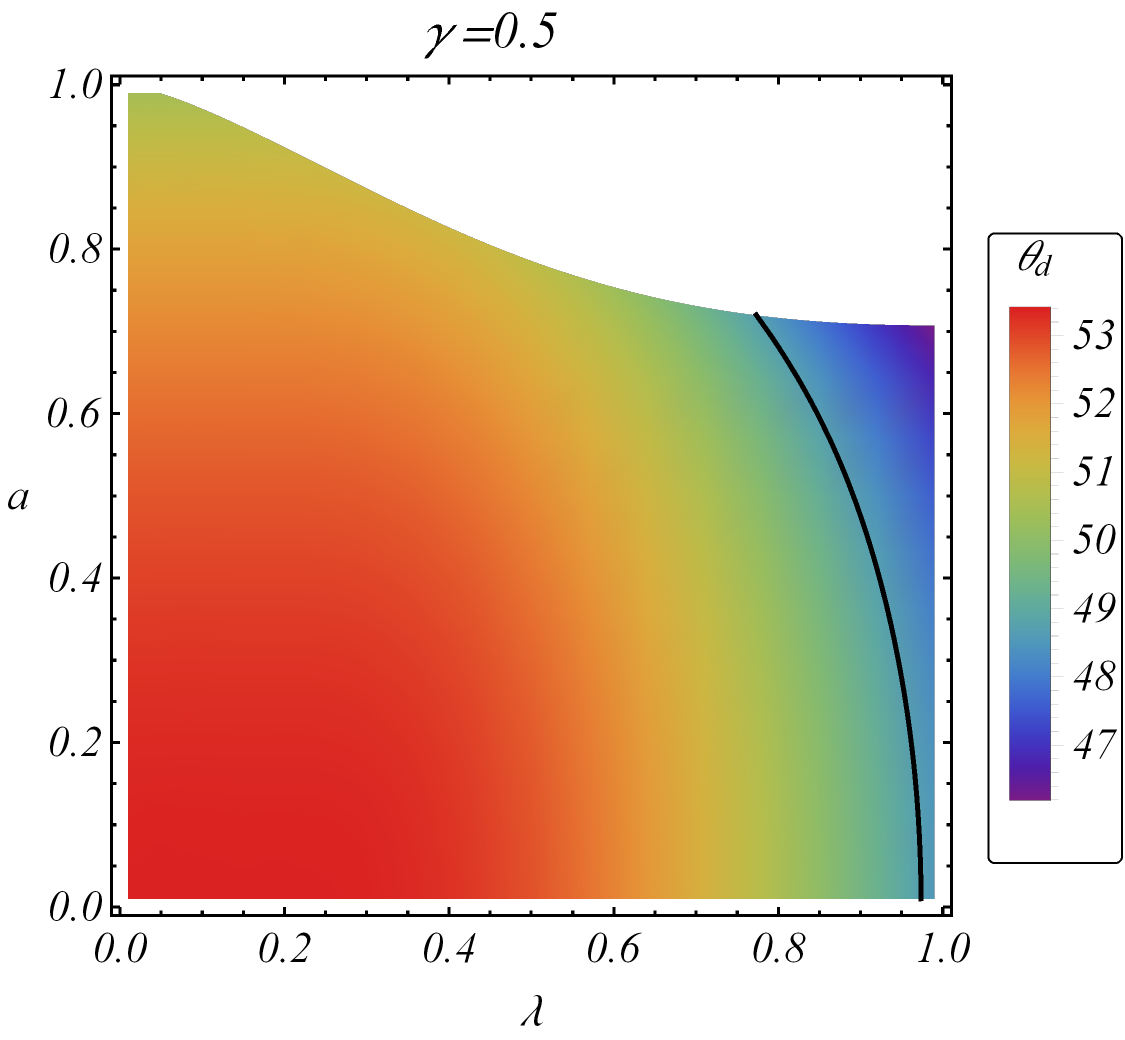}}~~~~~
\subfigure{
\includegraphics[width=0.41\textwidth]{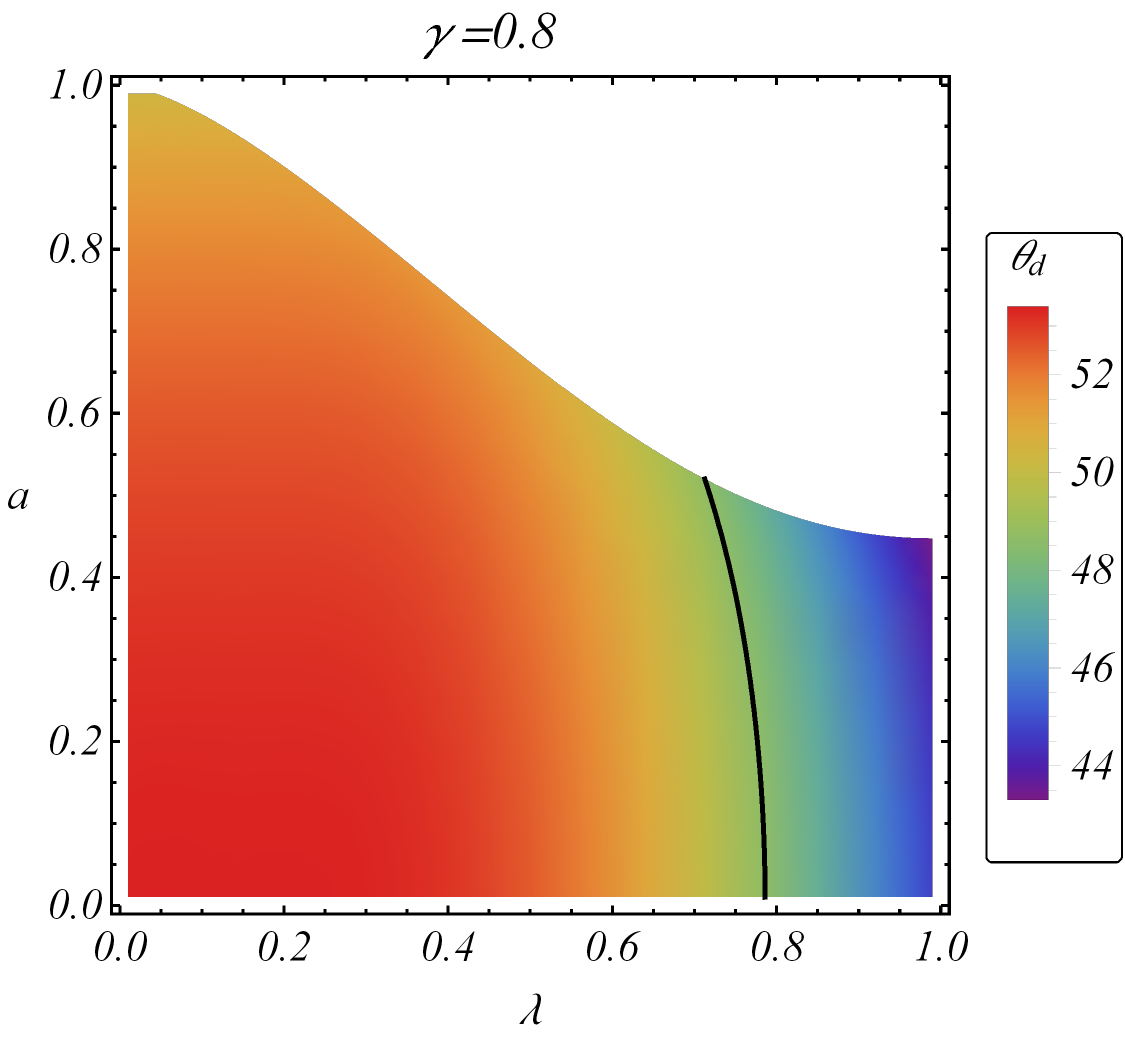}}\\
\subfigure{
\includegraphics[width=0.41\textwidth]{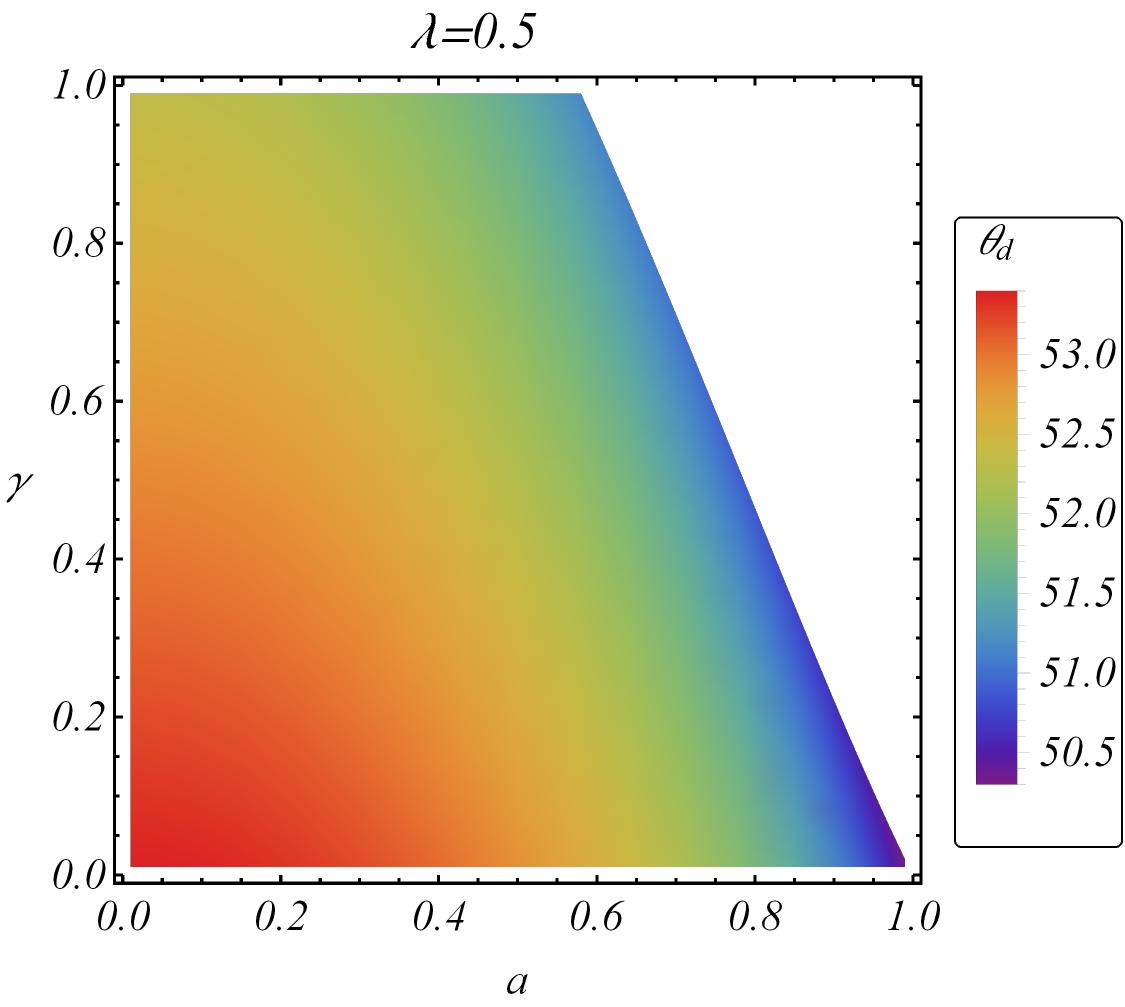}}~~~~~
\subfigure{
\includegraphics[width=0.41\textwidth]{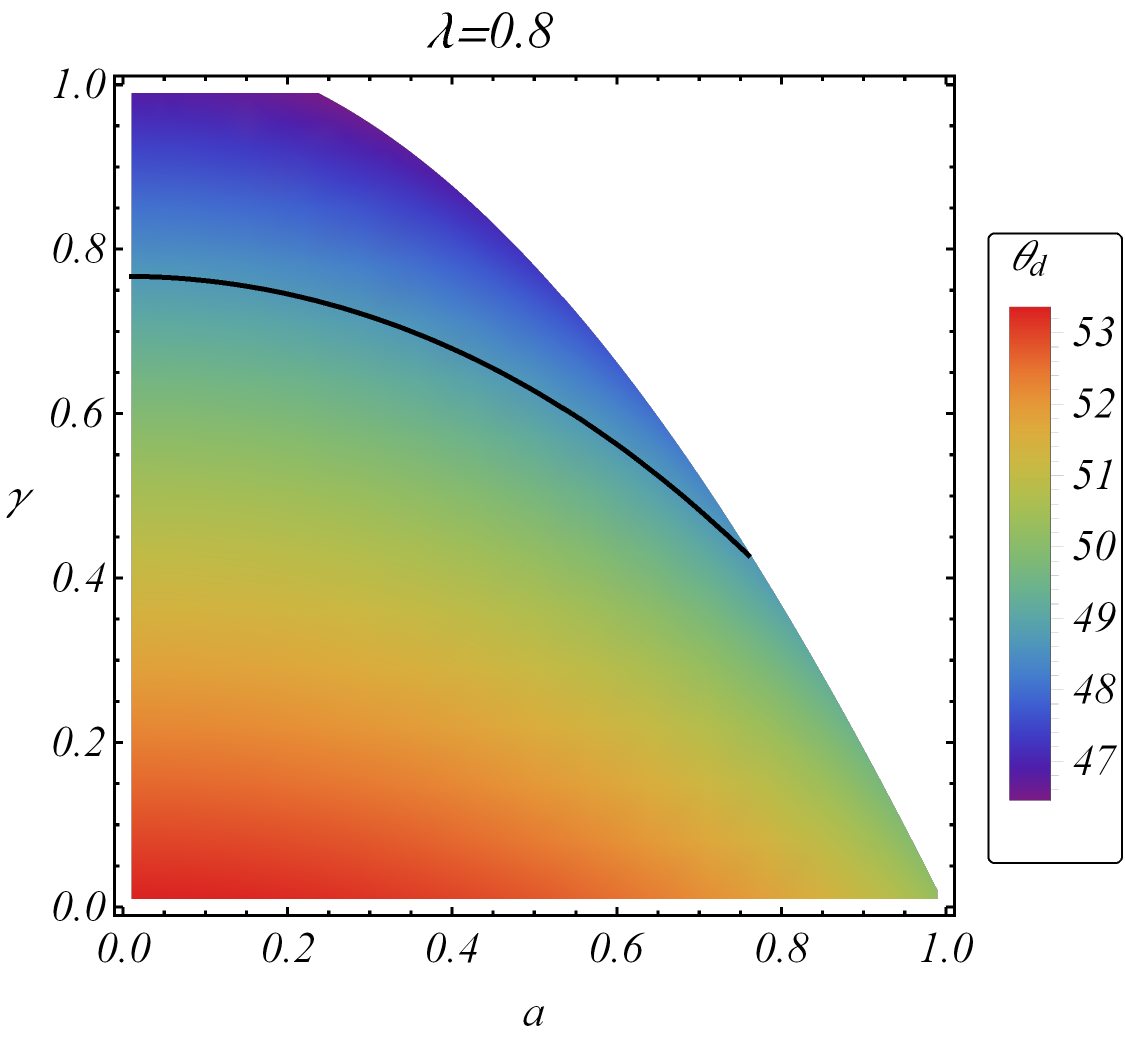}}
\end{center}
\caption{Plots for the angular diameter of rotating BH in KR gravity in terms of parametric spaces in comparison with Sgr A* BH. For the calculations, $\theta_0=50^\circ$ observer's angle has been considered.} \label{SgrA}
\end{figure}

The angular diameter $\theta_d$ has been plotted as density plots in Fig. \ref{M87} in terms of parametric spaces corresponding to an observer inclined at $17^\circ$. The dashed curves refer to the 1-$\sigma$ error interval of the angular diameter of M87* BH. Whereas, the dotted curves refer to the 2-$\sigma$ error interval of the angular diameter of M87* BH. For $a=0.6$ in the left plot in top panel, it can be seen that for the given values of $\lambda$ and $\gamma$ in the parametric space, the shadow size lies in the 2-$\sigma$ interval. Therefore, a bound for $\lambda$ and $\gamma$ for $a=0.6$ exists in the 2-$\sigma$ interval. Whereas, from the right plot, we can see that $\lambda$ and $\gamma$ has no bound for $a=0.9$ and thus for all values of $\lambda$ and $\gamma$ in the parametric space, the shadow size lies within 2-$\sigma$ interval. Since, for these parametric values, the shadow size does not lie within 1-$\sigma$ interval, therefore according to our choice, we do not regard the rotating BH in KR gravity as M87* BH. The middle panel shows that the shadow size lies in both 1-$\sigma$ and 2-$\sigma$ error intervals and thus the parameters $a$ and $\lambda$ have bounds for both plots. The parametric values enclosed by the dashed curves and the coordinate axes, give the shadow size within 1-$\sigma$ error interval for both plots. For these values of $a$ and $\lambda$ in the parametric spaces, the rotating BH in KR gravity is regarded as M87* BH. In the left plot in lower panel, it can be seen that the shadow size lies in both 1-$\sigma$ and 2-$\sigma$ error intervals. However, $a$ and $\gamma$ possess bound only in 1-$\sigma$ level. It means that for the values of $a$ and $\gamma$ with fixed $\lambda=0.5$, enclosed by the dashed curve and coordinate axes, the shadow size lies within 1-$\sigma$ level and thus the rotating BH in KR gravity can be regarded as M87* BH. Whereas, the right plot in lower panel shows that the shadow size lies in both 1-$\sigma$ and 2-$\sigma$ error intervals. Moreover, $a$ and $\gamma$ possess bounds in both 1-$\sigma$ and 2-$\sigma$ levels. However, the values of $a$ and $\gamma$ with fixed $\lambda=0.8$, lying below the dashed curve, give the shadow size within 1-$\sigma$ level and thus for these values, the rotating BH in KR gravity can be regarded as M87* BH.

\subsection{Constraints on Parameters from Sgr A*}
Following the same approach as for the case of M87* BH, we find the constraints on the BH parameters in comparison with the shadow of Sgr A*. The angular diameter of the shadow of Sgr A* BH is $\theta_d=48.7\pm7\mu as$ \cite{37}. The mass of Sgr A* and its distance from Earth is $M=4\times10^6M_\odot$ and $d=8kpc$, respectively \cite{37,77}. The results showing the bounds on the BH parameters have been presented in Fig. \ref{SgrA}. We determined these bounds for the BH parameters by comparing the shadow size of the rotating BH in KR gravity with the shadow size of Sgr A* corresponding to an observer located at $50^\circ$. In the top panel, we found that the shadow size lies within 1-$\sigma$ interval for all values of $\lambda$ and $\gamma$ given in parametric spaces for $a=0.6$ and $a=0.9$. It means that for all of these values of $\lambda$ and $\gamma$, the rotating BH in KR gravity can be regarded as Sgr A* BH. The black curve corresponds to $\theta_d=48.7\mu as$ that divides the 1-$\sigma$ and 2-$\sigma$ intervals into equal halves. Similarly, corresponding to all values of $a$ and $\lambda$ in the parametric spaces for both plots in the middle panel, the shadow size lies within 1-$\sigma$ interval suggesting that for all of these parametric values, the rotating BH in KR gravity is Sgr A* BH. Again similar to the top and middle panels, for the given values of $a$ and $\gamma$ in the parametric spaces for both plots in the lower panel, the shadow lies within 1-$\sigma$ interval which means that for all of these parametric values, the rotating BH in KR gravity is Sgr A* BH.

In Ref. \cite{55}, Kumar et al. also studied the comparison of the shadow sizes of rotating BH in KR gravity and M87* BH. They obtained the constraints on $a$ and $\gamma$ for only two values of $\lambda$. These two cases are equivalent to the lower panel of Fig. \ref{M87} in this manuscript. However, we considered the different fixed values of $\lambda$. Though as an analysis we can compare the right plot in the lower panel with the left plot in the Ref. \cite{55} as the values of $\lambda$ are close to each other. It can be seen that the bounds on $a$ and $\gamma$ are also close to the bounds in Ref. \cite{55} with a small difference due to the difference in $\lambda$. In this manuscript, we established the analysis for twelve different cases for M87* and Sgr A* BHs. Therefore, it is more detailed and comprehensive analysis as compared to the Ref. \cite{55} in which only two case are considered only for M87*. In Ref. \cite{44k}, an identical comparison has also been accomplished for KR and Sgr A* BHs. However, in this paper, the authors considered the static BH in KR gravity and for only one value of $\gamma$, they obtained the bound for $\lambda$ using the coordinate based numerical method. Moreover, we also mention that it is not a feasible study to compare the shadow size of a static BH in KR gravity with the shadow size of a rotating Sgr A* BH.
\begin{figure}[t]
\begin{center}
\subfigure{
\includegraphics[width=0.41\textwidth]{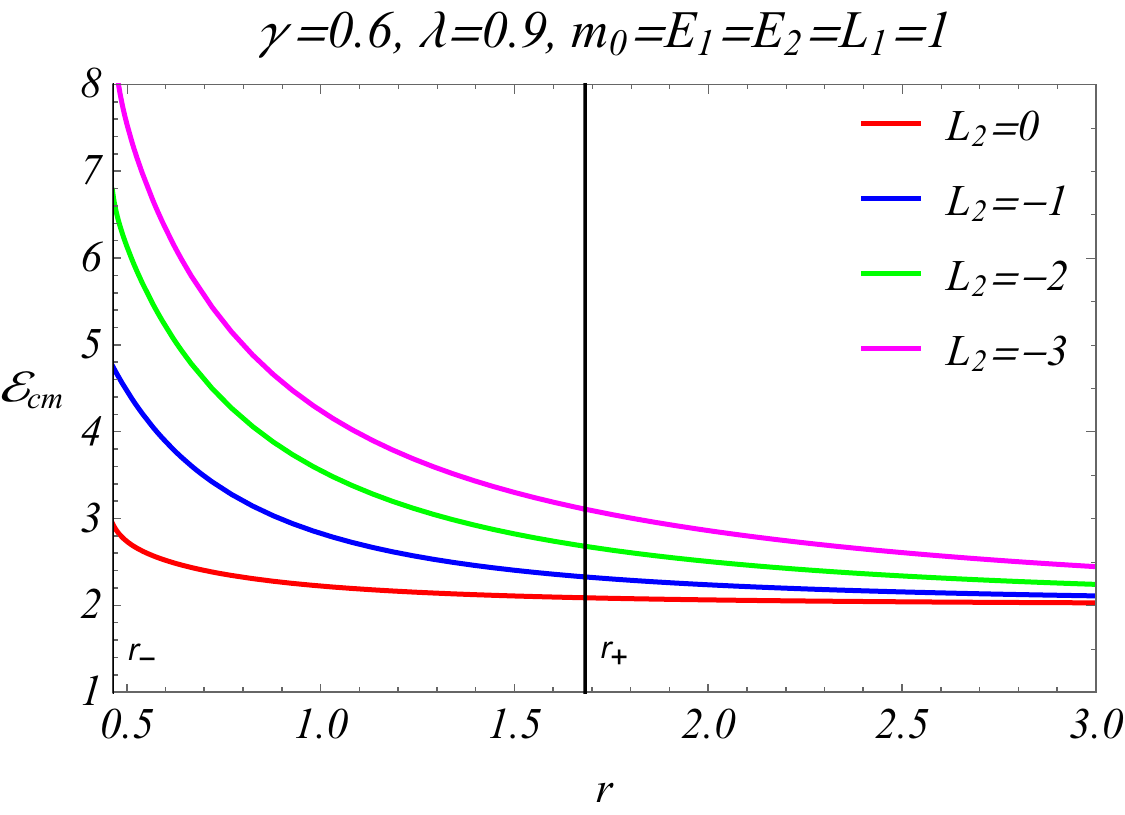}}~~~~~
\subfigure{
\includegraphics[width=0.41\textwidth]{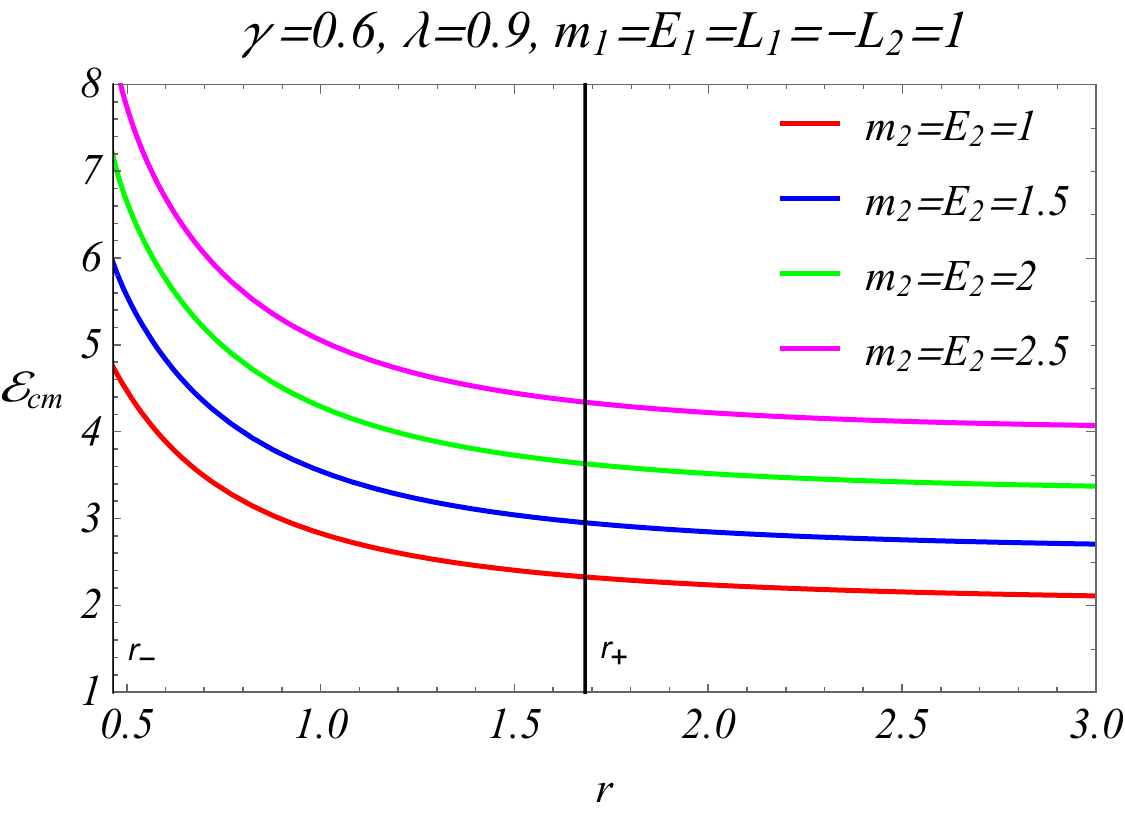}}\\
\subfigure{
\includegraphics[width=0.41\textwidth]{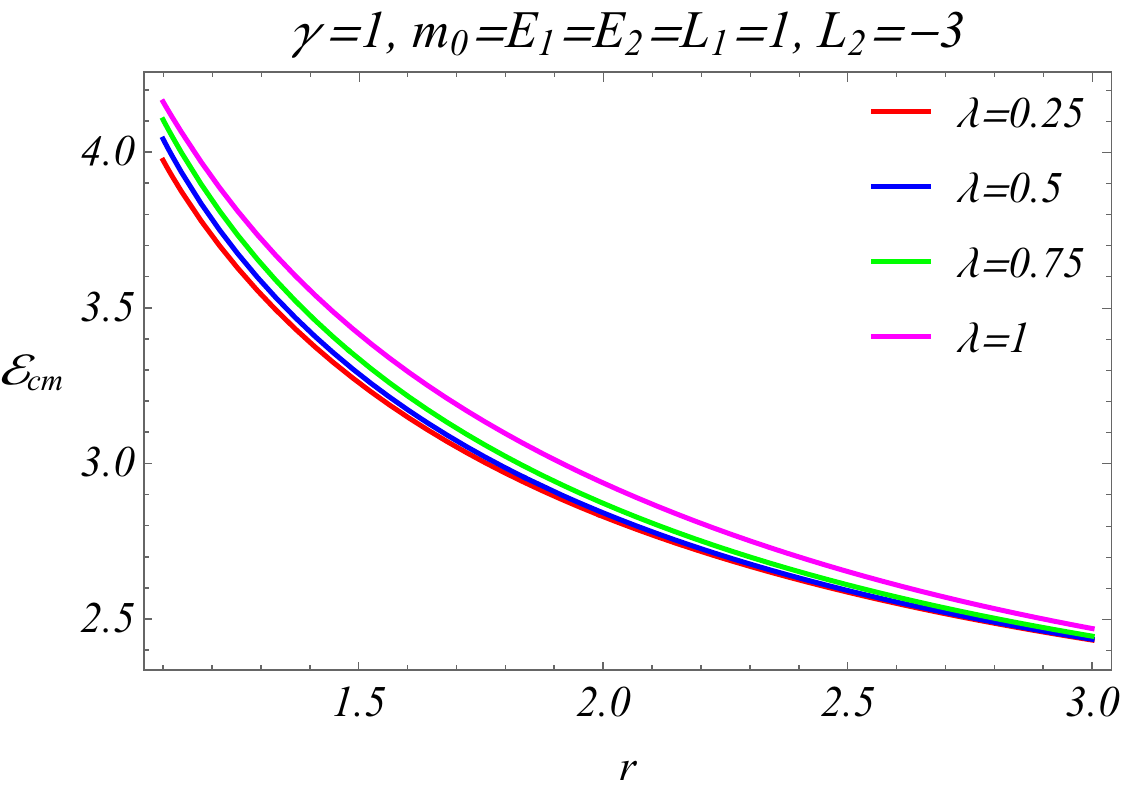}}~~~~~
\subfigure{
\includegraphics[width=0.41\textwidth]{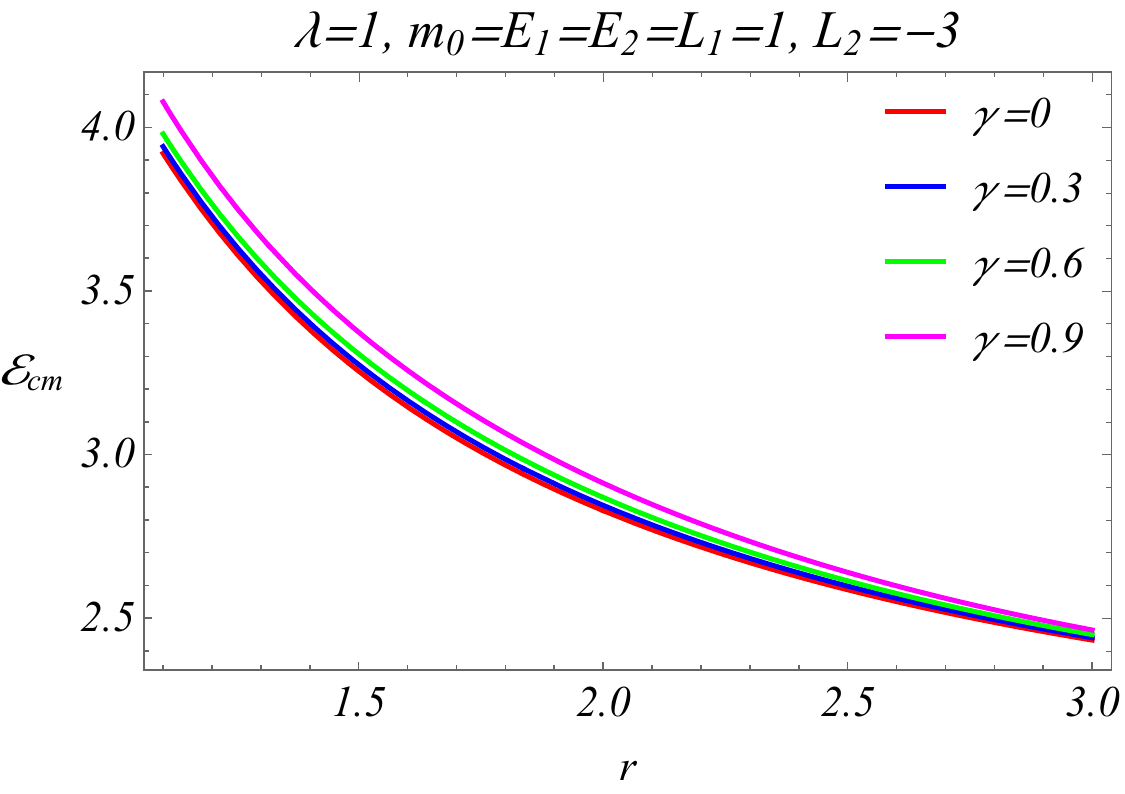}}
\end{center}
\caption{Plots for the CME of colliding particles outside a static BH in KR gravity.} \label{SCME}
\end{figure}
\begin{figure}[t]
\begin{center}
\subfigure{
\includegraphics[width=0.41\textwidth]{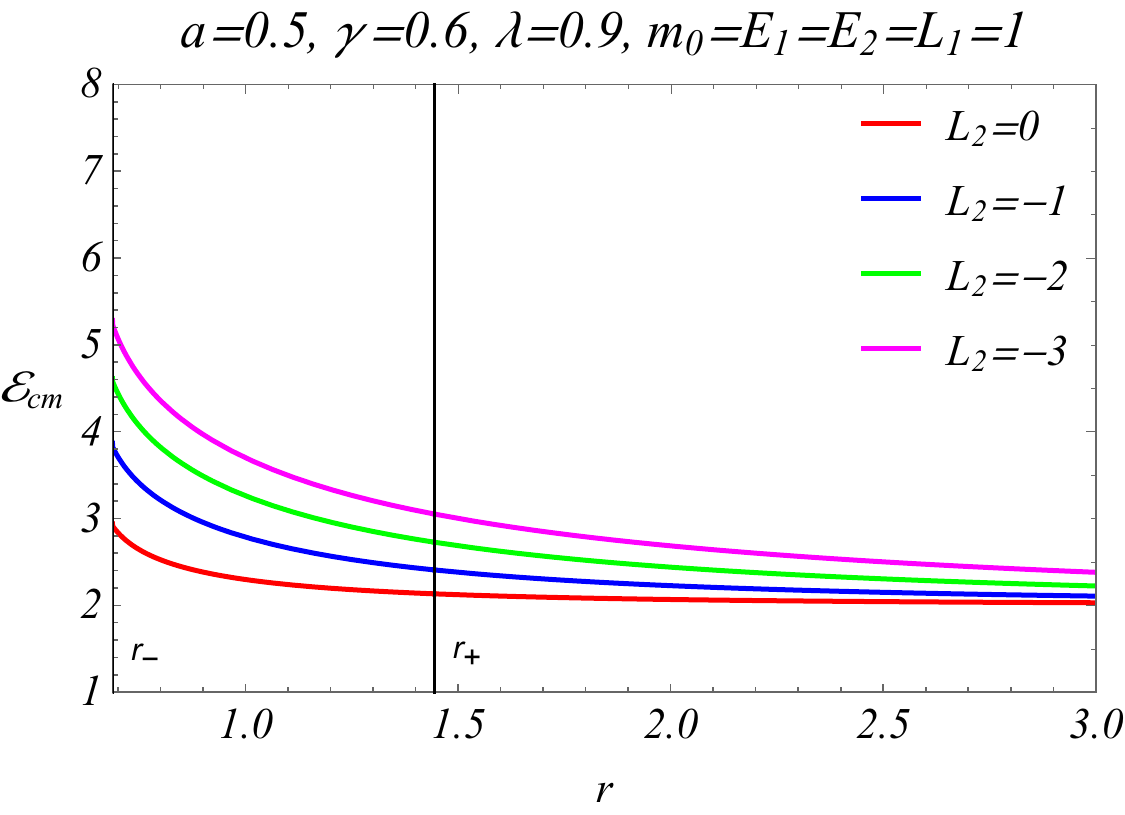}}~~~~~
\subfigure{
\includegraphics[width=0.41\textwidth]{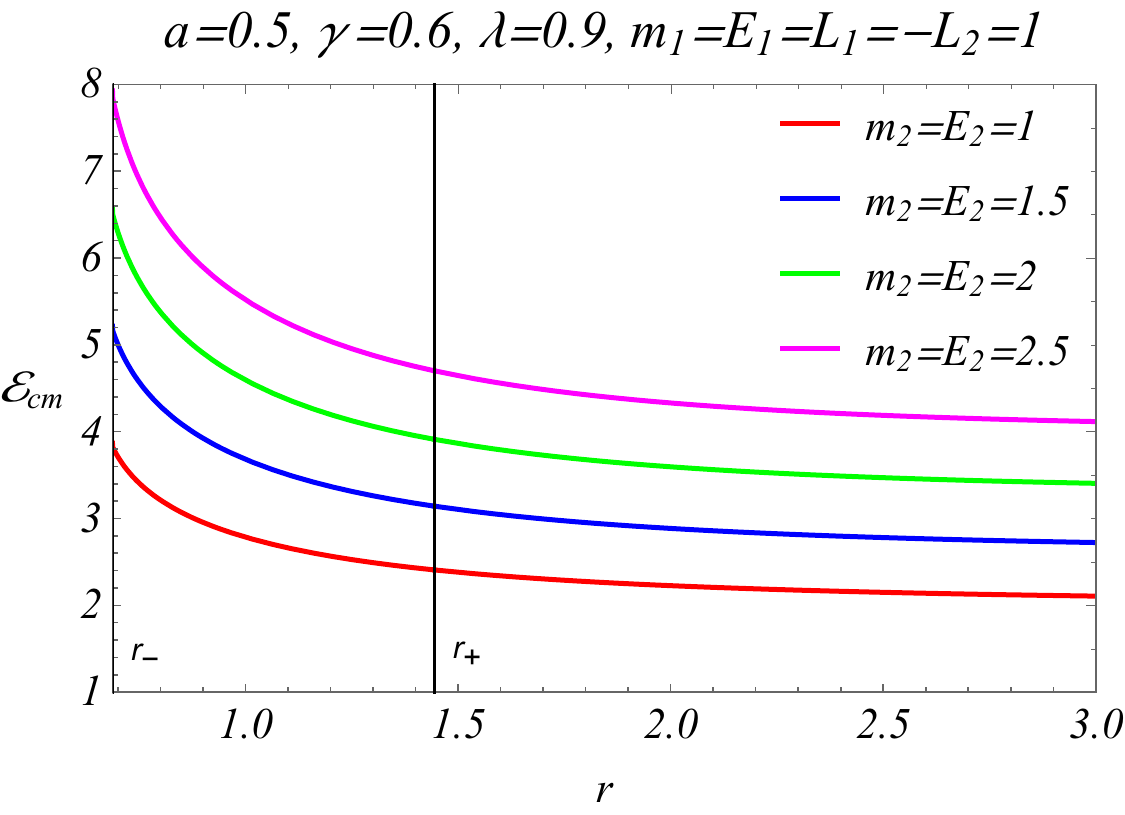}}\\
\subfigure{
\includegraphics[width=0.41\textwidth]{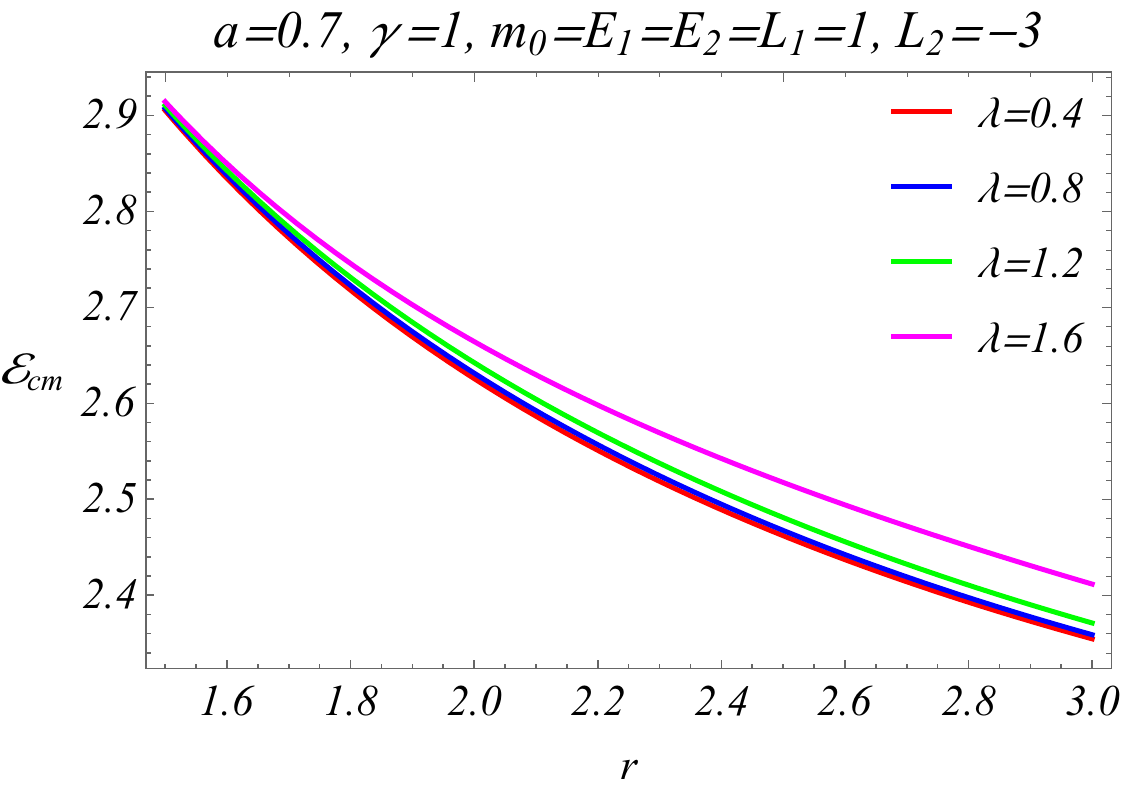}}~~~~~
\subfigure{
\includegraphics[width=0.41\textwidth]{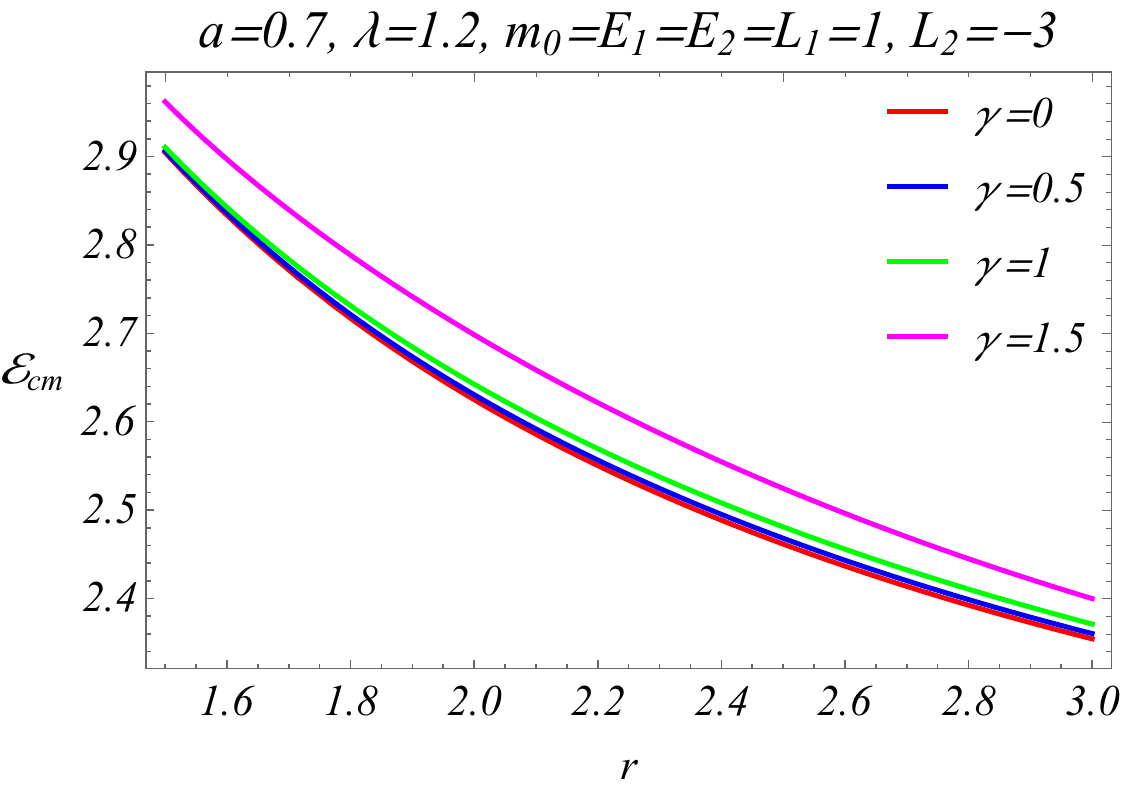}}
\end{center}
\caption{Plots for the CME of colliding particles outside a rotating BH in KR gravity.} \label{RCME}
\end{figure}

\section{Collision of Massive Neutral Particles}
So far we have discussed the motion of massless particles around the BHs in KR gravity. In this section, we will investigate the CME gained by the collision of two uncharged massive particles. Although, we can calculate CME of photons and charged particles, however, we will focus only on the neutral particles with non-zero mass. As we know that the particles possess energies and on collision some percentage of energies are interchanged and the other part is lost. It is complicated to calculate the resultant energy of the particles after collision. Therefore, we calculate the CME of the system after the collision. Let us consider two neutral particles with masses $m_1$ and $m_2$ initially at rest moving towards the BH from infinity in the equatorial plane with the condition $\big(\theta=\frac{\pi}{2}, \dot{\theta}=0\big)$. These particles have energies $E_1$ and $E_2$ and angular momenta $L_1$ and $L_2$. The $4$-momentum of the $i$th particle can be written as
\begin{equation}
p_i^{\mu}=m_iu_i^{\mu}, \label{62}
\end{equation}
where $i=1,2$ and $u^\mu_i$ is the 4-velocity of the $i$th particle. The total 4-momentum of the 2-particle system is given by
\begin{equation}
p_T^{\mu}=p_{(1)}^{\mu}+p_{(2)}^{\mu}. \label{63}
\end{equation}
As we know, the components $p^r$, $p^\theta$ and $p^\phi$ vanish in the frame of center of mass of the particles. Therefore, CME of the system can be taken as \cite{62}
\begin{equation}
\mathcal{E}_{cm}^2=-p_T^{\mu}p_{T\mu}=-(m_1u^{\mu}_{(1)}+m_2u^{\mu}_{(2)})(m_1u_{(1)\mu}+m_2u_{(2)\mu}). \label{64}
\end{equation}
By using the normalization condition $u^\mu_{(i)}u_{(i)\mu}=-1$ in Eq. (\ref{64}), we get
\begin{equation}
\mathcal{E}_{cm}^2=\big(m_1-m_2\big)^2+2m_1m_2\big(1-g_{\mu\nu}u_{(1)}^{\mu}u_{(2)}^{\nu}\big). \label{65}
\end{equation}
For the particles having same masses, we take $m_0=m_1=m_2$. Therefore the CME of the system can be written as
\begin{equation}
\mathcal{E}_{cm}^2=2m_0^2\big(1-g_{\mu\nu}u_{(1)}^{\mu}u_{(2)}^{\nu}\big). \label{66}
\end{equation}
The geodesic equations in Sec. \ref{nullgeod} correspond to the photon with mass $m_p=0$. However, for a massive particle of mass $m_p\neq0$, the geodesic equations become
\begin{eqnarray}
\rho^2\frac{dt}{d\tau}&=&a\big(L-aE\sin^2{\theta}\big)+\frac{r^2+a^2}{\Delta}\big(E\big(r^2+a^2\big)-aL\big), \label{67} \\
\rho^2\frac{dr}{d\tau}&=&\pm\sqrt{\big(E(r^2+a^2)-aL\big)^2-\Delta\big(\mathcal{Z}+(L-aE)^2+m_pr^2\big)}, \label{68} \\
\rho^2\frac{d\theta}{d\tau}&=&\pm\sqrt{\mathcal{Z}+\cos^2{\theta}\big(a^2E^2-L^2\csc^2{\theta}-m_pa^2\big)}, \label{69} \\
\rho^2\frac{d\phi}{d\tau}&=&\big(L\csc^2{\theta}-aE\big)-\frac{a}{\Delta}\big(aL-E\big(r^2+a^2\big)\big). \label{70}
\end{eqnarray}
We can obtain the 4-velocities of the particles moving around the BH in KR gravity from the above geodesic equations. For the particles moving in the equatorial plane, we get $\mathcal{Z}=0$. By considering $a=0$ for the static BH and setting $m_p=1$ in the above geodesic equations, the 4-velocity components $u^{\mu}_i=\big(\dot{t}_i,\dot{r}_i,0,\dot{\phi}_i\big)$ take the form
\begin{equation}
u^{\mu}_i=\bigg(\frac{E_i}{f(r)},\sqrt{E_i^2-f(r)\bigg(1+\frac{L_i^2}{r^2}\bigg)},0,\frac{L_i}{r^2}\bigg), \label{71}
\end{equation}
whereas for the rotating BH, the $4$-velocity components become
\begin{eqnarray}
\dot{t}_i&=&\frac{r^2+a^2}{r^2\Delta}\big(E_i\big(r^2+a^2\big)-aL_i\big)+\frac{a}{r^2}\big(L_i-aE_i\big), \label{72}\\
\dot{r}_i&=&\frac{1}{r^2}\sqrt{\big(E_i(r^2+a^2)-aL_i\big)^2-\Delta\big(\big(L_i-aE_i\big)^2+r^2\big)}, \label{73}\\
\dot{\phi}_i&=&\frac{a}{r^2\Delta}\big(E_i\big(r^2+a^2\big)-aL_i\big)+\frac{L_i-aE_i}{r^2}. \label{74}
\end{eqnarray}
The plots for CME $\mathcal{E}_{cm}$ versus $r$ for two neutral and massive colliding particles near the static BH in KR gravity have been depicted in Fig. \ref{SCME}. The upper panel corresponds to the fixed values of $\lambda$ and $\gamma$. Whereas, in the lower panel, one parameter in each plot is fixed and the different values of other parameter correspond to each curve. The upper left plot shows that if two neutral particles of equal masses and hence equal energies collide under the given values of $\gamma$, $\lambda$ and $L_1$, the CME increases as the particles approach towards the central singularity. The CME also increases with increase in $L_2$. It can be seen that the CME rises drastically near the Cauchy horizon as compared to the rise near the event horizon. The upper right plot shows the similar kind of results as compared to the left plot. It shows that as the mass of particle 2 increases, the CME also increases. However, in the left plot, as we go away from the event horizon, the difference between CME reduces for various curves. Whereas, the variation in CME is almost constant as $r$ increases outside the event horizon in the right plot. In the lower panel, for particles with equal masses and different angular momenta, the CME increases with decreasing $r$. The CME also increases with increase in $\lambda$ and $\gamma$ in each plot at a fixed value of $r$.

Figure \ref{RCME} shows the plots for CME $\mathcal{E}_{cm}$ versus $r$ for two massive and neutral colliding particles near the rotating BH in KR gravity. The parametric values of $a$, $\gamma$ and $\lambda$ in the upper panel have been fixed. Whereas, in the lower panel, $\lambda$ has been varied in the left plot and $\gamma$ has been varied in the right plot corresponding to each curve. The upper left plot shows that for the collision of two particles of equal masses and equal energies without having charge, the CME increases with decreasing $r$ and increasing $L_2$, under the fixed values of $a$, $\gamma$, $\lambda$ and $L_1$. The upper right plot shows that the CME increases with constant rate with rise in the mass of second particle. These two plots show a similar kind of behavior as in the case of static BH with the only difference of the shifting of horizons and the value of CME. In the lower panel, for particles with equal masses, equal energies and different angular momenta, it is seen that the CME increases with decreasing $r$ and increasing $\lambda$ and $\gamma$ in each plot. However, the variation in CME at a fixed $r$ with change in $\gamma$ and $\lambda$ is observed very small.

As we know that KR field is a quantum field closely associated with the String Theory. Therefore, the curved spacetime near the BH in KR gravity may possess quantum aspects and various features of elementary particles. Moreover, any BH behaves as a particle accelerator and therefore, it would be quite useful to study the motion and collision of elementary particles around the BH in KR gravity. Since, we considered massive neutral particles in this study, so the three types of neutrinos, Higgs and Z bosons being massive neutral elementary particles can be considered for the CME analysis and the results are comparable with our findings. One can establish this analysis by only changing the mass of these particles within our study. It may be found that the overall behavior of CME curves will be same as in Figs. \ref{SCME} and \ref{RCME} with the only difference of $\mathcal{E}_{cm}$ and variation between the curves due to change in mass. Furthermore, we found that for the small values of $\lambda$, the horizon and photon sphere is not much affected with the variation of $\gamma$. This is one important property of KR parameters. It may suggest that for these parametric values the spacetime structure near the BH is behaving identical at all radial distances. This is also proved in the case of CME. By considering small value of $\lambda$, there exist no variation in CME curves for different values of $\gamma$. It also means that the spacetime is not affecting the particles in this range of parameters.

\section{Conclusion}
We mainly studied the null geodesics and shadows of the static and rotating BHs in KR gravity and one aspect of timelike geodesics related to CME. A rigorous discussion has been presented on the effect of $a$, $\gamma$ and $\lambda$ on various features and properties of the BHs. We summarize as:

\begin{itemize}
\item As the first result, it is found that as $\gamma$ and $\lambda$ increase, the size of null sphere decreases. For $\lambda\lesssim0.2$, the size of null sphere is constant for all values of $\gamma$. The exactly same behavior is observed in the case of shadow radius. A reciprocal behavior of angular velocity has been identified as compared to the shadow radius. Lyapunov exponent remains constant for $\lambda\lesssim0.2$ for all values of $\gamma$. Then decreases up to a certain value of $\lambda$ and then increases rapidly for all values of $\gamma$.

\item With the increase in $a$, the event horizon reduces and for small values of $\lambda$, the event horizon remains constant for all values of $\gamma$. Then as $\lambda$ grows larger, the event horizon becomes smaller with increasing $\gamma$. The unstable circular null orbits shift towards central object with increase in $a$, $\lambda$ and $\gamma$ around rotating BH in KR gravity.

\item For the observer at $\theta_0=\pi/3$, the flatness in shadows is diminished as compared to the case of equatorial observer as in Ref. \cite{55}. The green curve in bottom right plot corresponds to extremal BH but with diminished flatness. However, an increase in $a$ does cause the oblateness in the shadows. Furthermore, with growing values of $\lambda$ and $\gamma$, the shadow size becomes smaller which is consistent with the results of null sphere and effective potential and for small $\lambda$, the shadow shape and size is constant for all values of $\gamma$. When the observer is shifted to $\theta_0=\pi/6$, the flatness is further diminished.

\item The distortion increases with increase in $\gamma$, $a$ and $\lambda$. However, under the effect of one parameter, the increase in another parameter causes a boost in the distortion. Moreover, as the observer moves away from the equator, the distortion decreases. The BH evaporation rate reduces with increase in $\gamma$, $a$ and $\lambda$ for all cases regardless of observer's position.

\item On comparison of shadow size with EHT data, we found that our BH in KR gravity is more likely to be Sgr A* instead of M87* for the chosen parametric values. It is also found that the CME rises more rapidly as $r\rightarrow0$ for all cases. Moreover, the CME increases with increase in angular momentum $L_2$, mass $m_2$, $\gamma$ and $\lambda$ for both rotating and static BHs. However, near the inner horizon, the growth in CME for static BH is higher than the rotating BH.
\end{itemize}

Finally, we conclude that the BH parameters have a significant influence on various observables and properties studied in this manuscript. All of the parameters, specifically, $\lambda$ and $\gamma$ cause the null orbits and shadows to shrink and therefore causing a slow evaporation rate of BH. For the particular choices of parameters, the BH in KR field is more likely to be Sgr A* instead of M87*. Following our work, there can be some potential future research directions such as the collision of elementary particles and extraction of CME around BHs in KR gravity. Furthermore, the LHC at CERN and other accelerator based labs might possibly verify such results to further open various research directions. As the BH in KR gravity is more likely to be Sgr A* BH, it is quite interesting to study further properties of Sgr A* especially the quantum aspects and then getting verified astrophysically.

\end{document}